\documentclass{emulateapj}
\usepackage{apjfonts}
\newcommand{\scaleup}{\epsscale{1.1}}
\newcommand{\plotter}{\plotone}
\newcommand{\plotterr}{\plotone}
\newcommand{\breaker}{}

\newcommand{\etal}{et al.}
\newcommand{\mbh}{M_{\rm BH}}

\newcommand{\mstar}{M_{\ast}}
\newcommand{\mhalo}{M_{\rm halo}}
\newcommand{\mgal}{M_{\rm gal}}
\newcommand{\msun}{M_{\sun}}
\newcommand{\lstar}{L_{\ast}}

\newcommand{\fgas}{f_{\rm gas}}
\newcommand{\mdyn}{M_{\rm dyn}}

\newcommand{\paperone}{Paper \textrm{I}}
\newcommand{\papertwo}{Paper \textrm{II}}
\newcommand{\paperthree}{Paper \textrm{III}}
\newcommand{\paperfour}{Paper \textrm{IV}}

\shorttitle{Evolution in Spheroid Scalings}
\shortauthors{Hopkins \etal}
\slugcomment{Submitted to ApJ, June 14, 2008}
\begin{document}

\title{Dissipation and Extra Light in Galactic Nuclei: \textrm{IV}.\ 
Evolution in the Scaling Relations of Spheroids}
\author{Philip F. Hopkins\altaffilmark{1}, 
Lars Hernquist\altaffilmark{1},
Thomas J. Cox\altaffilmark{1,2}, 
Dusan Keres\altaffilmark{1}, 
\&\ Stijn Wuyts\altaffilmark{1,2} 
}
\altaffiltext{1}{Harvard-Smithsonian Center for Astrophysics, 
60 Garden Street, Cambridge, MA 02138}
\altaffiltext{2}{W.~M.\ Keck Postdoctoral Fellow at the 
Harvard-Smithsonian Center for Astrophysics}

\begin{abstract}

We develop a model for the physical origin and redshift 
evolution of spheroid scaling relations. 
We consider spheroid sizes, velocity dispersions, dynamical masses, 
profile shapes (Sersic indices), stellar and supermassive 
black hole masses, and their related scalings. 
Our approach combines advantages of prior observational 
constraints in halo occupation models and libraries of 
high-resolution hydrodynamic simulations of galaxy mergers. 
This allows us to separate 
the relative roles of dissipation, dry mergers, formation time, 
and evolution in progenitor properties, and identify their impact on 
observed scalings at each redshift. 
We show that, at all redshifts, 
dissipation is the most important factor determining spheroid sizes 
and fundamental plane scalings, and can account (at $z=0$) for the observed 
fundamental plane tilt and differences between observed disk and 
spheroid scaling relations. Because disks (spheroid progenitors) 
at high redshift have characteristically larger gas fractions, this predicts 
more dissipation in mergers, yielding systematically more compact, 
smaller spheroids. In detail, this gives rise to a mass-dependent evolution 
in the sizes of spheroids of a given mass, that agrees well with observations. 
This relates to a subtle weakening of the tilt of the early-type fundamental 
plane with redshift, 
important for a number of studies that assume a non-evolving 
stellar mass fundamental plane. 
This also predicts evolution in the black hole-host mass relations, 
towards more massive black holes at higher redshifts. 
Dry mergers are also significant, but only for 
large systems which form early -- they 
originate as compact systems, but undergo a number of dry mergers 
(consistent with observations) such that they have sizes at any later 
observed redshift similar to systems of the same mass formed more recently. 
Most of the observed, compact high-redshift ellipticals will become the 
cores of present BCGs, and we show how their sizes, velocity dispersions, 
and black hole masses
evolve to become consistent with observations. We also predict 
what fraction might survive intact from early formation, 
and identify their characteristic $z=0$ properties. 
We make predictions for residual correlations as well: e.g.\ the 
correlation of size and fundamental plane residuals with formation time 
of a given elliptical, that can be used as additional tests of these models. 

\end{abstract}


\section{Introduction}
\label{sec:intro}

Understanding the scaling relations between the photometric and
kinematic properties of galaxy spheroids -- their masses, sizes,
velocity dispersions, and luminosities -- is fundamental to explaining
their origin. Any model which attempts to account for these
correlations must, of course, also determine how they evolve as a
function of redshift. In a $\Lambda$CDM universe, objects grow
hierarchically, implying that ellipticals do not comprise a monolithic
population formed at an early epoch, but have evolved and grown by
mergers from $z\gtrsim6$ to $z=0$.

Observations are beginning to probe
the populations of spheroids at these epochs, and reveal that there is
indeed evolution in such scaling relations \citep[see
e.g.][]{trujillo:size.evolution,mcintosh:size.evolution,
zirm:drg.sizes,vandokkum:z2.sizes}, generally in the sense of massive
ellipticals being more compact at high redshifts.  Some of these
systems, in fact, appear sufficiently compact \citep[see
e.g.][]{zirm:drg.sizes,vandokkum:z2.sizes} that there may be no
similar $z=0$ analogs whatsoever, leading to considerable debate 
regarding the fate of such objects.  Ultimately, the causes of such
evolution and its implication for e.g.\ the fundamental plane of
ellipticals at those redshifts (and today) are not well understood.

Moreover, observations now demonstrate that essentially all massive
spheroids host a supermassive black hole at their centers
\citep{KormendyRichstone95}. The mass of this black hole is tightly
correlated with a variety of structural parameters: the mass
\citep{magorrian}, velocity dispersion \citep{FM00,Gebhardt00}, and
concentration/Sersic index \citep{graham:concentration,graham:sersic}.

Recently, \citet{hopkins:bhfp.obs} and \citet{aller:mbh.esph}
demonstrated that these correlations could be understood in terms of a
``fundamental plane''-like relation, essentially one where 
black hole (BH) mass
tracks the binding energy of the stellar
bulge. \citet{hopkins:bhfp.theory} showed that this is the natural
expectation of theories where some form of feedback from accretion
leads to self-regulated growth of the BH.  This has important
consequences for e.g.\ quasar lightcurves \citep{hopkins:faint.slope}
and lifetimes \citep{hopkins:lifetimes.letter}, as well as for the
effects of ``feedback'' on the host galaxy, possibly necessary for
``quenching'' of star formation to create red elliptical galaxies at
$z=0$
\citep{croton:sam,hopkins:red.galaxies,hopkins:clustering,hopkins:groups.ell}.

Observations suggest that there may be evolution in the black hole
host-galaxy scalings \citep[e.g.][]{peng:magorrian.evolution}, but the
details remain ambiguous \citep[see
e.g.][]{woo06:lowz.msigma.evolution,
salviander:msigma.evolution,lauer:mbh.bias}.  As pointed out in
\citet{hopkins:bhfp.theory}, at least some such evolution should be a
natural consequence of evolution in host spheroid scaling relations,
provided that there is some fundamental quantity or property
traced by black hole mass. Again, if true, this is of basic
importance for models of black hole growth, AGN and quasar activity,
star formation, and galaxy formation and evolution.

A number of important questions therefore arise: If indeed the
observations are correct, why are these high-redshift spheroids
different from their analogs at $z=0$?  What can this tell us about
the process of spheroid formation? Can we use them to constrain and
test models for spheroid formation, quenching, and black hole growth?
Are such objects typical, or are they the result of complex selection
effects? What happens to them by $z=0$? Can we find the remnants or
left-overs of the population today?  Are evolution in spheroid
structural properties related to evolution in their hosted black
holes? And how is this connected to the cosmological process of quasar
fueling and black hole growth?

In a series of papers, \citet{hopkins:cusps.mergers,hopkins:cusps.ell,
hopkins:cores,hopkins:cusps.fp} (hereafter \paperone-\paperfour), 
we combined libraries of hydrodynamic simulations of galaxy merger remnants 
and observations of nearby ellipticals spanning the largest available dynamic 
range in e.g.\ spatial scale, surface brightness, and galaxy properties, 
and developed a methodology by which we could empirically separate 
spheroids into their two dominant physical components. First, a dissipationless 
component -- i.e.\ an ``envelope,'' formed from the violent relaxation/scattering 
of stars that are 
already present in merging stellar disks that contribute to the final 
remnant. Because disks are very extended, with low phase-space density 
(and collisionless processes in a merger cannot raise this phase-space density), 
these stars will necessarily dominate the profile at large radii, hence 
the envelope, with a low central density. 

Second, a dissipational component -- i.e.\ a dense central relic of
starbursts from gas that has been brought into the remnant, lost its
angular momentum, and turned into stars in a compact central starburst
similar to those observed in e.g.\ local ULIRGs (although these would
represent the most extreme cases).  Because gas can radiate, it can
collapse to very high densities, and will dominate the profile within
radii $\sim0.5-1\,$kpc, accounting for the high central densities in
ellipticals. The gas will probably reflect that initially brought in
from merging disks, but could in principle also stem from new cooling or
stellar mass loss in the elliptical \citep[see
e.g.][]{ciottiostriker:recycling}.  In subsequent mergers the two
components will act dissipationlessly (they are just stars and dark
matter), but the segregation between the two is sufficient that they
remain distinct even after multiple major dissipationless or ``dry''
re-mergers: i.e.\ one can still, in principle, distinguish the dense
central stellar component that is the remnant of the combined
dissipational starburst(s) from the less dense outer envelope that is
the remnant of low-density disk stars.

In \paperone\ we developed a methodology to empirically separate these components 
in observed systems with data of sufficient quality, and tested this on 
observed samples of nearby merger remnants from \citet{rj:profiles}. 
Using e.g.\ comparison with stellar populations, colors, and other properties 
as well as direct comparison of simulations with observed surface brightness profiles, 
galaxy shapes, and kinematics, we confirmed that this could reliably extract 
the dissipational component of the galaxy \citep[see also][and references therein]{jk:profiles}. 
In \papertwo\ we extended this to 
observed ellipticals with central ``cusps'' and in \paperthree\ to those with central 
``cores'' (believed to have undergone subsequent dry mergers, but as we argue, 
still retaining evidence for the initial dissipational/dissipationless stellar 
components), from \citet{jk:profiles} and \citet{lauer:bimodal.profiles}. 
In \paperfour, we used these results to argue that the relative 
mass fraction in the dissipational component -- i.e.\ the total mass fraction of 
the elliptical built up dissipationally, as opposed to brought in as stars in 
stellar disks, was the most important parameter determining the sizes 
and densities of ellipticals at $z=0$. We showed that this fraction depends 
systematically on mass: low mass ellipticals experienced more dissipation 
than high-mass ones (presumably reflecting the well-established 
observational fact that low-mass disks are more gas-rich than 
higher-mass disks). We then empirically 
demonstrated that this systematic dependence is 
both necessary and sufficient to explain e.g.\ the difference between  
the size-mass relation of ellipticals and that of their (ultimate) progenitor disks, 
and to explain the observed ``tilt'' in the fundamental plane relation of spheroids. 

There are, however, limitations in our ability to theoretically model these 
processes using idealized simulations and to understand them 
observing only $z=0$ ellipticals. Many galaxies are expected to have had complex 
merger histories (perhaps forming early in a rapid series of multiple, highly dissipational 
mergers, rather than a single disk-disk merger at late times), 
which might, in principle, ``smear our'' or bias some of these comparisons or 
systematically alter some predictions 
\citep[see e.g.][]{kobayashi:pseudo.monolithic,naab:etg.formation}. 
Moreover, the gas content in pre-merger disks (and possibly 
in ellipticals themselves) is both expected and observed 
\citep[e.g.][]{erb:lbg.gasmasses} to evolve strongly as a function of redshift. 
The disks building up ellipticals at $z\gtrsim2$ are much more gas 
rich, so their remnant ellipticals at these times would necessarily be more dissipational, 
and as such might have different structural properties and obey different 
scaling relations from their $z=0$ counterparts. 
It is therefore of particular interest to combine our empirical understanding of the 
role of dissipation from local samples, and our inferences from idealized simulations, 
with fully cosmological models that can account for the evolution in elliptical progenitors 
and merger histories with redshift. 

Some initial attempts at studying these correlations have been 
made using cosmological simulations \citep{naab:etg.formation,dimatteo:cosmo.bhs,
sijacki:radio,croft:cosmo.morph.density}.
It is, unfortunately, prohibitively expensive to simultaneously resolve the 
$\sim100\,$pc 
scales necessary to meaningfully predict e.g.\ spheroid scaling 
lengths, black hole masses, and central velocity dispersions 
in a fully cosmological simulation which would also contain a volume 
capable of modeling a large number of galaxies up to masses 
$\sim10^{12}\,\msun$. Moreover, disk formation remains an 
unsolved problem in such simulations, with the uncertain physics of 
e.g.\ star formation and feedback probably important to form disks with 
the appropriate scale lengths, thicknesses, and bulge-to-disk ratios 
for their masses \citep{robertson:cosmological.disk.formation,
donghia:disk.ang.mom.loss,governato:disk.formation,ceverino:cosmo.sne.fb,
zavala:cosmo.disk.vs.fb}. 
If initial disks do not have the appropriate observed 
structural properties, then even a perfect model for spheroid formation will 
have severe systematic inaccuracies. 

In order to get around some of these limitations, alternative attempts 
have been made to instead predict scaling resolutions 
from semi-analytic models of galaxy formation \citep{khochfar:size.evolution.model,
almeida:durham.ell.scalings}. 
Although they could not calibrate the details of their predictions with 
numerical simulations, 
\citet{khochfar:size.evolution.model} used this to illustrate, with very simple 
assumptions, that increasing dissipational fractions in ellipticals at 
higher redshift leads to the expectation that spheroids should be more 
compact. \citet{almeida:durham.ell.scalings} explicitly ignored dissipation, 
and as a result found that none of the models they considered could 
reproduce the observed $z=0$ fundamental plane correlations of 
spheroids (nor their redshift evolution). 

Such models, however, potentially suffer from the same problems
regarding the nature of disks (the ``initial conditions''):
uncertainties related to modeling spheroid formation are difficult to
disentangle from prescriptions for e.g.\ cooling, disk growth, star
formation, and feedback.  Moreover, while a few properties can easily
be estimated (in a mean sense) from {\em a priori} analytic arguments,
many important properties and physical factors affecting spheroid
evolution -- the role of dissipation in changing the shape and
velocity structure of a spheroid, the detailed density profiles
(especially at small radii and in extended envelopes) of ellipticals,
and the dark matter to stellar or baryonic mass ratios within various
radii in the galaxy -- require detailed, high-resolution numerical
simulations to robustly model.

One way of circumventing the problems in these cosmological models is
to adopt a halo occupation distribution (HOD) approach, in which
(beginning from a halo+subhalo distribution determined in numerical
simulations) halos are populated with galaxies according to empirical
constraints on their clustering properties. The success of such models
at reproducing the observed distributions at a variety of redshifts,
as a function of various galaxy properties, has now been demonstrated
\citep[see e.g.][]{conroy:monotonic.hod}.  This allows us to begin
with the empirical knowledge of where galaxies lie, and what their gas
properties are, without reference to some (potentially incomplete)
{\em a priori} model of e.g.\ star formation and feedback (which we
are not, in any case, interested in testing here). Using this approach
to identify and follow galaxies and their subsequent mergers, we can
then reference each merger to an idealized hydrodynamic simulation
with similar properties (e.g.\ initial morphologies, gas fractions,
sizes, masses, etc.), and use this to predict what the properties of
the remnant should be. Constructing a sample of such remnant spheroids
at any given redshift is straightforward, and allows us to compare
with observed scaling relations in a direct manner.

In this paper, we adopt this method to combine the advantages of 
high-resolution simulations, semi-analytic models, and observational 
constraints on spheroid progenitors in order to develop 
robust predictions for the evolution of spheroid structure and 
the correlation between black hole mass and host properties. 
We study how e.g.\ the evolution in disk gas fractions, structural properties, 
and the interplay between gas-rich and gas-poor mergers 
and multiple mergers in a hierarchical cosmology relate to  
observed correlations from $z=0-4$.

In \S~\ref{sec:model}, we describe our methodology, including our
cosmological models of merger histories
(\S~\ref{sec:model:cosmology}), empirical models for progenitor disks
(\S~\ref{sec:model:hod}), and library of simulations which we use to
determine remnant properties (\S~\ref{sec:model:sims}). In
\S~\ref{sec:z0}, we consider and compare observations to the
resulting predicted scalings of spheroids at $z=0$, and specifically
highlight the effects of dissipation (\S~\ref{sec:z0:diss}) and dry
mergers (\S~\ref{sec:z0:dry}).  In
\S~\ref{sec:evol} we make a number of predictions and comparisons to
observations regarding how these correlations evolve with redshift,
and again illustrate the effects of dissipation
(\S~\ref{sec:evol:diss}) and dry mergers (\S~\ref{sec:evol:dry}).  In
\S~\ref{sec:track} we follow the history of individual systems forming
at different redshifts to compare them at later redshifts, and discuss
the fates of early-forming galaxies which (initially) reflect these
evolved correlations. Finally, in \S~\ref{sec:track}, we summarize our
results and outline future observational tests.

Throughout, we adopt a WMAP5 
$(\Omega_{\rm M},\,\Omega_{\Lambda},\,h,\,\sigma_{8},\,n_{s})
=(0.274,\,0.726,\,0.705,\,0.812,\,0.960)$ cosmology 
\citep{komatsu:wmap5}, and normalize all observations and models 
shown to these parameters. Although the exact choice of 
cosmology may systematically 
shift the associated halo masses (and high-redshift evolution) 
at a given galaxy mass (primarily scaling with $\sigma_{8}$), 
our comparisons (in terms of stellar mass) are for the most part unchanged (many of these 
differences are implicitly normalized out in the halo occupation approach). 
Repeating our calculations for 
a ``concordance'' $(0.3,\,0.7,\,0.7,\,0.9,\,1.0)$ cosmology or 
the WMAP1 $(0.27,\,0.73,\,0.71,\,0.84,\,0.96)$ and WMAP 3
$(0.268,\,0.732,\,0.704,\,0.776,\,0.947)$ 
results of \citet{spergel:wmap1} and \citet{spergel:wmap3}
has little effect on our conclusions. 
We also adopt a \citet{salpeter64} stellar initial mass function (IMF), and convert all stellar masses 
and mass-to-light ratios accordingly. Again, the choice of the IMF systematically 
shifts the normalization of stellar masses herein, but does not substantially change 
our comparisons. 
All magnitudes are in the Vega system, unless otherwise specified.

\breaker
\section{The Simulations and Cosmological Model}
\label{sec:model}

\subsection{Overview}
\label{sec:model:sims:summary}

The model we will adopt consists of the following steps, summarized here 
and described in detail below: 
\begin{itemize}
\item We construct the halo+subhalo mass function, and populate this with galaxies 
of various masses according to a standard halo occupation model approach. We 
allow this population to vary within the range allowed by observational constraints, 
but find this to be a small source of uncertainty. 
\item We assign un-merged galaxies a scale length and gas fraction 
appropriate for their stellar mass and redshift according to observational 
constraints (with appropriate scatter). 
We also systematically vary assignments to span the range allowed by 
observational constraints, and find this is the dominant source of uncertainty in 
our predictions. 
\item We evolve this model forward in time (following the dark matter) and identify 
mergers. After a merger, the properties of the remnant galaxy are calculated 
as a function of the progenitor properties, according to the results from 
hydrodynamic simulations. 
\item At some later time, we construct a mock catalogue (uniformly sampling the 
mock population in stellar mass) and compare the predictions with 
observed galaxy properties. 
\end{itemize}

The exact implementation 
of the halo occupation model and merger 
rate calculations are outlined in the Appendix, for readers 
interested in reproducing the model calculations. Here we outline the 
relevant physics and model approach, and highlight the calibrations from 
simulations used to predict the properties of merger remnants, as 
well as the important sources of 
uncertainty for those predictions.

\subsection{Cosmological Model}
\label{sec:model:cosmology}

We wish to track merger histories in a cosmologically motivated
manner, in order to follow the growth of spheroids.  As discussed in
\S~\ref{sec:intro}, it remains prohibitive to simulate the relevant
hydrodynamic processes (with the necessary $\lesssim100\,$pc
resolution) in full cosmological simulations (where, in order to probe
the massive galaxies of interest here, we would require $\sim{\rm
Gpc^{3}}$ box sizes).  In principle, we could begin with a full
semi-analytic model: attempt to describe, in an {\em a priori} manner,
the entire process of cooling, disk growth, star formation, and
feedback.  This has been done by
\citet{khochfar:size.evolution.model}, for example, as well as
\citet{almeida:durham.ell.scalings}, and we compare with their results
below. We have also experimented with this methodology applied to the
approach of \citet{somerville:new.sam}, and find qualitatively similar
results.

However, a full semi-analytic model necessitates an attempt to predict
e.g.\ the star formation histories, gas fractions, disk galaxy
populations, and clustering of galaxies in an {\em a priori} manner.
Although these are certainly worthy of investigation, they are not the
quantities we are interested in here -- therefore attempting to model
e.g.\ disk formation, star formation, and stellar winds in this manner
will introduce additional uncertainties and dependences. 

For example,
\citet{khochfar:size.evolution.model} find that their most massive
galaxies are predicted to evolve very rapidly. However, their analysis
results in more rapid gas fraction evolution than observed, a problem
common to semi-analytic models without strong feedback \citep[see
e.g.][]{somerville:sam}.  In fact, over the redshift range $z\sim0-2$,
progenitor disk galaxy properties (at least the broad properties of
greatest interest to us here, namely their sizes (or the Tully-Fisher
relation) and gas fractions) are reasonably well constrained
empirically, but theory has had mixed success at best in reproducing
them from a fully {\em a priori} framework. If current {\em a priori}
models are incorrectly describing the cooling of gas and disk
formation (in particular the formation of massive galaxies), then the
predicted properties of ellipticals formed in mergers of those
progenitors are suspect.

We therefore adopt a halo occupation approach -- essentially beginning
with the observational constraints on e.g.\ disk gas fractions and the
Tully-Fisher relation, and predicting the properties of spheroids from
these empirically constrained progenitors. For our purposes, we are
not concerned with why a given disk population is e.g.\ gas-rich or
gas-poor, or has a particular effective radius -- we need those
numbers to compute the properties of the remnant of a major merger of
such disks.  This allows us to be as conservative as possible, and to
identify the robust prediction for spheroid scaling laws without
reference to uncertainties in e.g.\ disk formation models.  We can
(and do) include the uncertainties in the observational constraints in
our modeling, but we find that these are unimportant.

The details of our scheme are described in \citet{hopkins:groups.qso}
and \citet{hopkins:groups.ell}, but we briefly review them here. For
a particular cosmology, we identify the halo and subhalo populations,
and associate each main halo and subhalo with a galaxy in a Monte Carlo
fashion. Specifically: at a given redshift, we calculate the halo
mass function $n(\mhalo)$ for our adopted cosmology following
\citet{shethtormen}. For each halo, we calculate the (weakly mass and
redshift dependent) subhalo mass function (or distribution of
subhalos, $P[N_{\rm subhalo}\, | \, M_{\rm subhalo},\ \mhalo]$)
following \citet{zentner:substructure.sam.hod} and
\citet{kravtsov:subhalo.mfs}. Alternatively, we have adopted it
directly from \citet{gao:subhalo.mf,nurmi:subhalo.mf} or calculated it
following \citet{vandenbosch:subhalo.mf,valeostriker:monotonic.hod},
and obtain similar results: the uncertainties at this stage are
negligible. Note that the subhalo masses are defined as the masses
upon accretion by the parent halo, which makes them a good proxy for
the hosted galaxy mass \citep{conroy:monotonic.hod} and removes the
uncertainties owing to tidal stripping.

We then populate the 
central galaxies in each halo and subhalo 
according to an empirical halo occupation model. 
These empirical models determine both the mean stellar mass and dispersion in stellar masses of 
galaxies hosted by a given halo/subhalo mass $P(\mgal\,|\,M_{\rm subhalo})$, 
from fitting the observed galaxy correlation functions as a function of e.g.\ scale and 
stellar mass or luminosity at a given redshift. 
Other properties, such as e.g.\ the sizes and gas fractions of galaxies, are determined 
as a function of their stellar mass (and, according to our adopted model, merger 
history) as described in \S~\ref{sec:model:hod}-\ref{sec:model:sims:wet} below. 

Although such models are constrained, by definition, to reproduce the mean 
properties of the halos occupied by galaxies of a given mass/luminosity, there 
are known degeneracies between parameterizations that give rise to 
some (usually small) differences between models. 
We therefore
repeat all our calculations for our ``default'' model 
\citep[for which we follow the empirical constraints in][]{conroy:monotonic.hod} 
\citep[see also][]{valeostriker:monotonic.hod} and 
an alternate halo occupation model 
\citep{yang:clf} \citep[see also][]{yan:clf.evolution,zheng:hod}, which 
bracket the range of a number of calculations 
\citep[e.g.,][]{cooray:highz,cooray:hod.clf,zheng:hod,vandenbosch:concordance.hod,
brown:hod.evol} 
and direct observations of groups \citep{wang:sdss.hod}. 

We have also compared a variety of prescriptions for the 
redshift evolution of various components in the halo occupation model: 
we have adopted that directly fitted by the authors above at various redshifts, 
we have considered a complete re-derivation 
of the HOD models of \citet{conroy:monotonic.hod} and 
\citet{valeostriker:monotonic.hod} 
at different redshifts from each of the the observed mass functions of 
\citet{fontana:highz.mfs,bundy:mfs,borch:mfs,blanton:lfs} \citep[see][]{hopkins:groups.qso}, 
and have also found similar results assuming 
no evolution in $P(\mgal\,|\,M_{\rm subhalo})$ (for star forming galaxies). 
Indeed, a number of recent 
studies suggest that there is very little evolution in halo occupation 
parameters (in terms of mass, or relative to $L_{\ast}$) with 
redshift \citep{yan:clf.evolution,cooray:highz,
conroy:monotonic.hod,brown:hod.evol}, or equivalently that the masses of galaxies hosted in a 
halo of a given mass are primarily a function of that halo mass, not 
of redshift \citep{heymans:mhalo-mgal.evol,
conroy:mhalo-mgal.evol}. This appears to be especially true for 
star-forming galaxies \citep{conroy:mhalo-mgal.evol}, unsurprising 
given that quenching is not strongly operating in those systems to change 
their mass-to-light ratios, but it also appears to be true for 
red galaxies at least at moderate redshifts \citep{brown:hod.evol}.

Consequently, 
the differences between any of these choices are 
small (at least at $z\lesssim3$), and negligible compared to 
the uncertainties in our predictions from e.g.\ the uncertainties in disk 
sizes and gas fractions as a function of their stellar mass and redshift. 
Systems which have not undergone a major merger in the model are assumed 
to be disk-dominated, and we can populate them either according to a 
halo occupation model that treats all galaxies in an identical fashion, or as 
blue/star forming galaxies in a model where the constraints are derived 
for blue and red galaxies separately (we find it makes little or no difference). 

Finally, having populated a given halo and its subhalos 
with galaxies, we follow the evolution of those halos forward and identify 
major mergers\footnote{In a major merger, tidal forces are
sufficiently strong to drive nuclear inflows of gas and build
realistic spheroids.  The precise meaning of major merger in this
context is blurred by a degeneracy between the progenitor mass ratio
and the orbit
\citep{hernquist.89,hernquist.mihos:minor.mergers,bournaud:minor.mergers},
but both numerical \citep{younger:minor.mergers} 
and observational
\citep{dasyra:mass.ratio.conditions,woods:tidal.triggering} studies
indicate that massive inflows of gas and morphological transformation
are typical for mass ratios only below $\sim 3:1$.  This is ultimately
related to the non-linear scaling of the disk response to a merger
of a given mass ratio \citep{hopkins:disk.heating}.
Unless otherwise
noted, we generally take the term ``mergers'' to refer to major
mergers.}. The details of the treatment of subhalo-subhalo and subhalo-halo 
mergers are described in \citet{hopkins:groups.qso} and \citet{hopkins:groups.ell}, 
but in short the halo merger timescale is straightforward and well defined. When 
two subhalos are fully merged, we can either assume the galaxies have merged, 
or allow for some additional merger timescale within the merged halo. 
We have experimented with a variety of models for this, including e.g.\ the 
dynamical friction timescale and alternative timescales calibrated from 
simulations or calculated based on 
group capture or collisional cross section estimates and angular 
momentum (orbital cross section) capture estimates. But because these 
merger times within a halo are always much less than the Hubble time 
at the relevant redshifts, and we are interested only
in statistical properties across populations evolving over a Hubble time, 
it makes no difference to our calculations what choice we adopt. 

When galaxies merge, we estimate the properties of the remnant 
based on those of the progenitor galaxies, according to our 
numerical simulations, as described in \S~\ref{sec:model:sims}. 
This approach yields reasonable agreement with global quantities such as the 
spheroid mass function, the mass density of ellipticals as a function of 
mass, and the fraction of early-type galaxies as a function of mass 
\citep{hopkins:groups.ell}. 
Again, despite the assumptions involved thus far, we 
stress that the detailed choice of halo occupation model 
yields negligible differences in progenitor galaxy properties and in our predictions 
at all the masses and redshifts of interest, where the clustering and abundances 
of massive galaxies are reasonably well-constrained.

\subsection{Progenitors: Laying Galaxies Down}
\label{sec:model:hod}

Given a host halo mass and disk stellar mass, both determined 
from the halo occupation model, the two important parameters we must
assign to each disk are a size (effective radius $R_{e}$) and gas
fraction ($f_{\rm gas}\equiv M_{\rm gas}/(M_{\ast}+M_{\rm
gas})$).\footnote{Note that together the disk size, baryonic and halo
mass define another parameter of interest, the disk circular velocity
or maximum velocity; but this does not explicitly enter into the way
in which we calculate the properties of the remnant.}  This is where
the dominant uncertainties in our modeling arise -- we will show that
e.g.\ the gas fraction in pre-merger disks is an important parameter
determining the characteristics of the remnant, so we can only predict
those properties to within the uncertainties in disk gas fractions at
the appropriate redshifts.

To minimize the uncertainties in modeling this, 
we therefore begin with an empirical estimate of the gas fraction distribution in 
spiral galaxies as a function of mass and redshift. At $z=0$, there is 
no significant 
uncertainty; various measurements 
\citep[e.g.][]{kennicutt98,belldejong:disk.sfh,kannappan:gfs,mcgaugh.tf.old,
mcgaugh:tf} and indirect constraints 
from star formation and the baryonic Tully-Fisher relation \citep{bell:baryonic.mf,
noeske:sfh,calura:sdss.gas.fracs} give 
consistent estimates of the median gas fraction in spirals as a function of their 
stellar mass, and the scatter at each mass. We obtain identical results if we 
use any of these constraints (or the data points themselves), but 
note for convenience that all of these observations can be well-fitted by the trend 
\begin{equation}
\langle\fgas(z=0)\rangle \approx \frac{1}{1+(M_{\ast}/10^{9.15} M_{\sun})^{0.4}}
\label{eqn:fgas.z0}
\end{equation}
with a constant $\sim0.2-0.25$\,dex scatter at each mass. 

Similar, albeit less robust, estimates exist at 
$z=1$ and $z=2$ \citep[see e.g.][respectively]{shapley:z1.abundances,erb:lbg.gasmasses}. 
We therefore adopt a simple functional form that interpolates between these 
measurements. Motivated by the power-law form of the observed Kennicutt-Schmidt 
star formation relation ($\dot{M}_{\ast}\propto \Sigma_{\rm gas}^{1.4}$), 
we adopt the following interpolation formula: 
\begin{equation}
\langle\fgas(z)\rangle = \fgas(z=0)\,{\Bigl [}1+\tau_{\rm LB}(z)\,{(1-\fgas(z=0)^{\beta})} {\Bigr]}^{-2/3}
\label{eqn:fgas.z}
\end{equation}
where $\beta \approx 3/2$ and $\tau_{\rm LB}(z)$ is the fractional 
lookback time to redshift $z$. 
This can be derived, for example, by assuming a simple model where the star formation 
rate scales according to the observed relation, the disk scale length varies as a 
power of the baryonic mass, and the net mass accretion rate (inflows minus 
outflows) is also a power-law 
function of the mass of the system; given the requirement that the system begin 
at an initial time $t=0$ with $f_{\rm gas}=1$ and have the observed $f_{\rm gas}$ at 
$z=0$. 

In any case, the values above provide a good fit to the observations at $z=0-2$ 
\citep{belldejong:disk.sfh,shapley:z1.abundances,erb:lbg.gasmasses} 
and a reasonable approximation to results of cosmological simulations 
\citep{keres:hot.halos,keres:prep} and semi-analytic models \citep{somerville:new.sam}, 
and interpolate smoothly between $z=0$ and 
high redshift. Lowering $\beta$ systematically weakens the 
implied redshift evolution in $f_{\rm gas}$ -- in order to represent the uncertainty in the 
observations, we consider a range $\beta\sim0.5-2.0$ (bracketing the observational 
errors). Similar prescriptions for $f_{\rm gas}$ evolution can be obtained from 
various toy model star formation histories including e.g.\ commonly adopted 
exponential $\tau$-models \citep[see e.g.][]{belldejong:disk.sfh,noeske:sfh} 
and integration of specific tau models enforcing proportionality between inflow, 
star formation, and outflow rates \citep{erb:outflow.inflow.masses}; these 
are all discussed in more detail in \citet{hopkins:bhfp.theory,hopkins:groups.qso}, 
but they generally yield similar results and lie within the uncertainties we consider 
(differences being much less than the typical observed scatter in $f_{\rm gas}$ 
at each mass and redshift).

The other important property 
we must estimate for progenitor disks is their effective radius. 
At $z=0$, this is well-measured for all disk masses of interest. We 
employ the fits from \citet{shen:size.mass} to late-type galaxy sizes 
as a function of stellar mass (from $\lesssim10^{9}\,\msun$ to 
$\gtrsim10^{12}\,\msun$), and the scatter in those sizes. 
Adopting different local estimates or even completely ignoring this 
scatter makes little difference. Observations are somewhat more 
ambiguous regarding the redshift evolution of sizes, but fortunately 
the observational constraints \citep[e.g.][]{trujillo:size.evolution,barden:disk.size.evol,
ravindranath:disk.size.evol,rix:gems.gal.evol,ferguson:disk.size.evol,
akiyama:lbg.weak.size.evol} and state-of-the-art models 
\citep{somerville:disk.size.evol} suggest that any evolution is relatively weak. 
This is also reflected in e.g.\ the baryonic Tully-Fisher 
relation, which appears to evolve negligibly from 
$z=0-2$ \citep{conselice:tf.evolution,flores:tf.evolution,
kassin:tf.evolution,vandokkum:tf.evolution}. 

We therefore consider two extremes. In the first case, 
disk sizes and the baryonic Tully-Fisher relation are completely 
fixed with redshift, according to their $z=0$ values. In the second, 
disk sizes evolve according to observational estimates 
from the various measurements above. A simple power law
\begin{equation}
R_{e,\rm disk}[\mstar\, | z] = (1+z)^{-\beta_{d}}\, R_{e,\rm disk}[\mstar\, | z=0] 
\label{eqn:rdisk.evol}
\end{equation}
where $\beta_{d}\approx0.4$ provides a good fit to the observations and 
is consistent with all of the measurements, spanning the
redshift range $z=0-3$. We adopt this estimate, but obtain a similar 
result if we use the more detailed cosmologically motivated model 
in \citet{somerville:disk.size.evol}, which allows disks to form conserving 
specific angular momentum from halo gas \citep[see][]{momauwhite:disks}, 
yielding (approximately) $R_{d}\propto R_{\rm vir}/c$, where 
$c\propto1/(1+z)$ is the halo concentration -- although $R_{\rm vir}$ is smaller 
for halos of a given mass at high redshift, they are less concentrated, yielding 
a weak size evolution similar to that observed 
\citep{bullock:concentrations,
wechsler:concentration,neto:concentrations,comerford:obs.concentrations,
buote:obs.concentrations}. 

Together this gives us an estimate of our input disk parameters, and we allow for 
the described range in both the estimated gas fraction and disk size evolution. 
At low redshift, this introduces little uncertainty (most spheroids at $z=0$ 
have assembled relatively recently, so their structural properties reflect those of 
low-redshift disk progenitors, where the uncertainties are minimal and the relevant 
sizes, etc.\ can be directly measured from observations). However, these uncertainties 
begin to grow rapidly at $z\gtrsim 2-3$, where 
there are no more direct observational constraints. 
Moreover, at these redshifts, it is not clear that ``un-merged'' galaxies will 
necessarily be disks analogous to those well-understood at $z<2$ 
\citep[see e.g.\ the clumpy morphologies and increased dispersions 
observed in][]{reddy:z2.lbg.spitzer,flores:tf.evolution,
bournaud:chain.gal.model}, although higher-resolution observations suggest that 
at least a significant fraction of this population does exhibit 
regular rotation and smoother morphologies (albeit most likely still  
puffier and more dispersion-supported than low-redshift disks) in the rest-frame 
optical \citep{puech:highz.vsigma.disks,akiyama:lbg.weak.size.evol,
genzel:highz.rapid.secular,shapiro:highz.kinematics}. 

In either case, the uncertainties as this limit is approached 
can be considered part of the uncertainty in e.g.\ progenitor sizes and 
structural properties, as observational estimates of this quantity often 
do not make strict morphological cuts at these redshifts but rather 
assign gas-rich star-forming systems to a single category 
(and this is how such uncertainties would enter into our 
model). In this aspect, the uncertainties from assigning them implicit ``disk'' properties 
may actually be less than one might expect, because at the point where 
a galaxy is very gas-rich (increasingly 
common at these high redshifts), it does not matter what initial configuration this 
gas is in (whether e.g.\ clumpy, filamentary, or infalling in direct collapse), as 
it is dissipational and in any case will (in mergers) lose angular momentum 
and form a dense central starburst. 
In any case, these uncertainties lead us to limit our predictions to $z<4$. 
Fortunately, because at any lower redshift most spheroids have assembled 
relatively recently, with decreasing mass fractions contributed by 
galaxies assembled at very early times, the uncertainties associated with 
the lack of constraints at these high redshifts rapidly become 
less important to our predictions.

\subsection{The Simulations}
\label{sec:model:sims}

In order to determine the effects that a given merger will have on galaxy properties, we 
require some estimates of how e.g.\ the gas content of initial galaxies (and other properties) 
translates into properties of the remnant. 
We derive these prescriptions from a large library of hydrodynamic 
simulations of galaxy encounters and mergers, described in detail 
in \citet{robertson:fp,cox:kinematics,younger:minor.mergers} and \paperone. 
These amount to several hundred unique simulations, spanning a wide 
range in progenitor galaxy masses, gas fractions, orbital parameters, 
progenitor structural properties (sizes, concentrations, bulge-to-disk ratios), 
and redshift. 

Most of the simulations are major (mass ratios 
$\sim1:3$-$1:1$), binary encounters between 
stellar disks (which provide an idealized scenario useful for calibrating 
how details of the remnant depend on specific progenitor properties), 
but they include a series of minor mergers (from mass ratios $\sim1:20$ to 
$\sim1:3$), as well as spheroid-spheroid ``re-mergers'' or ``dry mergers'' (i.e.\ mergers of the 
elliptical remnants of previous merger simulations), 
mixed-morphology (spiral-elliptical) mergers \citep[see also][for a 
detailed study of these mergers]{burkert:mixed.morph.boxiness,
johansson:mixed.morph.mbh.sims}, multiple 
mergers, and rapid series of hierarchical mergers. 
Our adopted prescriptions are robust to these choices. 
The simulations usually include accretion and feedback from supermassive 
black holes, as well as feedback from supernovae and stellar winds. 
However, we have performed parameter studies in these 
feedback prescriptions, and find that the 
structural properties of interest here are relatively insensitive to 
these effects \citep{cox:winds,hopkins:bhfp.theory,hopkins:cusps.mergers}.

The predicted simulation scalings are discussed in detail 
in \paperone-\paperthree, where we 
take advantage of high-resolution observations of local merger remnants 
\citep[from][]{rj:profiles} as well as both nuclear cusp and core ellipticals 
\citep[from][]{lauer:bimodal.profiles,jk:profiles} to test them observationally. 
We find good agreement between the predicted and observed scalings 
with e.g.\ stellar mass and gas content (of the progenitor galaxies) 
at the time of the merger. If, as is commonly believed, core ellipticals 
are the product of spheroid-spheroid re-mergers, then 
our comparisons in \paperthree\ also confirm the scalings 
derived from simulated re-mergers.

\subsubsection{Mergers without Gas}
\label{sec:model:sims:dry}

In mergers, the stars and dark matter form a collisionless system, and hence
gas free (dissipationless or dry) mergers 
are particularly easy to handle. Furthermore, direct comparison 
with our simulations suggests that, to the desired accuracy, 
it is a good approximation to treat the dissipational and dissipationless 
components of a merger separately, allowing us to apply simple rules to the 
{\em stellar} remnant 
in mergers of both gas-rich disks and gas-poor spheroids. 

It is well established in numerical simulations that, to lowest order, the 
effect of any merger of dissipationless components 
is to ``puff them up'' by a uniform factor, 
while roughly conserving profile shape \citep[at least for 
certain ``equilibrium'' profile shapes such as 
the \citet{nfw:profile}, \citet{hernquist:profile}, or \citet{devaucouleurs} profiles; 
see e.g.][]{barnes:disk.disk.mergers,boylankolchin:mergers.fp,boylankolchin:dry.mergers,
hopkins:cores}. 
The conservation of both energy and phase space 
density means that, modulo a small normalization offset owing to possibly 
different profile shapes 
the remnant of e.g.\ two initial stellar disk or spheroid mergers will have a similar final 
effective radius.

In detail, if we temporarily ignore the 
halos of the galaxies (a good approximation in most simulations, since 
the specific binding energy of the galaxy baryonic mass is much larger than 
that of the halo) and consider isotropic systems, we obtain for the parabolic merger 
of systems of mass $M_{1}$ and $M_{2}=f\,M_{1}$ ($f\le1$) the energy conservation equation 
\begin{equation}
E_{f} = k_{f}\,(M_{1}+M_{2})\,\sigma_{f}^{2} = E_{i} = k_{1}\,M_{1}\,\sigma_{1}^{2} + 
k_{2}\,M_{2}\,\sigma_{2}^{2}
\label{eqn:egy}
\end{equation}
where $k$ is a constant that depends weakly on profile shape \citep[for greater 
detail, see][]{ciotti:dry.vs.wet.mergers}. If profile 
shape is roughly preserved and $R\propto M/\sigma^{2}$, 
then we expect 
\begin{equation}
R_{f} = R_{1}\,\frac{(1+f)^{2}}{(1+f^{2}\,\frac{R_{1}}{R_{2}})}.
\label{eqn:re.dry}
\end{equation}
A merger of two perfectly identical spheroids 
will double $R_{e}$ and conserve $\sigma$. 
It is straightforward to solve the appropriate energy conservation equation 
numerically, allowing for arbitrary profile shapes and 
for halo components (assuming the halo profile follows 
\citet{nfw:profile} or \citet{hernquist:profile} profiles), but in practice we find that 
these subtleties make no almost no difference compared to
using Equation~(\ref{eqn:re.dry}). 

In a merger of two stellar disks, violent relaxation will transform the 
stellar distribution from initially exponential disks into 
\citet{devaucouleurs}-like profiles, more generally \citet{sersic:profile} profiles (of the form 
$I_{e}\propto \exp{[-(r/r_{0})^{1/n_{s}}]}$ with $n_{s}\sim2.5-3.5$ 
after a single major disk-disk merger
\citep[see \paperone-\papertwo\ and][]{naab:profiles}, 
while the halo profile shape is approximately 
preserved \citep[\paperfour,][]{boylankolchin:mergers.fp}. These trends in simulations can both 
be roughly explained by
allowing the merger to scatter stars and dark matter 
(in addition to uniformly puffing up the 
profile) in their final three-dimensional radii by some lognormal broadening 
factor $\sigma_{r}\sim0.3-0.4\,$dex (in a $1:1$ merger). In other words, 
while the median final (post-merger) effective radius of stars at some initial radius $r_{i}$ 
is given by the uniform puffing up or stretching of 
the profile in Equation~(\ref{eqn:re.dry}), 
there is a lognormal scatter with dispersion $\sigma_{r}$ in the final radii of stars 
from that initial radius. 
Over the relevant dynamic range for 
most practical observational purposes (from $\sim 0.01\,R_{e}$ to $\gtrsim10\,R_{e}$, 
or as much as 15 magnitudes in surface brightness), 
this will effectively transform an initial exponential profile into an 
$n_{s}\approx3$ ($\sigma_{r}=0.3$) or $n_{s}=4$ ($\sigma_{r}=0.4$) 
profile. 

Subsequent spheroid-spheroid or spheroid-disk mergers continue to scatter stars out 
to larger radii, building an extended envelope of material, 
and raising the best-fit Sersic index of the dissipationless component of the remnant. 
We study this in \paperthree, and find to a good approximation that a re-merger of 
mass ratio $f\leq1$ will raise the $n_{s}$ of the dissipationless component 
by $\Delta n_{s}\sim f$, or equivalently ``broaden'' the three-dimensional stellar 
distribution by a lognormal factor $\sim 0.3\,f$ (the prescription we adopt). At 
$z=0$ we can then determine an ``observed'' Sersic index by projecting this mock 
profile (with that of the dissipational component) and fitting to the system. 
If we consider the Sersic indices directly fitted to our simulations -- 
i.e.\ draw a Sersic index for a disk-disk merger remnant randomly from 
that fitted to our simulated mergers of the same mass ratio, and so on -- we 
obtain similar predictions. 

Finally, having the profile of the dissipationless stellar and halo components, we 
can calculate the velocity dispersion $\sigma$ of stars within some radius 
(here we take the mass-weighted velocity dispersion of stars within 
$R_{e}$, and assume isotropic orbits, although this is not a dominant 
uncertainty in our predictions). 
We can instead (without any reference to the isotropy assumption) 
assume that the dynamical mass estimator, $\sigma^{2}\,R_{e}$, 
is a good tracer of the true enclosed total mass within the stellar effective 
radius $R_{e}$ (modulo a normalization constant that is the same for 
most ellipticals) -- a fact seen in our simulations (\paperfour) and 
observations \citep{cappellari:fp,bolton:fp,bolton:fp.update} -- 
and use the known total mass ($M(<R_{e})$) to predict $\sigma$. 
The results are similar. 

We note that halos are tracked with stars in these processes, 
but between mergers, the halos still grow. We assume at all times 
that the dark matter halo follows a \citet{hernquist:profile} profile (essentially 
identical to an \citet{nfw:profile} profile at the radii that matter for the 
baryonic galaxy, but with finite mass and analytically tractable properties, a 
more useful approximation since we are considering halos as if in isolation), with a 
concentration from the mean concentration-halo mass-redshift 
relation in simulations and observations (see \S~\ref{sec:model:hod}
and \citet{springel:models}). 
We then calculate the halo contribution to quantities such as e.g.\ the 
dynamical mass within $R_{e}$ and $\sigma$ based on this profile. 
However, we could also assume the dark matter within the stellar effective 
radius ``freezes out'' with each major merger (new dark matter within $R_{e}$ 
only being what is brought in by subsequent mergers, within the $R_{e}$ of 
those galaxies at the time of their first merger) -- it makes relatively 
little difference, since the mass densities in the cores of halos do not 
strongly evolve with redshift \citep{bullock:concentrations}.

\subsubsection{Mergers with Gas}
\label{sec:model:sims:wet}

In a gas-rich merger, the gas loses angular momentum and, being dissipative, 
radiates its energy and collapses to the center of the galaxy. The collapse will, 
in general, continue until the gas becomes self-gravitating, at which point it 
rapidly turns into stars as it further contracts \citep[for details, see][]{hopkins:disk.survival}. 
The remnant will 
exhibit a ``dissipational component'' -- the dense central stellar concentration 
which is the relic of these dissipational merger-induced starbursts. 

\citet{covington:diss.size.expectation} consider the dissipation of energy in 
mergers and use it to estimate the effect this will have on the size of the merger 
remnant \citep[see also][]{ciotti:dry.vs.wet.mergers}, 
which we test in \papertwo\ and \paperthree\ and show is a good approximation to 
our simulations and observed systems, and can be reduced 
to the approximation
\begin{equation}
%R_{e}({\rm net}) = \frac{R_{e}({\rm dissipationless})}{1+(f_{\rm gas}/0.1)^{0.8}}, 
%R_{e}({\rm net}) \approx \frac{R_{e}({\rm dissipationless})}{[1+(f_{\rm gas}/0.2)^{1.2}]^{0.8}}, 
R_{e} \approx \frac{R_{e}({\rm dissipationless})}{1+(f_{\rm gas}/f_{0})}, 
\label{eqn:re.wet}
\end{equation}
where $f_{0}\approx0.25-0.30$, $R_{e}({\rm dissipationless})$ is the effective 
stellar radius of the final merger remnant if it were entirely dissipationless, and 
$f_{\rm dissipational}$ is the baryonic 
mass fraction from the starburst. For most major mergers, 
gas in the disks at the time of the 
merger will rapidly lose energy, and star formation is very efficient, 
so $f_{\rm dissipational}$ reflects the gas fractions of the progenitor 
disks immediately before the merger $f_{\rm dissipational}({\rm merger})= 
f_{\rm gas} = M_{\rm gas}({\rm disks})/M_{\ast}({\rm remnant})$ 
(gas blown out by feedback from black holes and stellar winds, even with extreme 
feedback prescriptions, makes little difference here: it predominantly effects $f_{\rm gas}$ 
of the pre-merger disks over much longer timescales, rather than the consumption 
in mergers). We show in \citet{hopkins:disk.survival} that this is not strictly true in 
extremely gas-rich mergers ($f_{\rm gas} > 0.5$), where torques are inefficient at 
driving dissipation, and ellipticals may not be formed from even major mergers: these 
extreme cases however are seen only at the lowest observed masses, and do not affect 
most of our predictions (though we note where this may be important). 

Given the typical exponential-like ($n_{s}\sim1-2$) character of these dissipational components 
when they first form (owing to their formation from infalling gas), we can invert this to 
infer what the effective radius of the dissipational component 
specifically must be (given that effective radius and profile, we 
can then construct a mock profile of the entire galaxy). 
Based on our study of re-merger simulations in \paperthree, we find that 
the two components (those which were originally dissipational at their formation, and 
those which were not -- i.e. which originally came from stellar disks) tend to 
be separately conserved even after several dry or gas-poor spheroid-spheroid re-mergers. 
In a series of mergers, then, it is a good approximation to treat the two separately: 
dissipationless components grow and add as described above. Dissipational components 
form with a mass given by the gas content of merging disks with sizes appropriate for the 
above relation, and then evolve either by gas poor (spheroid-spheroid) merging 
(where they will add with 
other stellar dissipational components formed in earlier mergers, following the rules 
for dissipationless merging above since both components are stellar), or 
by more gas-rich (spheroid-disk) merging (in which case new dissipational mass 
is added given the prescriptions above). 

Dissipation will also fuel growth of the 
central supermassive black hole. Based on our theoretical modeling of 
self-regulated black hole growth in simulations \citep{dimatteo:msigma,
hopkins:faint.slope,hopkins:seyferts,hopkins:bhfp.theory} 
and that of others 
\citep[e.g.][]{silkrees:msigma,murray:momentum.winds, ciottiostriker:recycling}, essentially 
any model where the black hole regulates its growth by halting accretion (allowing 
some fraction of the energy or momentum of accretion to couple to the surrounding 
media) will give rise to a fundamental correlation 
between black hole mass and spheroid binding energy \citep{hopkins:bhfp.theory}, 
more specifically with the binding energy of the gas in the central regions whose 
inflow must be suppressed to regulate black hole growth. 

There is direct evidence for this in the observations as well, with \citet{hopkins:bhfp.obs} 
finding a $\sim3\,\sigma$ preference in the data for a correlation of the 
form $M_{\rm BH}\sim \mstar^{0.5-0.7}\,\sigma^{1.5-2.0}$, similar 
to and consistent with the ``fundamental'' correlation being one with bulge 
binding energy, which is roughly traced by $E_{b}\sim\mstar\,\sigma^{2}$ 
\citep[$M_{\rm BH}\propto E_{b}^{0.71}$, see also][]{aller:mbh.esph}. 
Indirectly, this can then 
explain e.g.\ the observed $M_{\rm BH}-\sigma$ \citep{FM00,Gebhardt00} and 
$M_{\rm BH}-\mstar$ \citep{magorrian} correlations. 

Motivated by our simulations and these observational results, we model black 
hole growth as follows: in each merger, we allow the black hole to grow by an 
amount proportional to the binding energy of 
the new dissipational components of the merger (i.e.\ the starburst gas), in addition 
to summing the masses of the two progenitor black holes \citep[we do not model 
any gravitational kicks that might expel black holes, but theoretical estimates suggest 
that by $z=0$ in the massive galaxies we are interested in here, such effects 
are small if they are present at all;][]{volonteri:xray.counts,volonteri:recoils}. 
We include an intrinsic scatter $\sim0.2\,$dex, again motivated by simulations 
(although there will be some scatter in the resulting correlations regardless, 
reflecting different gas content and potential depth at the time of merger, with 
subsequent evolution to $z=0$). 
The exact proportionality constant 
is a reflection of the physics of feedback efficiencies and 
coupling, but we adopt a constant calibrated in simulations to 
match the normalization of the observed $z=0$ $M_{\rm BH}-\sigma$ relation 
 \citep[][]{dimatteo:msigma,hopkins:lifetimes.methods,hopkins:bhfp.theory}. 
Any other results (e.g.\ the slope 
of these correlations, residual correlations or correlations with other 
structural parameters, and/or redshift evolution in the correlations) 
are genuine predictions of the model, not this calibration.

\breaker
\section{Predicted Scaling Laws at $z=0$}
\label{sec:z0}

\begin{figure}
    \centering
    %\scaleup
    %\plotone{MC_fp_proj.ps}
    \plotone{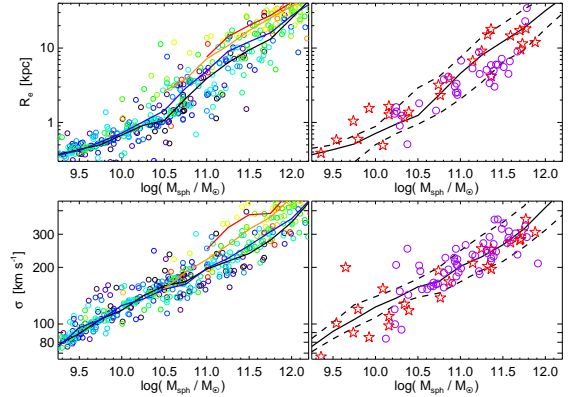}
    \caption{Predicted spheroid scaling laws at $z=0$. {\em Top:} Effective 
    radius as a function of stellar mass. {\em Left:} Our simulated systems are shown 
    (equal numbers of systems per logarithmic interval in mass are simulated and 
    plotted, to represent the full dynamic range), 
    with color encoding the redshift of their last gas-rich merger 
    (from black for $z=0$ to red for $z\gtrsim3$); solid 
    lines show the median trends for systems with 
    this redshift: $z=0-1$ (black), $1-2$ (blue), $2-3$ (orange), and $>3$ (red). 
    {\em Right:} Solid (dashed) line shows the median ($\pm1\sigma$) trend 
    from the simulations. 
    Observed systems 
    from the samples of \citet[][red stars]{jk:profiles} and 
    \citet[][purple circles]{lauer:bimodal.profiles} are shown for comparison ({\em right}). 
    This notation is used throughout. 
    {\em Bottom:} Projected central velocity dispersion as a function of 
    stellar mass.
    \label{fig:fp.proj}}
\end{figure}

\begin{figure}
    \centering
    %\scaleup
    %\plotone{MC_fp.ps}
    \plotone{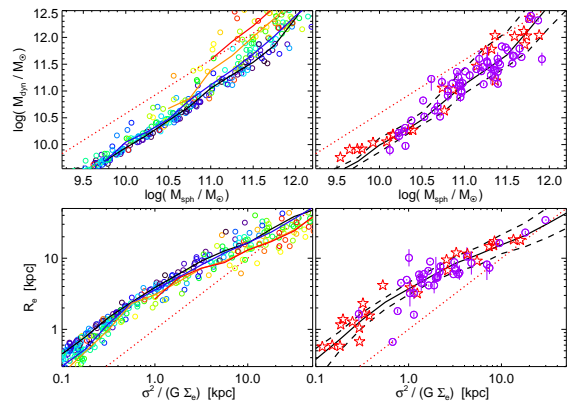}
    \caption{The fundamental plane. {\em Top:} Dynamical mass as a function of 
    stellar mass (style as in Fig. \ref{fig:fp.proj}). {\em Bottom:} Effective radius as a 
    function of $\sigma$ and $\mu$ (here converted to stellar surface density $\Sigma$). 
    Dotted red line shows the virial relation (no tilt in the fundamental plane). 
    \label{fig:fp}}
\end{figure}

Figures~\ref{fig:fp.proj} \&\ \ref{fig:fp} plot the fundamental plane correlations of 
ellipticals at $z=0$, produced by this simple model. Here 
(and in subsequent figures) we draw systems randomly 
from the halo occupation distribution with the restriction that we draw equal numbers 
of simulated systems per logarithmic interval in halo mass (in order to represent 
the full dynamic range of interest in the plots). This differs slightly from selecting 
e.g.\ a volume-limited sample, which would be dominated by low-mass systems 
(less useful for studying the distribution in a parameter at a given mass, 
which is our intention here). 

We compare 
with a large sample of observed ellipticals: specifically the combination of 
the \citet{jk:profiles} compilation of Virgo ellipticals 
and the bright elliptical sample of \citet{lauer:bimodal.profiles} (which 
extends this dynamic range at the cost of slightly higher uncertainties in the data), 
with nuclear {\em HST} observations and ground-based data at large radii 
(allowing accurate surface brightness profile measurements from 
$\sim 50$\,pc to $\sim50$\,kpc)\footnote{The profiles cover sufficient dynamic 
range that effective radii can be determined directly from the integrated surface 
brightness profiles; adopting those from fitted profiles changes the results 
by a negligible amount. Velocity dispersions are compiled by the authors, 
Sersic indices are fitted in \papertwo, and stellar masses are determined from 
the integrated photometry using the color-dependent mass-to-light ratios 
from \citet{bell:mfs}. We have compared the results given 
different observed profiles for the same objects 
and different stellar mass estimators, and find that (in a statistical sense) they 
are unchanged.}. 
The correlations traced by these samples throughout this paper are 
statistically indistinguishable from the best-fit correlations (over the range observed) 
determined from volume-limited samples of (primarily field) early-type galaxies in 
the SDSS, in e.g.\ \citet{shen:size.mass,bernardi:fp,bernardi:bcg.scalings,
gallazzi06:ages,vonderlinden:bcg.scaling.relations}. There may be a slight (not highly 
significant) offset between the data-sets, in the sense that the 
objects in the Virgo sample may have slightly larger $R_{e}$ and smaller $\sigma$ 
than the \citet{lauer:bimodal.profiles} sample 
at fixed mass; if real, this may relate to the former being a cluster sample, 
or to the likely inclusion of a non-negligible S0 population in the latter. 
But in any case, the possible offset between the two is smaller than the uncertainties 
in our modeling (in terms of normalizing e.g.\ absolute sizes of systems, requiring 
accurate priors on their progenitor disk and halo sizes), and within the 
predicted and observed scatter, and so reasonably brackets the range of 
different observations.  (For more details of the observational samples 
and their analysis, we refer to those papers and \papertwo.)

The correlations are accurately reproduced at $z=0$. Specifically, 
the size-mass relation has a steep logarithmic slope $R_{e}\sim \mstar^{0.6}$, compared 
to $R_{e}\sim \mstar^{0.3}$ for the progenitor disks -- the sense of this is specifically 
that low-mass ellipticals are more compact, relative to their progenitor disks. The 
velocity dispersion-mass (Faber-Jackson) relation is also reproduced, although 
this correlation is less dramatically different from that obeyed by disks (the 
baryonic Tully-Fisher relation). The resulting fundamental plane, expressed 
as either $\mdyn(\mstar)$ or $R_{e}(\sigma,\ \mstar)$, is also reproduced; i.e.\ the 
predicted ratios of enclosed dark matter to stellar mass within the stellar 
$R_{e}$ agree with observations as a function of mass (note that with the adopted 
definition of $\mdyn\propto\sigma^{2}\,R_{e}$, $\mdyn$ can be less 
than $M_{\ast}$). Specifically, 
the tilt of the fundamental plane is reproduced -- rather than $\mdyn\propto\mstar$, 
the expectation if systems were perfectly homologous, we predict 
$\mdyn\propto\mstar^{1+\alpha}$ (where $\alpha\approx0.2$ is the FP tilt), 
namely that lower-mass systems (as a consequence of their being more compact) 
are more baryon-dominated within their stellar effective radii. 
(The nature of the FP scalings are discussed in much greater detail in 
\paperfour; however, we outline the dominant physical effects in 
\S~\ref{sec:z0:diss} \&\ \ref{sec:z0:dry} below.)

We plot the simulated systems, with colors denoting the redshift of their 
last gas-rich merger -- i.e.\ approximately the time at which they should 
have stopped forming stars. It is clear that, at fixed stellar mass (or 
velocity dispersion, given that it more tightly correlates 
with stellar mass than e.g.\ effective radius), systems with larger radii (or larger $\mdyn/\mstar$) 
tend to be older -- they are often systems that merged at early times and 
have grown dissipationlessly since then. 
This appears to be borne out in recent observational estimates based on stellar population 
constraints in local ellipticals \citep{gallazzi06:ages,graves:prep}. 
(We discuss this further below.)

\begin{figure}
    \centering
    %\scaleup
    %\plotone{MC_ns.ps}
    \plotone{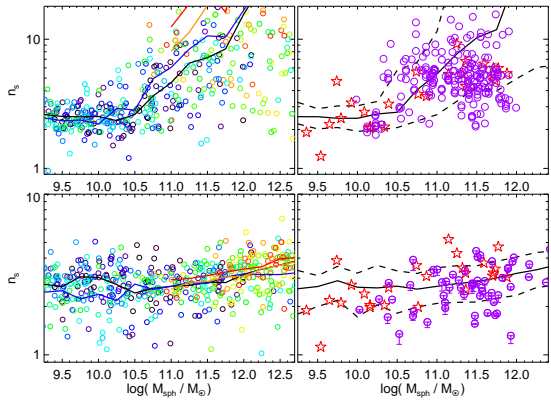}
    \caption{Correlation between stellar mass and estimated best-fit galaxy 
    Sersic index $n_{s}$ (style as in Figure~\ref{fig:fp.proj}). 
    {\em Top:} Sersic indices obtained fitting the entire galaxy profile 
    to a single Sersic law (over typical observed dynamic range). 
    {\em Bottom:} Sersic indices of just the dissipationless (outer, violently relaxed) 
    component (not including the dissipational central starburst). Observational 
    estimates are from the two-component modeling in \papertwo\ and \paperthree. 
    Owing to the 
    cosmological dependence of merger history on mass, there is an effective correlation 
    between stellar mass and $n_{s}$. This is most directly reflected in the 
    outer, dissipationless component. Fitting the systems to a single Sersic index 
    represents a more complex combination of merger history and the amount of 
    dissipation (both effects can mimic one another). 
    %{\bf use a different label for $n_{s}$(outer)? ???}
    \label{fig:ns}}
\end{figure}

Figure~\ref{fig:ns} plots the 
Sersic index of the galaxy light profiles predicted at $z=0$. We stress that these are 
rough estimates at best -- individual galaxy profiles show a degree of diversity 
representative of many details in their formation and merger history that only 
high-resolution numerical simulations can capture. We therefore refer to 
\papertwo\ and \paperthree\ for a more detailed study and comparison 
of Sersic indices in simulated and observed systems; however, comparison of 
our estimates with numerical simulations suggests that the adopted prescriptions 
capture at least the median behavior in simulations. 

First, we  
consider the Sersic index fitted to the {\em entire} galaxy profile -- this is of course a 
useful observational parameter, but it reflects a mix of different {\em physical} 
components of the galaxy. In order to transform a galaxy from an exponential 
disk profile (a Sersic index $n_{s}=1$) to a typical massive elliptical with a high Sersic 
index $\gtrsim4$, both dissipation and violent relaxation are necessary. Violent 
relaxation will scatter stars at large radii, building up an extended envelope 
(characteristic of high-$n_{s}$ systems), but high-$n_{s}$ profiles rise steeply 
at small radii, so dissipation is needed to build up a dense central stellar concentration. 
The effects of repeated mergers on the profile at large radii are relatively straightforward, 
and (more or less) monotonically increase $n_{s}$, but the 
consequences of dissipation are more complex (related not just to the dissipational 
mass fraction but to the relative concentration of the dissipative component and 
its profile shape) and non-monotonic. 

We construct the best-fit Sersic index 
fitting a dynamic range comparable to the observations plotted (representative of 
local galaxies for which the highest-quality data is available), 
from $\sim50\,$pc to $\sim 30$\,kpc. Because the galaxy profiles are 
multi-component, the resulting $n_{s}$ reflects a complex combination of 
the observed dynamic range, the relative mass, size, and slope of the dissipational 
component, and the merger history. 
Together, these effects yield an apparent 
steep dependence of Sersic index on mass, in agreement 
with the observations, but with large scatter (whereas if the profiles were entirely 
dissipationless, the lack of dense stellar concentrations at small radii would restrict them 
to smaller Sersic indices and lead to a weaker dependence of $n_{s}$ on 
merger history and mass). 
Most of the strength of the dependence 
here is ultimately a coincidence reflecting particular combinations of parameters 
moving into or out of the dynamic range for which the fits are sensitive.

If we consider 
the Sersic indices of just the dissipationless component (i.e.\ the sum of the 
stellar components of pre-merger disks that were violently relaxed in the 
initial gas-rich mergers that formed each spheroid progenitor), 
the dependence on mass is weaker, but it more directly reflects the merger 
history of the systems (and depends much less on e.g.\ the dynamic 
range of the fit). 
We compare this with multi-component decompositions of the same observed 
systems, based on a more detailed, 
physically motivated approach to studying galaxy profiles which 
we discuss in detail in \paperone-\paperthree, where we demonstrate that 
the ``dissipationless'' component of observed ellipticals can, in fact, be recovered. 
By definition in our simple model, the dependence we predict now is 
entirely driven by the mean dependence of merger history on mass -- systems 
which have experienced more mergers (and more violent mergers) will have had 
more stars scattered to large radii in those mergers, raising the $n_{s}$ of the 
outer component.  
At low masses, 
most systems formed in $\sim1$ major merger at relatively low redshifts 
$z<1-2$, so a typical $n_{s}\sim2-3$ is obtained. At high masses, systems often have their 
first merger at very high redshifts, involving a large degree of dissipation, forming a 
dense core, and then experience a number of dry mergers, building up a more 
extended dissipationless outer component and envelope and raising $n_{s}$. 

The apparent trends of Sersic index with mass are 
much stronger if dwarf spheroidals are included, as 
in many observational samples \citep[e.g.][]{graham:sersic,ferrarese:profiles}. These 
systems have most likely not experienced any significant mergers -- as a structural 
family, they are clearly related to disks, not ellipticals, and their Sersic indices 
(typical $n_{s}\sim1$) demonstrate this; that they dominate at low masses reflects e.g.\ the 
dominance of disks and ellipticals, and is again driven by the cosmological 
dependence of merger history on mass. 

\begin{figure}
    \centering
    %\scaleup
    %\plotone{MC_bh_corr.ps}
    \plotone{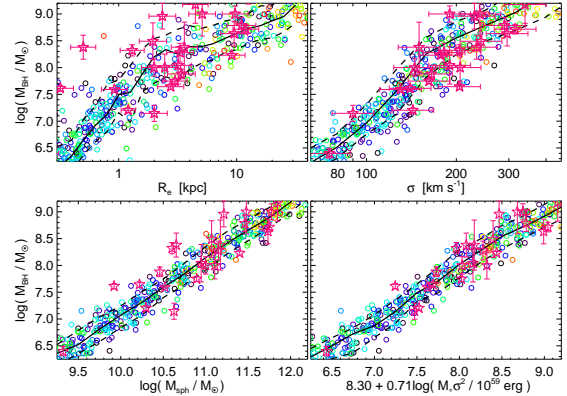}
    \caption{Correlation between black hole mass and host spheroid properties 
    (effective radius, velocity dispersion, stellar mass, and the observable proxy for 
    bulge binding energy, $\mstar\,\sigma^{2}$, effectively the same as the ``black hole 
    fundamental plane'' observed). Style as in Figure~\ref{fig:fp.proj}. 
    Magenta stars show observed systems with direct black hole mass measurements, 
    compiled in \citet{hopkins:bhfp.obs} \citep[see][]{magorrian,merrittferrarese:msigma,
    tremaine:msigma,marconihunt,haringrix}. 
    A simple prescription for black how growth based 
    on dissipational binding energy reproduces the $z=0$ correlations. 
    \label{fig:mbh}}
\end{figure}

Figure~\ref{fig:mbh} compares the predicted and observed correlations between 
nuclear black hole mass and host bulge/spheroid properties (effective radius, 
velocity dispersion, stellar mass, and binding energy). Recall, motivated 
by observational and theoretical expectations of a black hole ``fundamental plane'' 
\citep{hopkins:bhfp.obs,hopkins:bhfp.theory} the fundamental ``assumed'' 
correlation here is between black hole mass and binding energy of the 
dissipational component of the bulge, determined at the time of that dissipation 
(time of gas-rich merger). Despite the complex interplay between this and the 
total bulge mass and size, dry mergers, and evolution in halo bulge properties, 
the $z=0$ correlations 
\citep[including the residual correlations observed in][]{hopkins:bhfp.obs}) 
are accurately reproduced. The curvature seen in the $M_{\rm BH}-R_{e}$ 
and $M_{\rm BH}-\sigma$ relations reflects a combination of the curvature 
seen in the correlations between these properties and stellar mass, and 
the increasing importance of dry mergers at higher masses. 

\begin{figure}
    \centering
    %\scaleup
    %\plotone{MC_mbh_ns.ps}
    \plotone{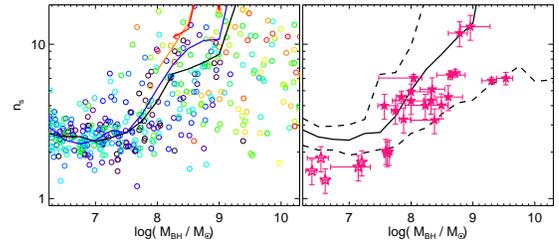}
    \caption{Correlation between black hole mass and estimated best-fit galaxy 
    Sersic index $n_{s}$ (style as in Figure~\ref{fig:fp.proj}). 
    Magenta stars show observed systems as in Figure~\ref{fig:mbh}. 
    Owing to the correlation between $\mbh$ and $\mstar$, 
    there is a correlation between $\mbh$ and $n_{s}$, but the 
    scatter is significant at large $M_{\rm BH}$ (where there is an
    extended tail towards very high $n_{s}$, sensitive to the dynamic range fitted). 
    \label{fig:mbh.ns}}
\end{figure}

We also consider the correlation between black hole mass and Sersic index in 
Figure~\ref{fig:mbh.ns}, comparing to the data compiled in 
\citet{hopkins:bhfp.obs} \citep[an update of the fits in][]{graham:concentration,
graham:sersic}. Owing to the 
dependence of $n_{s}$ on stellar mass, there is an indirect correlation 
between $M_{\rm BH}$ and $n_{s}$, consistent with that claimed
in \citet{graham:sersic}. Those authors note that the correlation is not well-fitted 
by a single power-law; here we see such curvature. The reason is that 
most objects at $\mstar\lesssim10^{10}\,\msun$ have similar 
merger histories, there is little dependence of $n_{s}$ on $\mstar$ 
or $\mbh$ in this mass range, then a steep dependence, then 
again a relatively flat dependence at high $M_{\rm BH}$. 
The same caveats regarding these Sersic indices apply as in 
Figure~\ref{fig:ns}. In this model, the correlation between Sersic index 
and black hole mass may appear reasonably tight over a narrow 
range, but this is accidental, and there is no 
additional information on the black hole mass 
derived from the single Sersic index fits.

\subsection{The Effects of Dissipation}
\label{sec:z0:diss}

\subsubsection{What Drives the Amount of Dissipation?}
\label{sec:z0:diss:driver}

\begin{figure}
    \centering
    %\scaleup
    %\plotone{MC_fdiss.ps}
    \plotone{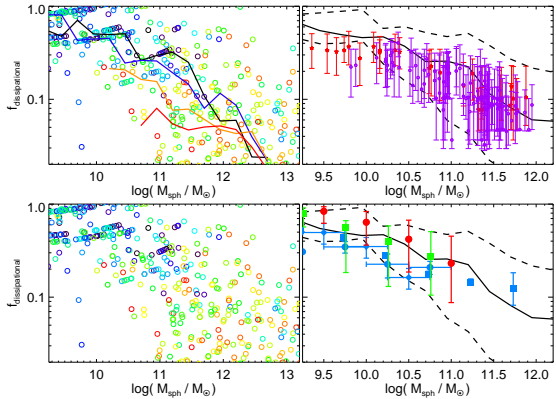}
    \caption{Mass fraction formed dissipationally (integrated mass fraction of the 
    $z=0$ elliptical formed from gas in progenitor disks at the time of 
    gas-rich mergers, in dissipational starbursts) as a function of stellar mass 
    (style as in Figure~\ref{fig:fp.proj}). 
    We compare ({\em top right}) with empirical estimates of 
    this quantity in observed ellipticals from the study of a large sample of observed 
    elliptical surface brightness 
    profiles and dissipation in \papertwo\ and \paperthree. 
    We also compare ({\em bottom}) with observed disk galaxy gas fractions as a function of 
    stellar mass, at $z=0$ \citep[][blue diamonds, 
    squares respectively]{kannappan:gfs,mcgaugh:tf}, 
    $z=1$ \citep[][green squares]{shapley:z1.abundances}, 
    and $z=2$ \citep[][red circles]{erb:lbg.gasmasses}. 
    The dissipational fractions at $z=0$ agree well with those observed, and
    reflect the gas fractions of their progenitor disks over the redshift range $z\sim0-2$ 
    where most systems had their last merger(s). 
    \label{fig:diss}}
\end{figure}

Dissipation is one of the most important factors 
controlling the properties of remnant ellipticals. 
We demonstrate this robustly with detailed observations and simulations in 
\papertwo\ and \paperthree. Here, we discuss how its effects are manifest in a 
cosmologically representative population, including systems which have experienced 
a number of dry mergers. 
Figure~\ref{fig:diss} shows the integrated fraction of our $z=0$ ellipticals which formed 
in dissipative starbursts (from gas in disks at the time of each merger), as opposed to 
dissipationlessly scattered stellar disks. Recall, dissipation allows gas to collapse to 
small scales, building up a dense central component in the stellar mass distribution and 
changing the phase-space distribution of the remnant. 
We compare with the observationally estimated 
dissipational fractions in local ellipticals, determined in \papertwo\ and 
\paperthree. The agreement is good at each mass - in other words, the 
amount of dissipation observed in ellipticals can be explained by our 
simple model.

What drives the trend of dissipation with mass? We find that it reflects the typical 
gas fractions of disks as a function of mass. 
That trend (in disks) may, of course, be a complex 
consequence of poorly understood processes such as star formation, cold 
gas accretion, and stellar feedback; but for our purposes here we are not interested 
in how this property arises in disks (simply what the observed properties of disks are, 
which are captured by construction in our halo occupation approach). 

Figure~\ref{fig:diss} 
compares these integrated dissipational fractions with observationally estimated 
gas fractions of disks (as a function of stellar mass) 
at $z=0$ \citep[see e.g.][]{belldejong:disk.sfh,
kannappan:gfs,mcgaugh:tf}, $z=1$ \citep{shapley:z1.abundances}, 
and $z=2$ \citep{erb:lbg.gasmasses}. 
There is a systematic evolution in gas fractions with redshift, as expected, but 
the gas content of observed disks from $z\sim0-2$ appears to bracket the 
range in both our predicted dissipational fractions and the observed dissipational 
fractions of ellipticals. 

At the lowest masses, where the 
predicted dissipational fractions and observed disk gas fractions approach unity, 
the observationally inferred dissipational fractions are somewhat lower, appearing 
to reach a maximum near $\sim0.4$. This most likely relates to the fact that extremely 
gas-rich mergers will not efficiently dissipate angular momentum and will leave 
disk-dominated, rather than elliptical remnants \citep[for details, see][]{hopkins:disk.survival}. 
These subtleties, not included in our model here, are important for the evolution of 
bulge-to-disk ratios in low-mass galaxies, but do not significantly affect our predictions over 
most of the mass range of interest ($M_{\ast}\gtrsim10^{10}\,M_{\sun}$, where 
$f_{\rm gas}\lesssim0.5$). 

At fixed mass, systems with earlier formation times tend to have 
slightly lower dissipational 
fractions; we emphasize that this is true when the systems are observed 
at $z=0$. At their time of formation, early-forming systems may have had 
very large dissipational fractions, reflecting the characteristically higher 
gas fractions of disk progenitors at these redshifts (we show this in 
greater detail in \S~\ref{sec:track} below). However, these systems will 
also be in the most biased environments, and so (in a $\Lambda$CDM 
context) undergo relatively more growth at later times, assembling 
much of their $z=0$ mass via mergers with less gas content (systems 
that have largely gas-exhausted and/or quenched) and correspondingly 
lower dissipational fractions. 

\begin{figure}
    \centering
    %\scaleup
    %\plotone{MC_strongz_n_mgr.ps}
    \plotter{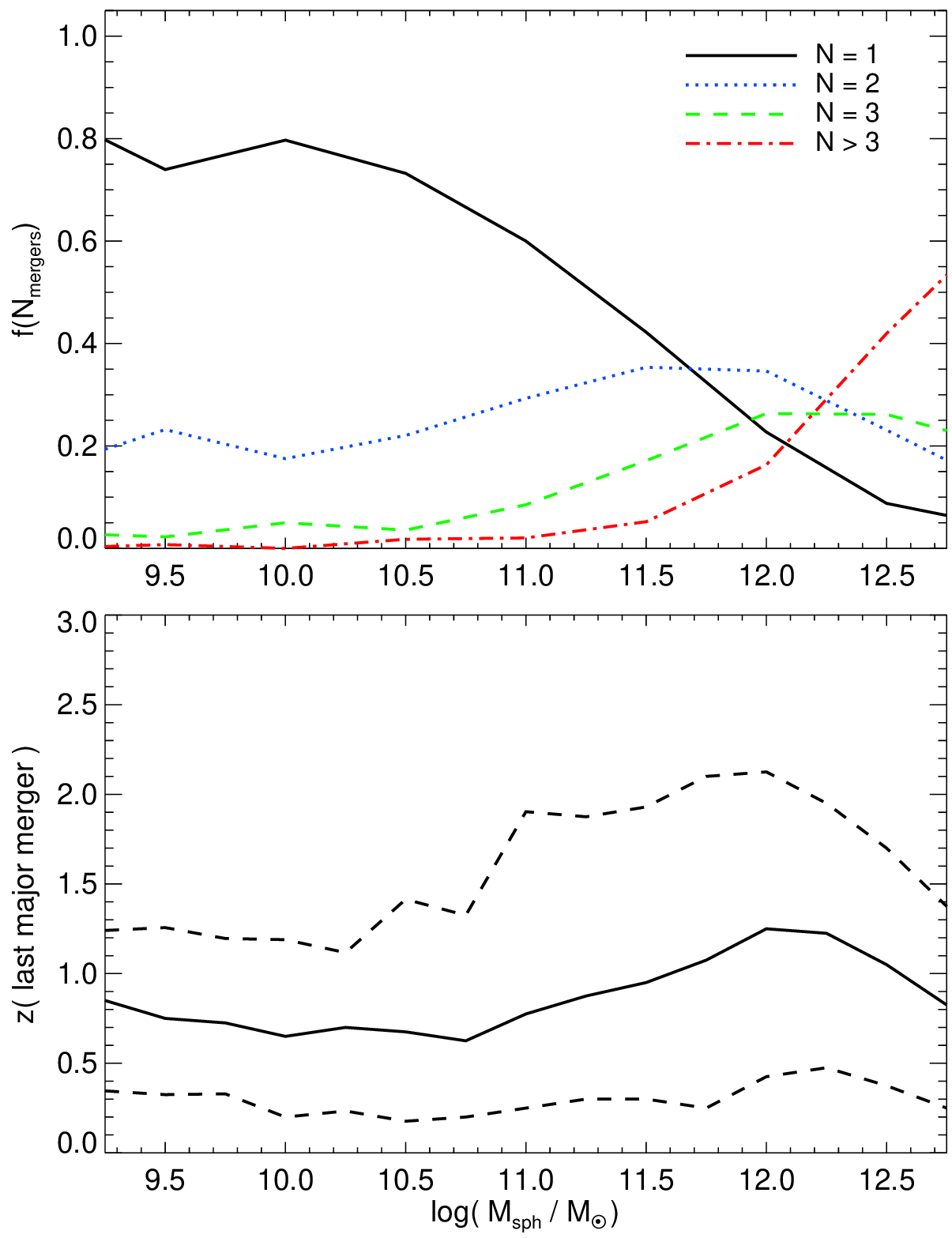}
    \caption{{\em Top:} Fraction of ellipticals at a given stellar mass that have 
    had $N$ major mergers from $z=0-6$. {\em Bottom:} Median (solid) 
    and $\pm1\sigma$ (dashed) range of redshifts of the last major merger 
    for ellipticals of a given stellar mass 
    (not the last gas-rich merger; where the number of mergers 
    $N\gtrsim2$, this last merger is often dry). Most of the mass in ellipticals (at least at 
    $L\lesssim$ a few $\lstar$)
    is assembled by a relatively small number of mergers at relatively low redshifts 
    $z\sim0-2$. As a result, dissipational fractions roughly reflect those of disks in 
    this redshift interval. 
    \label{fig:diss.reasons}}
\end{figure}

This simple correspondence is expected, because most of the mass in ellipticals is 
assembled in a relatively small number of major mergers, at relatively low redshifts 
corresponding to this observed range ($z\sim0-2$). Figure~\ref{fig:diss.reasons} 
shows the typical number of mergers of ellipticals as a function of stellar mass, 
as well as the typical redshifts of the most recent major merger. 
The predicted trend is similar to our estimate in \citet{hopkins:groups.ell} as well 
as other cosmological models \citep[see e.g.][]{delucia:sam}; most ellipticals 
at $\lesssim L_{\ast}$ have had just one major merger, with even massive 
ellipticals at $\sim$a few $L_{\ast}$ still having only a small number of such mergers. 

In \citet{hopkins:groups.ell} we show that the fraction of systems with e.g.\ 
only one major merger in their history versus multiple major mergers, or 
for which the last major merger was gas-rich, agrees well with the observed 
trend in the fraction of ellipticals with central cusps (power-law nuclear slopes), 
rapid rotation, or disky isophotal shapes as opposed to central cores, slow rotation, 
or boxy isophotal shapes \citep[indicators of wet versus dry mergers, discussed 
in detail in e.g.\ \papertwo\ and][]{faber:ell.centers,naab:gas,cox:kinematics}. 
The most recent major mergers, where the final stellar mass 
(at least $\sim1/2$ of it) is assembled, occurs at relatively low redshifts, with a 
median $z\sim1$ and spanning the range $z\sim0-2$. Of course, the most massive BCGs 
form a tail in this distribution with much more complex histories, and for massive galaxies 
undergoing $\gtrsim1-2$ major mergers, the most recent mergers are dry, 
but in any case the broad point remains: most of the $z=0$ mass in ellipticals is assembled 
in a relatively small number of major mergers at $z\lesssim2$, and therefore 
their dissipational fractions reflect those of disks over the same interval. 

This general conclusion (that major mergers with mass ratios 
$\lesssim$1:3, as opposed to more 
numerous minor mergers, dominate the 
mass assembly in mergers once gas is exhausted and 
star formation ``quenched'') 
is supported by calculations from 
other halo occupation models \citep{zheng:hod.evolution} 
and cosmological simulations \citep{maller:sph.merger.rates}. 
However, these calculations, as well as 
clustering estimates \citep{masjedi:cross.correlations} and 
hydrodynamic simulations \citep{naab:etg.formation} 
suggest that as galaxies approach the most massive $M_{\ast}\gtrsim 10^{12}\,M_{\sun}$ 
BCG mass regimes, growth by a large series of minor mergers becomes more 
important than growth by major mergers. 

Roughly speaking, these 
calculations suggest that, since most of the mass is in $\sim L_{\ast}$ systems, 
mergers with such systems dominate mass growth: so when galaxies 
are at masses $\lesssim$ a few $L_{\ast}$, their growth is dominated 
by major mergers, and at higher masses, their growth is dominated by 
progressively more minor mergers. 
Our predictions in this limit should 
therefore be treated with some degree of caution, in contrast with 
the bulk of the spheroid population at $\lesssim L_{\ast}$ 
\citep[typically observed in field or low-density environments, 
where the expectation for growth by a small number of major 
mergers since $z\sim2-4$ is reasonable; see][]{blanton:env,wang:sdss.hod,
masjedi:merger.rates}. On the other hand, it is still unclear whether 
such minor mergers actually contribute significantly to mass growth 
even in the high-mass regime, or whether 
satellites are disrupted in this regime or kicked into orbits where the 
merger time is longer than the Hubble time \citep[see e.g.][]{hashimoto03:varying.culomb.log.in.dynfric,
tormen:cluster.subhalos,benson:heating.model,zheng:hod.evolution,
kazantzidis:mw.merger.hist.sim,brown:hod.evol}.

We, in any case, have experimented with extending our modeling to include 
minor mergers, and find that this yields qualitatively similar 
conclusions (with moderate quantitative but no qualitative differences 
in the predicted 
properties of the most massive galaxies, and no significant differences otherwise). 
However, our predictions (and those of 
other halo-occupation based approaches) are less certain for 
minor mergers owing to ambiguities in e.g.\ 
satellite disruption and weaker constraints on the satellite mass function at small 
mass ratios.

\subsubsection{What Effects Does this Dissipation Have?}
\label{sec:z0:diss:effects}

How does this amount of dissipation change the properties of the $z=0$ ellipticals? 
We consider this question in detail in idealized simulations and observed ellipticals 
in \paperfour: in short, increasing the amount of dissipation yields remnants 
with more of their stellar mass in a compact central starburst remnant. This yields a 
stellar remnant with a smaller effective radius $R_{e}$, and therefore overall a more 
baryon-dominated object (lower $M_{\rm dyn}/M_{\ast}$). We demonstrate that these 
general conclusions remain both qualitatively and quantitatively valid in a more general 
scenario here. In order to compare the effects with and without dissipation, we 
re-run our model, but without including the effects of dissipation: this effectively amounts to 
treating gas identically to stars in mergers, or equivalently treating initial disks as if 
they are all gas-free. 

\begin{figure}
    \centering
    %\scaleup
    %\plotone{MC_nodiss_fp_proj.ps}
    \plotone{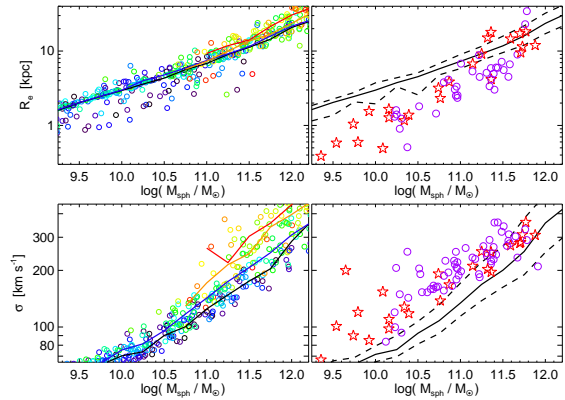}
    \caption{Same predicted spheroid scaling laws at $z=0$ as Figure~\ref{fig:fp.proj}, 
    but in a model with no dissipation (gas is treated the same as stars in mergers). 
    Without dissipation, the correlations are similar to those observed for disks, 
    with a too-shallow size-mass relation ($R_{e}\propto\mstar^{0.3}$). 
    Low-mass systems require dissipation to be as compact as observed -- other effects 
    (redshift evolution in disk sizes and dry mergers) are insufficient to reproduce this 
    effect, and (even tuned) cannot simultaneously reproduce the slope {\em and} 
    normalization of the size-mass relation. 
    \label{fig:fp.proj.nodiss}}
\end{figure}

\begin{figure}
    \centering
    %\scaleup
    %\plotone{MC_nodiss_fp.ps}
    \plotone{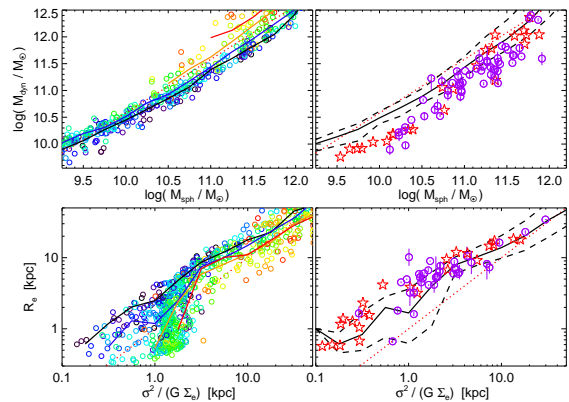}
    \caption{The fundamental plane, as Figure~\ref{fig:fp}, with no dissipation. 
    Without dissipation, there is no tilt (the low-mass ellipticals here 
    have too-large an $M_{\rm dyn}/M_{\ast}$, related to their too-large sizes 
    in Figure~\ref{fig:fp.proj.nodiss}).
    Fitting $\mdyn\propto\mstar^{1+\alpha}$, we obtain $\alpha \sim -0.1$ to $0$, 
    similar to disks, as opposed to $\alpha=0.2$ observed. In this projection, 
    dry mergers and redshift evolution in disk/halo sizes cannot give significant 
    tilt.  The pile-up in the lower-left panel is related to an effective upper limit 
    in the phase space density of disks/halos without dissipation. 
    \label{fig:fp.nodiss}}
\end{figure}

Figures~\ref{fig:fp.proj.nodiss} \&\ \ref{fig:fp.nodiss} 
plot the fundamental plane correlations of 
ellipticals at $z=0$, yielded from this alternative model. 
Low-mass ellipticals, without dissipation, now have sizes similar to those of their 
progenitor disks, and much larger than the sizes observed. 
The slope of the size-mass relation, $R_{e}\propto\mstar^{0.3}$, reflects that in 
progenitor disks and halos, and disagrees dramatically with 
the observed slope for ellipticals $R_{e}\propto\mstar^{0.6}$. As we show in 
\papertwo, the trend of increasing dissipation at low masses, 
owing to increasing gas fractions in lower-mass disks, drives smaller mass 
ellipticals to relatively lower masses, steepening the size-mass relation 
and giving rise to the trend in Figure~\ref{fig:fp.proj}. Other trends, 
such as e.g.\ the 
relation of dry mergers to mass, are insufficient to explain the observed 
size-mass relation (this is expected: dissipationless and dry mergers 
can only lower the density of objects; they can increase radii at large mass but 
cannot make low-mass ellipticals smaller or more dense as needed). 
Invoking evolution in disk sizes and high-redshift mergers can make 
ellipticals more compact overall, but affects the size-mass slope in the {\em opposite} 
sense required (since the highest-mass ellipticals will have their first mergers 
at the earliest times, they would be the most compact relative to disks today; 
the opposite is observed). 

This difference in our predicted size-mass relation is 
also reflected in the velocity dispersion-mass 
and fundamental plane relations, albeit more weakly, since the effective 
radius of the stars enters only indirectly: much of $\sigma$ is set by the potential 
of the halo, which is not altered much by dissipation. Nevertheless, the 
Faber-Jackson relation disagrees with that observed (again, in the sense 
that ellipticals are not dense enough), and the fundamental plane tilt is 
erased (we obtain $\mdyn\propto\mstar^{0.9-1.0}$, essentially identical to 
that observed in disks). 

These discrepancies are also apparent from the 
predictions in \citet{almeida:durham.ell.scalings}, 
who use the semi-analytic models from both \citet{baugh:sam} and \citet{bower:sam} 
to predict galaxy properties in a similar manner as we have done, 
but treating gas identically to stars 
(in fact, their prescriptions for galaxy sizes and dark matter scalings 
are almost identical to ours in this ``no dissipation'' case). As a direct result, at $z=0$, 
their predicted size-mass relation is much too flat ($R_{e}\propto\mstar^{0.3}$)
relative to that observed for 
spheroids, and the predicted fundamental plane has no tilt (in 
fact it is slightly tilted in the incorrect 
sense $M_{\rm dyn}\propto\mstar^{0.8-0.9}$). 
They consider the differences between the two semi-analytic models, 
as well as the inclusion or removal of adiabatic contraction in the 
galaxy halos, supernova and AGN 
feedback, self gravity in the baryons, and systematically higher or lower-energy orbits, 
and find that none of these significantly change the predicted fundamental 
plane scalings. In short, at $z=0$, dissipation (making galaxy stellar 
mass distributions more compact) is much more important and more significant to 
the fundamental plane correlations than any 
differences between various models for galaxy formation and evolution -- 
only dissipation works as a viable explanation of the observed scalings.

\begin{figure}
    \centering
    %\scaleup
    %\plotone{MC_nodiss_ns.ps}
    \plotone{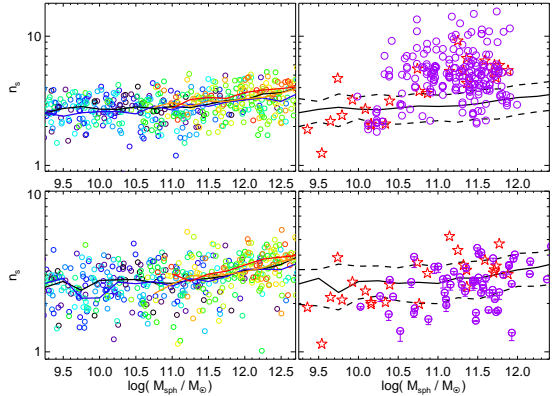}
    \caption{Correlation between stellar mass and estimated best-fit galaxy 
    Sersic index $n_{s}$, as Figure~\ref{fig:ns}, with no dissipation. 
    Without 
    dissipation, Sersic indices remain too low at all masses (central densities 
    cannot rise to the values observed) -- the Sersic indices fitted to the whole galaxy 
    are the same as those of just the dissipationless outer component ({\em bottom}); 
    i.e.\ they can match the outer regions of elliptical profiles but fail to match the 
    central surface brightness of ellipticals (not dense enough at the center). 
    \label{fig:ns.nodiss}}
\end{figure}

The Sersic profiles in the absence of dissipation will be those of the 
outer/dissipationless components, already shown in Figure~\ref{fig:ns} 
(in this case, both the Sersic index of the entire profile and the Sersic index 
of just the dissipationless component will be the same, and be 
identical to that of just the dissipationless component in our full model). We 
directly compare this with the observed Sersic indices in Figure~\ref{fig:ns.nodiss}. 
Although this matches the outer profiles of observed systems, it clearly is 
not the same as the Sersic indices obtained fitting the observed galaxies to a 
single Sersic law over the entire dynamic range -- the high Sersic values (for 
the entire profile) $n_{s}\sim5-10$ observed can only be reproduced with 
some dissipation to make a dense central concentration of light (matching 
the steep rise at small radii of large-$n_{s}$ Sersic profiles). 
This ultimately points to the same issue as the size-mass relations: dissipation 
is needed to raise the central stellar densities in order to match observed 
correlations. 

Without dissipation, there is of course no way to grow the central 
black hole; there is no correlation between black hole mass and host galaxy 
properties for us to predict. We could in principle assume that black hole 
mass is dissipationlessly conserved in mergers, but initially set by disk masses -- 
however, this would predict that it is disk, not bulge properties that set the 
black hole mass, in direct contradiction with observations. We could 
assume that $\mbh$ strictly traces the stellar mass or velocity dispersion 
in the bulge (or in the dissipationless component of the bulge), 
but as demonstrated in \citet{hopkins:bhfp.obs}, this fails to explain 
residual correlations between black hole mass and bulge binding energy 
at fixed $\mstar$ or $\sigma$.

%\begin{figure}
%    \centering
%    %\scaleup
%    \plotone{MC_nodiss_ns.ps}
%    \caption{Correlation between stellar or black hole mass and estimated best-fit galaxy 
%    Sersic index $n_{s}$, as Figure~\ref{fig:ns}, with no dissipation. Without 
%    dissipation, Sersic indices remain too low at all masses (central densities 
%    cannot rise to the values observed). 
%    \label{fig:ns.nodiss}}
%\end{figure}
%Figure~\ref{fig:ns.nodiss} shows the Sersic indices 

\subsection{The Effects of Dry Mergers} 
\label{sec:z0:dry}

Having analyzed the effects of dissipation, we now turn to the effects of dissipationless 
or dry mergers. Since essentially all disk-dominated galaxies 
do, in fact, have significant gas fractions, 
this in practice refers to subsequent spheroid-spheroid re-mergers. What effect do 
such mergers have on the $z=0$ scaling relations of ellipticals? 
In order to study this, we re-run our model, including dissipation  
but excluding such dry mergers 
(we have also considered allowing them to occur in the sense that they 
add stellar mass, but not allowing them to alter the structural properties of the 
galaxy; the results are similar).

\begin{figure}
    \centering
    %\scaleup
    %\plotone{MC_nodry_fp_proj.ps}
    \plotone{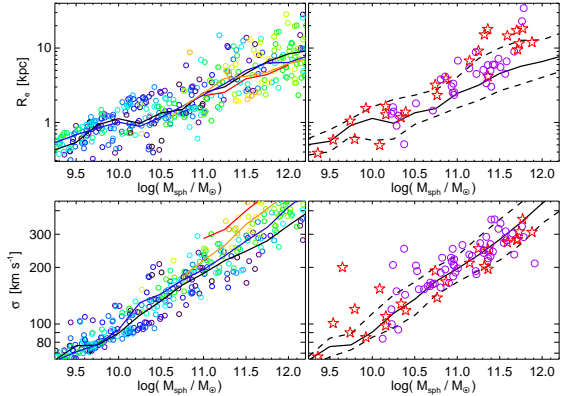}
    \caption{Same predicted spheroid scaling laws at $z=0$ as Figure~\ref{fig:fp.proj}, 
    but in a model with no dry mergers (structural change in 
    spheroid-spheroid mergers is suppressed). 
    Dry mergers do not affect the predictions 
    at low mass, but at high mass ($\mstar\gg 10^{11}\,\msun$), 
    a lack of dry mergers means that ellipticals are not as large as observed.
    The sense of residual correlations at high mass is now reversed -- the oldest systems 
    at fixed mass or $\sigma$ are now the smallest, in contrast to observations 
    \citep{gallazzi06:ages,graves:prep}
    \label{fig:fp.proj.nodry}}
\end{figure}

\begin{figure}
    \centering
    %\scaleup
    %\plotone{MC_nodry_fp.ps}
    \plotone{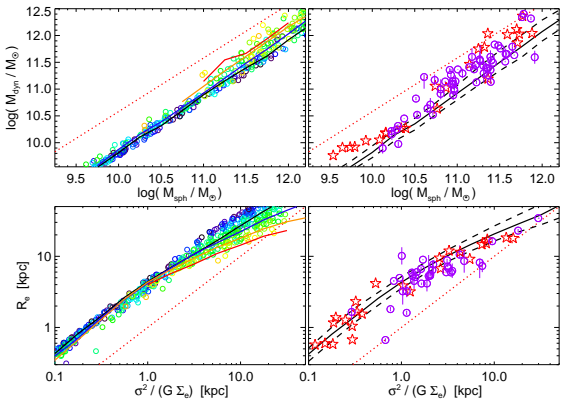}
    \caption{Fundamental plane, as Figure~\ref{fig:fp}, with no dry mergers. 
   The difference is minimal (although there is slightly less tilt), 
   since dry mergers tend to move systems 
   parallel to the FP. 
    \label{fig:fp.nodry}}
\end{figure}

Figures~\ref{fig:fp.proj.nodry} \&\ \ref{fig:fp.nodry} plot the fundamental plane correlations of 
ellipticals at $z=0$ from this alternative model. At intermediate and low 
masses, the number of mergers in our full model is small (see Figure~\ref{fig:diss.reasons}), 
and dissipation is dominant, so the correlations predicted here are essentially 
identical to those in Figures~\ref{fig:fp.proj} \&\ \ref{fig:fp}. At high masses or 
luminosities ($\gtrsim$a few $\lstar$), however, there is an appreciable effect 
on the projections of the FP. The highest mass systems have their first mergers at early times, 
when their progenitors are more gas-rich, and are therefore compact when first formed. 
Without dry mergers to puff them up at later times, they do not have effective radii 
as large as observed. The difference in the scaling laws over the dynamic range observed 
in Figure~\ref{fig:fp.proj.nodry} is subtle, becoming dramatic only at the highest masses. 

There is, however, a significant effect in the residual correlations with e.g.\ size and 
merger history at fixed stellar mass or velocity dispersion. Without dry mergers, 
then (because of the same effect), the oldest systems at fixed mass would be the 
smallest. This is the opposite of observed trends 
\citep{gallazzi06:ages,graves:prep}: older BCGs 
are extremely clustered, and so (unless CDM models are seriously incorrect) 
had their first mergers at early times, where spheroids are observed to be 
compact (see \S~\ref{sec:evol}). The observations that e.g.\ older 
BCGs tend to be the most extended systems 
\citep{vonderlinden:bcg.scaling.relations,
bernardi:bcg.scalings}
therefore demands some dry mergers 
in their history \citep{hausman:mergers,hernquist:phasespace}.
However, as shown in Figure~\ref{fig:diss.reasons}, the number of 
such dry mergers (especially at $z\lesssim1-2$) is still relatively small, so 
this is consistent with the $\sim1$ major dry merger since $z\sim1$ which direct observations 
of dry merging pairs \citep{vandokkum:dry.mergers,bell:dry.mergers,lin:mergers.by.type} 
and constraints from mass function evolution 
\citep{bundy:mfs,borch:mfs,perezgonzalez:mf.compilation,
brown:mf.evolution,hopkins:transition.mass,zheng:hod.evolution} suggest. 

The fundamental plane itself is less affected by dry mergers. This is not surprising; 
numerical simulations demonstrate that both major and 
minor dry mergers tend to move systems parallel to the fundamental plane 
\citep[see \paperfour\ and][]{boylankolchin:mergers.fp,robertson:fp}. Where an individual 
galaxy ends up on the FP is therefore affected by its dry merger history, but 
the FP itself is not very much altered. There is a very weak effect because of the 
evolution in the FP discussed in \S~\ref{sec:evol}, but it is much less apparent 
here than in e.g.\ the size-mass relation. 

\begin{figure}
    \centering
    %\scaleup
    %\plotone{MC_nodry_ns.ps}
    \plotone{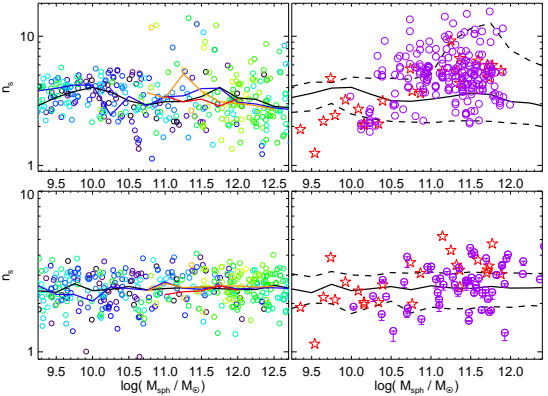}
    \caption{Correlation between stellar mass and estimated best-fit galaxy 
    Sersic index $n_{s}$, as Figure~\ref{fig:ns}, with no dry mergers. Without 
    dry mergers, Sersic indices
    remain flat at the values characteristic of gas-rich merger 
    remnants. The large Sersic indices at high masses (reflecting the extended envelopes 
    of the most massive ellipticals) are not reproduced.
    \label{fig:ns.nodry}}
\end{figure}

Figure~\ref{fig:ns.nodry} shows the Sersic indices as a function of mass 
in the absence of dry mergers. Without repeated dry mergers to scatter 
stars at large radii and build up a more extended envelope, the profile 
of the outer/dissipationless component is self-similar at all masses (failing 
to reproduce the weak, but significant dependence observed). Fitting the 
entire profile to a single Sersic index, the effects of dissipation still 
contribute to a dependence of this index on mass (again emphasizing that 
with such a single-component fit, the Sersic index does not robustly track merger 
history as it does in a two-component decomposition). 
Both the effects of dry mergers and dissipation (in particular 
combinations) are needed to reproduce the observed trends.

\begin{figure}
    \centering
    %\scaleup
    %\plotone{MC_nodry_bh_corr.ps}
    \plotone{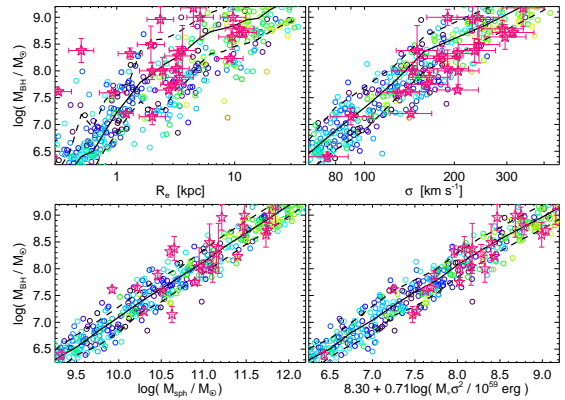}
    \caption{Correlation between black hole mass and host spheroid properties, 
    as Figure~\ref{fig:mbh}, with no dry mergers. There is fairly little effect here -- 
    dry mergers conserve both spheroid binding energy and total black hole mass, 
    so move systems parallel on the black hole fundamental plane correlations. 
    The (weak) change along projected correlations (e.g.\  
    $M_{\rm BH}-\sigma$) reflects the small structural changes of the 
    remnants in Figure~\ref{fig:fp.proj.nodry}.
    \label{fig:mbh.nodry}}
\end{figure}

Figure~\ref{fig:mbh.nodry} shows the correlations between 
BH mass and host properties in this model. The 
effects here are quite weak. 
Since stellar mass and black hole mass are conserved in dry mergers 
(and the projected correlation is roughly $M_{\rm BH}\propto\mstar$) there 
is almost no change in that correlation with or without dry mergers. 
Total spheroid binding energy is also conserved (which leads to 
velocity dispersion being nearly constant) in dry mergers, but the 
correlation here is not linear (although it is the more fundamental correlation 
in our models). If we assume the ``initial'' correlation is the 
power-law $M_{\rm BH}\propto E_{\rm bul}^{0.7}$ (or $M_{\rm BH}\propto \sigma^{4}$ 
for the $M_{\rm BH}-\sigma$ relation), then (assuming $\mstar\propto\sigma^{4}$,
which gives  
$E_{\rm bul}\propto \mstar^{3/2}$) after a merger of mass ratio 
$f$ ($f\le1$) two systems initially on the correlation will move a 
factor $(1+f)/(1+f^{3/2})^{0.71}$ off the correlation (likewise for 
the $M_{\rm BH}-\sigma$ relation. 
This amounts to a $\sim20\%$ ($\sim0.1$\,dex) effect in major mergers -- 
so we expect (given the $\sim0.3$\,dex intrinsic scatter in the correlations) 
it will only be noticeable at the most massive end where systems 
have undergone a large number of dry mergers. 
We do see this, but over the dynamic range of the present data, 
there is little distinction that can be clearly drawn -- the scatter 
in various properties (and the fact that e.g.\ neither $M_{\rm BH}-\sigma$ 
nor $M_{\rm BH}-\mstar$ is the most fundamental correlation in this model) 
mostly washes it out. 
In terms of the correlations with effective radius $R_{e}$, 
there are more noticeable changes, directly reflecting 
the different size-mass relation in Figure~\ref{fig:fp.proj.nodry}.

\breaker
\section{Evolution of Scaling Laws as a Function of Redshift}
\label{sec:evol}

We have shown how dissipation and dry mergers alter the correlations of 
ellipticals at $z=0$. Of course, gas fractions of their progenitors evolve with 
redshift, and dry mergers preferentially occur at low redshift. In addition, other 
properties of the progenitors, such as disk and halo sizes, will evolve. 
We therefore expect there could be evolution in these scaling laws, and 
make predictions for this here. We adopt our standard model, but construct 
samples as they would be observed at a given non-zero redshift. 

\subsection{Predicted Evolution In Our Standard Model}
\label{sec:evol:pred}

\subsubsection{Spheroid Sizes}
\label{sec:evol:sizes}

\begin{figure}
    \centering
    %\scaleup
    %\plotone{r_z_mbins.ps}
    \plotone{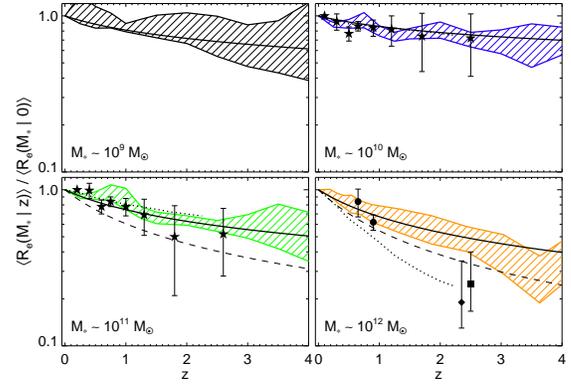}
    \caption{Predicted evolution in spheroid sizes with redshift. We show the 
    predicted median size of spheroids at a given stellar mass, relative to the 
    median size of objects of that mass at $z=0$. Shaded range is the uncertainty owing 
    to both cosmology and how rapidly progenitor (disk and halo) properties evolve. 
    Solid lines are simple power-law fits (Equation~\ref{eqn:re.z}). 
    Dashed line shows the power-law fit corrected (approximately) to the $R_{e}$ 
    that would be measured in optical/UV bands (as opposed to the half-mass radius: 
    stellar population effects tend to make very young ellipticals with 
    recent central starbursts more concentrated 
    in short-wavelength light). 
    Dotted line is 
    the prediction from the semi-analytic model of \citet{khochfar:size.evolution.model}. 
    Observational estimates for galaxies in the appropriate stellar mass intervals 
    are shown from 
    \citet[][circles]{mcintosh:size.evolution}, \citet[][stars]{trujillo:size.evolution}, 
    \citet[][square]{zirm:drg.sizes}, and \citet[][diamond]{vandokkum:z2.sizes}. 
    Progenitors are more gas-rich 
    at high-$z$, yielding mergers with more dissipation and therefore smaller remnants. 
    At high masses, low-$z$ disks are gas poor (and dry mergers have been more important), 
    so the relative difference is larger; at low masses, there is little room for gas fractions
    to increase, hence weak evolution. 
    \label{fig:reff.z}}
\end{figure}

Figure~\ref{fig:reff.z} shows how the median effective radii of spheroids 
at fixed stellar mass are predicted to evolve with redshift. We vary the assumptions 
described in \S~\ref{sec:model}, most importantly (other uncertainties being 
relatively negligible at the redshifts of interest here) whether (and how strongly, 
within the observational limits) disk sizes gas fractions evolve with redshift, 
and show the resulting allowed range. We compare with observations of 
the spheroid size-mass relation from \citet{trujillo:size.evolution}, 
\citet{mcintosh:size.evolution}, \citet{vandokkum:z2.sizes}
and \citet{zirm:drg.sizes} \citep[see also][]{franx:size.evol,damjanov:red.nuggets}. 
Although the observational 
constraints are weak, they agree with our predictions at redshifts 
$z\sim0-3$. 

Evolution in galaxy gas fraction, which directly translates to evolution in the amount 
of dissipation in mergers, drives this behavior.
This is expected -- in \S~\ref{sec:z0:diss}, 
and in greater detail in \papertwo, we show that the degree of 
dissipation is the dominant factor determining the stellar effective radius. 
Disk gas fractions are larger at high redshift, so in mergers, a larger mass fraction will 
be formed in a compact dissipational starburst, yielding a smaller stellar remnant. 
Dry mergers have also had less time to act and puff up remnants at high redshift. 
If disks are more compact at high redshift, this drives further evolution; however, 
observational constraints suggest that this is relatively weak 
(see \S~\ref{sec:model:hod}), and so it is less 
important than the evolution in gas fractions. 

This also predicts that the relative size evolution 
should be stronger in high-mass systems. 
Fitting these predictions to simple scaling laws of the form 
\begin{equation}
\langle R_{e}(M_{\ast}\,|\,z) \rangle = (1+z)^{-\beta}\,\langle R_{e}(M_{\ast}\,|\,z=0) \rangle
\label{eqn:re.z}
\end{equation}
we obtain $\beta=(0.0-0.2,\,0.24,\,0.48,\,0.64)$ ($\pm0.05$ or so for each) for 
$M_{\ast}\sim(10^{9},\,10^{10},\,10^{11}\,10^{12})\,\msun$ 
(roughly $\beta\approx {\rm MAX}[0.23\,\log{(M_{\ast}/10^{9}\,\msun),\ 0.2]}$). 
% fit to exp(-z^beta) -> 0.00, 0.11, 0.23, 0.31 %
Low mass disks are still gas-rich at $z=0$ (and have generally 
not experienced many dry mergers), so there is limited room for 
evolution (the evolution we do predict is almost entirely driven by 
whatever evolution in progenitor disk sizes is assumed, hence the 
quoted range $0.0-0.2$ in the lowest-mass bin). 
At high mass, however, dry mergers have been increasing spheroid sizes 
for a significant redshift interval. Furthermore, disks at high mass at low redshift 
are relatively gas-poor; by $z\sim3$, they have doubled or tripled their gas 
fractions \citep{erb:lbg.gasmasses}. The relative size 
difference is therefore much more pronounced. 

These predictions are similar to those made by \citet{khochfar:size.evolution.model}, 
who adopted a full semi-analytic model to estimate the history of disk galaxies, 
accretion, and star formation, and then 
modeled disk sizes by assuming effective radius is proportional to 
the mass fraction in the dissipationless (non-starburst) component: 
$R_{e}(M_{\ast},\,z)=R_{e}(M_{\ast},\,0)\,\times\,(f_{\rm dissipationless}[z]/f_{\rm dissipationless}[0])$. 
For comparison, we show their predictions in Figure~\ref{fig:reff.z}. 
At intermediate and low masses, they are similar to ours. 
This is because their simple model captures the key qualitative aspect which drives 
our predicted evolution, namely the evolution in disk gas fractions with redshift, 
making high-redshift ellipticals more dissipational than their low-redshift counterparts. 
Ultimately, their determination of the size evolution with dissipational fraction is 
not a bad approximation over intermediate gas fractions (to the accuracy here). 
However, there is a significant difference in the predictions at the highest masses -- 
\citet{khochfar:size.evolution.model} predicted much stronger evolution than we do. This is 
the regime where our attempts to improve on their estimates are of particular importance. 

First, their predictions are tied to some assumptions and modeling of disk galaxies 
in their semi-analytic model; this is precisely why we have tried to be as empirical 
as possible in constructing our progenitor galaxies. They 
note that their strong evolution in this mass bin is the 
result of a high-redshift overcooling problem in 
describing massive galaxies in their 
semi-analytic model (a well-known difficulty without strong feedback). 
As a consequence, their disk gas fraction evolution to $z\sim2$ is much steeper than observed -- 
i.e.\ ``progenitor'' disks at these masses and redshifts are demonstrably too gas-rich and 
in halos of too low a mass. \citet{erb:lbg.gasmasses} see typical gas 
fractions for $M_{\ast}\gtrsim10^{11}\,\msun$ 
galaxies at $z\sim2$ of $\sim20\%$ (whereas \citet{khochfar:size.evolution.model} 
predict $\sim70-80\%$) 
 -- even if the ellipticals observed at $z=2$ formed from 
higher-redshift mergers with gas fractions say, double this value, this predicts 
factor $\lesssim2$ size evolution at $z=2$, similar to our estimates. By using a halo 
occupation approach, we avoid this problem. 

Second, 
our size estimates are calibrated directly to numerical simulations: we track the 
evolution of multiple galaxy components in merger simulations and use 
this to design our prescriptions. Of particular importance, we include the effects of 
dry mergers appropriately and extrapolate our prescriptions to arbitrary dissipational 
fractions: at high $f_{\rm dissipational}$, the simple size estimator from 
\citet{khochfar:size.evolution.model} clearly must break down, as galaxies will still have 
finite sizes. This direct calibration with simulations 
avoids ambiguity in e.g.\ degeneracies between 
cosmological effects and assumptions about how spheroid sizes scale, and 
allows for more robust predictions. 

Nevertheless the reason for our predicted evolution, and its sense, 
is essentially the same as that identified in \citet{khochfar:size.evolution.model}. 
Ellipticals at 
high redshift have, on average, formed a larger fraction of their mass in dissipational 
starbursts (owing to their progenitors being, on average, more gas-rich) 
at these redshifts. Dissipation allows this component to be significantly more compact 
than the dissipationlessly violently relaxed stars from the progenitor disks, 
giving rise to a more compact remnant.

At the highest masses ($\gg 10^{11}\,\msun$), the evolution we predict is somewhat weaker 
than that observationally inferred by 
\citet{zirm:drg.sizes,vandokkum:z2.sizes,franx:size.evol}. However, as demonstrated in \paperone, 
this can be explained by 
an additional bias -- if systems at high $z$ are very young 
(ages $\lesssim0.5-1\,$Gyr) 
remnants of 
gas-rich mergers \citep[as we predict precisely these systems are, 
and as observations of their stellar populations suggest, see][]{kriek:drg.seds} and observed in 
even rest-frame optical wavelengths (let along rest-frame UV), 
the gradients in $M_{\ast}/L$ introduce a bias towards smaller $R_{e}$ 
\citep[there may also be some related, but weak, biases in these stellar population parameters 
and stellar masses, see e.g.][]{wuyts:irac.drg.colors}. 
The younger stellar population from the starburst -- i.e.\ the dissipational 
component -- is significantly brighter in $B$-band at ages $\lesssim1\,$Gyr, 
so this dominates the fit, yielding a smaller $R_{e}$ (equivalent 
to artificially further increasing the dissipational mass fraction -- this is why 
it actually looks as if the \citet{khochfar:size.evolution.model} prediction provides an 
acceptable match to these objects). 

In line with this expectation, the sizes 
we obtain if we ignore the dissipationless components of our predicted 
$z>2-3$ systems are consistent with the observations 
from \citet{zirm:drg.sizes,vandokkum:z2.sizes,franx:size.evol,damjanov:red.nuggets}. 
Essentially, this effect can yield a bias of an additional factor $\sim2$ 
(see Figure~20 in \paperone) for gas-rich systems. 
It will not be a problem at lower redshifts, both because systems are older 
(the effect is negligible at starburst ages $>1\,$Gyr, or even earlier, as 
induced metallicity gradients from the starburst can partially cancel it out), 
and because gas fractions are lower (we expect significant bias only
for $f_{\rm dissipational} \gtrsim 0.3-0.4$). Furthermore, there is no bias {\em within} 
the starburst light (there is not a strong gradient within the starburst itself), 
so for low-mass systems (which are largely dissipational at every redshift), 
there is no significant effect. It is precisely for 
large mass systems at high redshift 
that we expect this to be an important concern, and measurements that could 
ultimately estimate e.g.\ color gradients and stellar population gradients in 
these systems will be necessary for more detailed constraints.

\subsubsection{Velocity Dispersions}
\label{sec:evol:sigma}

\begin{figure}
    \centering
    %\scaleup
    %\plotone{s_z_mbins.ps}
    \plotone{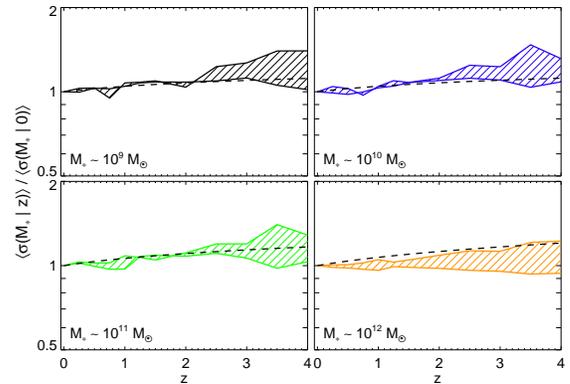}
    \caption{Predicted evolution in spheroid velocity dispersions with redshift 
    (style as Figure~\ref{fig:reff.z}). Despite significant evolution in $R_{e}$ at 
    fixed mass, the role of dark matter halos (which evolve more weakly) in setting 
    $\sigma$ yields much less evolution in $\sigma$ for the same systems. 
    Lines are the simple prediction given the fitted scaling of 
    $R_{e}$ with $z$ (Figure~\ref{fig:reff.z}) and Equation~(\ref{eqn:sigma.z}).
    \label{fig:sigma.z}}
\end{figure}

Figure~\ref{fig:sigma.z} shows the corresponding evolution in velocity 
dispersion with redshift. The evolution is much weaker than might be 
intuitively expected: if $\sigma^{2}\sim G\,M/R_{e}$, then the factor 
$\sim3$ evolution in size in massive systems in Figure~\ref{fig:reff.z} would 
translate to factor $\sim\sqrt{3}$ evolution in $\sigma$. Instead, we predict 
such systems have roughly the same or at most a factor $\sim1.3$ larger 
$\sigma$. This is because, at $z=0$, dark matter plays an important role 
in setting the velocity dispersion -- the potential at $r=0$ has a contribution 
$\sim G\, \mhalo/R_{\rm halo}$ (here $R_{\rm halo}$ is the effective radius of 
the halo, {\em not} the virial radius; $R_{\rm halo}\sim R_{\rm vir}/c$, where $c$ 
is the halo concentration), which is comparable to or larger than 
$\sim G\,\mstar/R_{e}$ 
in most systems (although $R_{\rm halo}\gg R_{e}$, 
$M_{\rm halo}\gg \mstar$ by about the same factor). This is especially true 
for the high-mass ellipticals (for which the evolution in $R_{e}$ is strongest) 
which, at $z=0$, are the most dark-matter dominated 
within their stellar $R_{e}$.
At a given halo mass, $R_{\rm halo}$ evolves weakly -- the evolution in halo 
concentrations 
almost completely offsets the evolution in virial radii \citep[this reflects 
the fact seen in most simulations that massive halos build inside out -- the 
central potential is set first and then the outer halo builds up; 
see][]{bullock:concentrations, wechsler:concentration}, 
so assuming the ratio $\mstar/\mhalo$ does not 
evolve much with redshift (as inferred in most observations and 
halo occupation models; see \S~\ref{sec:model:cosmology}), then the contribution of the halo to 
$\sigma$ ($\sigma_{\rm halo}$) evolves weakly at fixed $\mstar$. If anything, it 
can decrease -- if the baryonic component is more compact at high-$z$, 
then the enclosed dark matter mass in the stellar $R_{e}$ is smaller. 

Roughly, then, if we consider the observed velocity dispersion to reflect a 
galaxy potential and halo potential which add linearly, we obtain 
$\sigma^{2}\propto (\mstar/R_{e} + \mhalo/R_{\rm halo})$. If we consider 
systems of fixed $\mstar$, and $\mhalo(\mstar)$ and $R_{\rm halo}(\mhalo)$ 
evolve weakly, then we obtain 
\begin{equation}
\frac{\langle\sigma(z\, |\,\mstar)\rangle}{\langle\sigma(0\, |\,\mstar)\rangle}
=\frac{1}{\sqrt{1+\gamma}}
\,\sqrt{\gamma + \frac{\langle R_{e}(0) \rangle}{\langle R_{e}(z) \rangle}}, 
\label{eqn:sigma.z}
\end{equation}
where $\gamma\equiv (\mhalo/R_{\rm halo})/(\mstar/R_{e})\sim 1-2$ is the 
relative fraction of the central potential contributed by the dark matter at $z=0$. 
For the values of $\gamma$ we estimate ($\sim1$ at $\mstar\sim10^{11}\,\msun$ 
and $\sim2$ at $\mstar\sim10^{12}\,\msun$), this simple expectation fits 
the observed trends well.

\subsubsection{The Fundamental Plane}
\label{sec:evol:fp}

\begin{figure}
    \centering
    %\scaleup
    %\plotone{md_ms_z_mbins.ps}
    \plotone{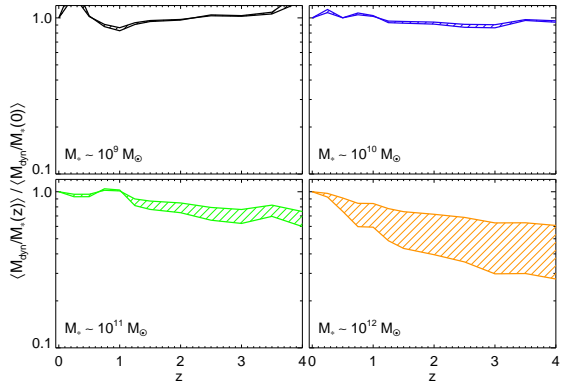}
    \caption{Predicted evolution in the ratio of dynamical mass estimator 
    $\mdyn=k\,\sigma^{2}\,R_{e}/G$ to stellar mass $\mstar$ with redshift 
    (as Figure~\ref{fig:reff.z}). 
    Comparing Figures~\ref{fig:reff.z} \&\ \ref{fig:sigma.z}, the evolution is 
    primarily driven by evolving $R_{e}$ -- more dissipational, compact 
    stellar remnants are increasingly baryon-dominated inside their 
    stellar $R_{e}$. Again, low-mass systems are already highly dissipational 
    and baryon-dominated inside $R_{e}$ at $z=0$, so there is little room 
    for evolution. 
    \label{fig:mdyn.z}}
\end{figure}

Combining these trends, we anticipate the evolution in $\mdyn/\mstar$ shown 
in Figure~\ref{fig:mdyn.z}. Since $\mdyn\equiv k\,\sigma^{2}\,R_{e}/G$, 
the stronger evolution in $R_{e}$ largely drives the evolution in $\mdyn$. 
We emphasize that this reflects a real difference in the enclosed matter 
within the stellar $R_{e}$ (as we have defined it, there are no significant structural or 
kinematic non-homology effects). The decrease in $\mdyn$ at fixed $\mstar$ in 
the highest-mass systems is a consequence of the fact that at low-$z$, they have 
small dissipational content and have larger stellar $R_{e}$, so have larger 
dark matter masses enclosed in $R_{e}$ (and thus higher enclosed mass in $R_{e}$). 
At high-$z$, their progenitors are gas-rich, so they are formed in highly dissipational 
mergers and have compact stellar distributions, which enclose less dark matter 
in the stellar $R_{e}$. 

\begin{figure}
    \centering
    %\scaleup
    %\plotone{fp_z.ps}
    \plotter{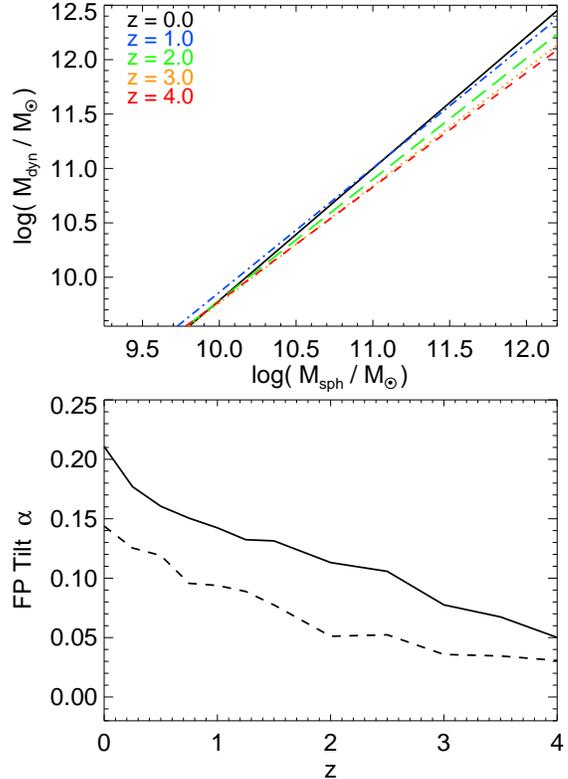}
    \caption{Best-fit fundamental plane ($\mdyn\propto\mstar^{1+\alpha}$, where 
    $\alpha$ is the FP tilt) as a function of redshift. 
    The evolution in Figure~\ref{fig:mdyn.z} 
    is reflected here: at high-$z$, even large-mass disks are gas-rich so high-mass 
    systems experience significant dissipation, bringing their stellar $R_{e}$ in and 
    making the remnant more baryon-dominated within $R_{e}$ (lowering the 
    dark matter fraction and therefore $\mdyn$ within $R_{e}$, at fixed $\mstar$). 
    The tilt therefore weakens slightly with redshift (there is 
    less difference in the degree 
    of dissipation -- the driver of tilt -- at high-$z$). Solid and dashed lines 
    allow (do not allow) for progenitor disk size and stronger (weaker) 
    disk gas fraction evolution, bracketing the range allowed by observations 
    (representing the uncertainty in our predictions). 
    \label{fig:fp.z}}
\end{figure}

Figure~\ref{fig:fp.z} shows the implications of this evolution for the FP 
as a function of redshift. 
At low redshift, the observed tilt ($\alpha\approx0.2$) is recovered. At high redshift, 
however, all progenitors are gas-rich, 
so there is not much difference in the dissipational content at low and high masses -- meaning 
their compactness (relative to e.g.\ their dark matter halos or progenitor disks) is no longer a 
strong function of mass. The high-mass systems, being nearly as dissipational as low-mass 
systems, are similarly baryon-dominated in their stellar $R_{e}$, and there is 
less tilt. 

At the level predicted here, it may be difficult to observe the predicted FP evolution. 
Robust velocity dispersions and effective radii in the same rest-frame band (avoiding a 
bias towards smaller $R_{e}$ as observations probe closer to the rest-frame UV) may be 
obtainable in large samples at $z\lesssim1$, but the predicted evolution in that interval 
is weak -- within the systematic uncertainties at $z\sim0$. 

This is consistent with observational 
constraints in this redshift interval from 
weak lensing \citep{heymans:mhalo-mgal.evol} 
and optical studies \citep[][]{alighieri:fp.evolution,
treu:fp.evolution,vanderwel:fp.evolution,vandokkum:fp.evol,brown:hod.evol}. These observations do see 
evolution in optical bands, (in the opposite sense to that predicted here, namely an increasing 
apparent tilt), but find that this owes to stellar population effects 
\citep{alighieri:fp.evolution} -- the evolution observed in terms of $\mdyn$ and $\mstar$ is 
either negligible or slightly negative (as predicted here). 
Any model where the stellar populations of 
lower-mass ellipticals are systematically younger 
(as is predicted here; see \citet{hopkins:groups.ell} for details) will predict the 
observed trend in optical bands: if, say, a 
low-mass system at $z=0$ has a stellar population age of $\sim7\,$Gyr whereas a 
high-mass system has an age of $\sim12\,$Gyr, as observed 
\citep[see e.g.][]{trager:ages,nelan05:ages,thomas05:ages,gallazzi06:ages}, 
then inverting passive evolution to $z=0.7$, the low-mass (younger) system would have 
a lower $M_{\ast}/L_{B}$ by a factor $\gtrsim10$, whereas the high-mass system 
would have increased $M_{\ast}/L_{B}$ by a factor of just $\approx2$. It is therefore 
challenging to observe the weak evolution here ($\sim0.1$\,dex), given expected 
stellar population effects of magnitude $\sim0.5-1$\,dex. 

However, this degree of evolution is important for a number of subtle effects: 
for example, \citet{vandokkum:imf.evol} uses the evolution of the optical FP with redshift to 
infer evolution in elliptical mass-to-light ratios (under the assumption that 
the physical -- i.e.\ stellar mass -- FP is invariant). This is broadly fine, as the optical 
$M/L$ for any reasonable stellar population age evolves much more rapidly than 
the physical FP evolution in Figure~\ref{fig:fp.z}. But the authors then compare this 
change in $M/L$ with time to the mean colors of ellipticals as a function of time, 
and argue that the relation between the two suggests evolution in the stellar 
initial mass function. 
At this level (as the authors acknowledge), the comparison is sensitive to 
evolution in the physical (stellar mass) FP at the $\sim0.1\,$dex level -- if massive galaxies 
at $z\sim1$ have lower $M_{\rm dyn}/M_{\ast}$ by $\sim0.1\,$\,dex (similar to what we 
predict in Figure~\ref{fig:mdyn.z}) this will bias the inferred $M/L$ ratios 
(under the assumption of no physical FP evolution), but not, obviously the galaxy colors, 
by an amount comparable to the significance of the effect seen (in other words, allowing 
for this evolution, the evidence for any evolution in the stellar IMF may be less strong). 

At this level of detail, physical evolution in the FP 
in the manner predicted here may change this comparison, and should be 
considered in future, more detailed comparisons with observations. However, 
the effects here may not be important so long as the observed sample is 
appropriately selected: \citet{vandokkum:imf.evol} focus on satellites in rich 
clusters, which are not likely to undergo significant subsequent merging 
(especially as compared to the central galaxies modeled here, particularly those 
in moderately dense field and group environments that contribute 
much of the ``lever arm'' to the FP tilt evolution predicted). This emphasizes the 
importance of appropriate sample selection when comparing these quantities, 
and the need for further study to determine exactly how our predictions 
generalize to populations with more complex or distinct histories (such as 
satellites) and therefore pertain to the observations.

\subsubsection{Black Hole Masses}
\label{sec:evol:bh}

\begin{figure}
    \centering
    %\scaleup
    %\plotone{mbh_ms_z_mbins.ps}
    \plotone{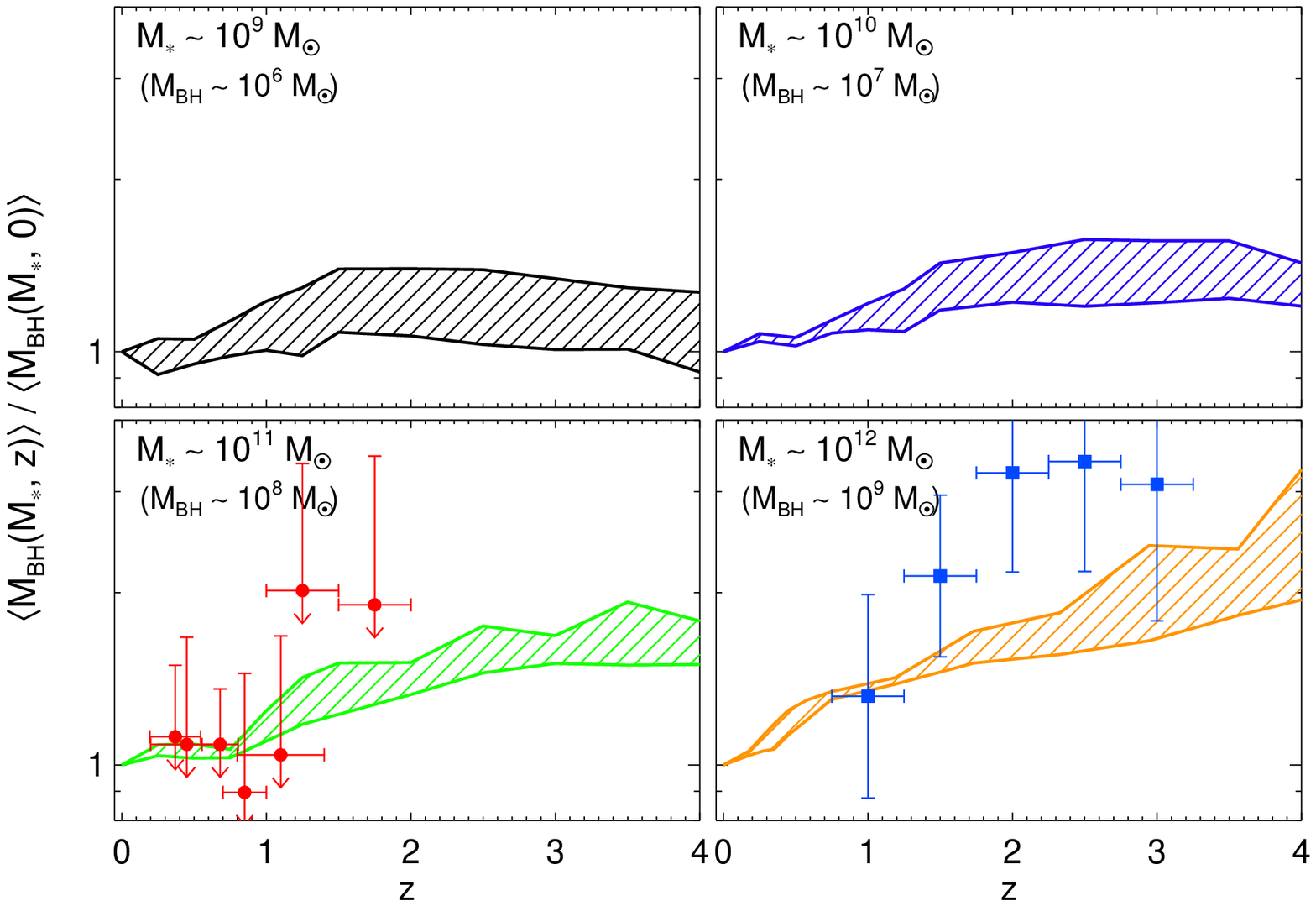}
    \caption{Predicted evolution in $\mbh$ at fixed stellar mass with redshift 
    (style as Figure~\ref{fig:reff.z}). Red circles with arrows show the upper limits 
    from observed evolution in spheroid mass 
    density in \citet{hopkins:old.age} \citep[see also][]{merloni:magorrian.evolution}, 
    applicable to $\sim M_{\ast}$ 
    galaxies. Blue squares show the observational estimates of evolution for 
    massive black holes and galaxies from \citet{peng:magorrian.evolution}. More dissipation 
    at high redshifts means more work for the black hole to do before its 
    growth self-regulates, giving rise to larger $M_{\rm BH}$ (for fixed 
    feedback efficiency); however, the net effect is relatively weak. 
    Selection effects \citep{lauer:mbh.bias} may explain 
    the small difference between $M_{\rm BH}/M_{\ast}$ observed and predicted. 
    The trend is mass-dependent in the same manner 
    as Figure~\ref{fig:reff.z}. 
    \label{fig:mbh.ms.z}}
\end{figure}

\begin{figure}
    \centering
    %\scaleup
    %\plotone{mbh_sig_z_mbins.ps}
    \plotone{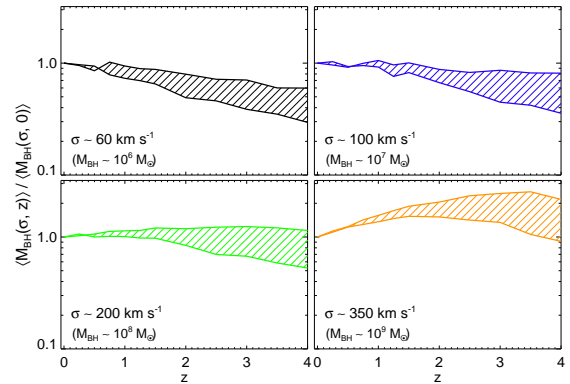}
    \caption{Predicted evolution in $\mbh$ at fixed velocity dispersion 
    $\sigma$ with redshift (style as Figure~\ref{fig:reff.z}). 
    The complex interplay between evolution in $M_{\rm BH}$, $R_{e}$, 
    and $\sigma$ yields weak trends here that are not necessarily 
    monotonic. In general the predicted $M_{\rm BH}/\sigma$ evolution is 
    weaker than (or inverse to) that in $M_{\rm BH}/ M_{\ast}$, 
    but at $z\sim2-3$ massive black holes may be overmassive 
    relative to $\sigma$ by a factor $\lesssim2-3$ (owing to the evolution 
    in $M_{\rm BH}/ M_{\ast}$ and weak evolution in $\sigma/M_{\ast}$). 
    \label{fig:mbh.sigma.z}}
\end{figure}

Figure~\ref{fig:mbh.ms.z} \&\ \ref{fig:mbh.sigma.z} show the evolution in 
BH masses at fixed stellar mass and velocity dispersion, respectively. 
Because systems are more dissipational at high redshifts, the binding energy of 
the baryonic material which must be supported or expelled to halt accretion is 
higher (at fixed stellar mass). As a consequence, the typical BH masses 
are larger. We discuss this evolution, its physical causes, and consequences, 
in greater detail in \citet{hopkins:bhfp.theory}. Here, we note that the same conclusions 
hold, adopting a more complex cosmological model than was considered in that 
paper.

The evolution is similar to that suggested by a number of recent observations 
\citep[e.g.][]{shields06:msigma.evolution,
walter04:z6.msigma.evolution,peng:magorrian.evolution,woo06:lowz.msigma.evolution,
salviander:msigma.evolution} 
and indirect estimations \citep{hopkins:old.age,merloni:magorrian.evolution,
adelbergersteidel:magorrian.evolution,
wyithe:magorrian.clustering,hopkins:clustering,lidz:clustering}
but we emphasize that it is not strong (it is in fact on the 
lower $\sim1\sigma$ end of the estimated evolution) -- typical BHs are 
a factor $\lesssim2$ larger at $z\sim2-3$. Larger differences from the 
local relations (in particular in very luminous quasar populations) may 
result from selection biases \citep{lauer:mbh.bias}, where the most 
luminous quasars (most massive BHs) are likely to represent the upper 
end of the scatter in the BH-host correlations. The combination of such a 
selection effect with our predicted evolution is a very good match to the 
evolution in \citet{peng:magorrian.evolution}. 

We also compare with the constraints 
in \citet{hopkins:old.age}, derived from observations of the evolution in the 
spheroid mass density with redshift (since the black hole mass density cannot, in 
any reasonable model, decrease, the maximum evolution in the 
typical ratio $\mbh/\mstar$ is limited by the fact that the resulting black hole 
mass density, predicted from the observed spheroid mass functions, must 
not be higher than that observed at $z=0$). This constraint 
is most directly applicable to $\sim L_{\ast}$ galaxies, since this is where most of the 
mass density of the universe resides. These constraints limit evolution to a 
factor $\lesssim 2$ at $z=2$, consistent with our predictions. Comparing 
to similar estimates from \citet{merloni:magorrian.evolution}, who adopt a similar approach 
but model the evolution in the black hole mass density from quasar luminosity functions, 
yields a similar constraint and expectation of 
weak evolution, $M_{\rm BH}/M_{\ast}\sim(1+z)^{0.4-0.6}$, 
consistent with our predictions. 

The evolution in $M_{\rm BH}/\sigma$ is somewhat more complex, reflecting the 
full interplay between evolution in $M_{\rm BH}$ with spheroid binding energy 
and evolution in $\sigma$ with redshift (discussed above). Because 
deeper potentials will, in general, be reflected in higher $\sigma$ values, 
we expect weaker (or even inverse) evolution here, relative to $M_{\rm BH}/M_{\ast}$ 
(where $M_{\ast}$ was not a measure of the central potential depth, so 
deeper potentials at fixed $M_{\ast}$ translated to higher $M_{\rm BH}$). 
However, because $\sigma$ reflects a combination of the halo and galaxy over a 
significant dynamic range, the results are non-trivial and can be non-monotonic in 
redshift. For example, in our most massive bin, the predicted evolution in 
black hole masses is sufficiently strong (and evolution in $\sigma$ relatively weak 
as a reflection of the tradeoff between dark matter and baryonic potential) 
that evolution to larger $\mbh/\sigma$ (albeit at the factor $\lesssim2-3$ level) 
may be expected, consistent with recent observations 
\citep{shields06:msigma.evolution,salviander:msigma.evolution}, although 
again selection effects are probably important and may explain 
cases where the inferred evolution appears to be much larger.

\subsection{Effects of Dissipation}
\label{sec:evol:diss}

\begin{figure}
    \centering
    %\scaleup
    %\plotone{r_z_mbins_nodiss.ps}
    \plotone{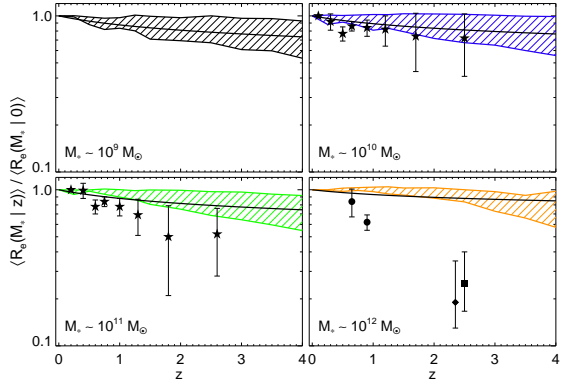}
    \caption{Predicted evolution in spheroid sizes with redshift, as 
    Figure~\ref{fig:reff.z}, but in a model with no dissipation (gas is treated 
    the same as stars in mergers). Even allowing for evolution in disk 
    sizes with redshift, there is essentially no evolution predicted in 
    spheroid sizes without dissipation 
    (other effects can make the evolution in spheroid sizes 
    {\em weaker} than that in disk sizes, which is already observed to be weak). 
    \label{fig:reff.z.nodiss}}
\end{figure}

We have argued that these effects are largely driven by varying degrees of dissipation 
as a function of redshift. We briefly illustrate this here by repeating our predictions 
for the size evolution of ellipticals (which, as we showed in \S~\ref{sec:evol}, is the 
most pronounced evolution) in the version of our model where we ignore 
dissipation or treat gas identically to stars (as in \S~\ref{sec:z0:diss:effects}). 
Figure~\ref{fig:reff.z.nodiss} shows the results. Without 
evolution in gas fractions driving evolution in 
the degree of dissipation, only the (weaker) evolution in progenitor disk sizes 
has a noticeable effect, and the evolution is substantially suppressed. 
In fact, the predicted evolution can, in this case, be even weaker than 
that in the progenitor disks, because ellipticals with a given final mass 
tend to undergo their mergers over a similar redshift range -- so their final sizes 
will reflect whatever the sizes of disks were at that time, whether or not 
those disk sizes evolved later.

\begin{figure}
    \centering
    %\scaleup
    %\plotone{mbh_ms_z_mbins_nodiss.ps}
    \plotone{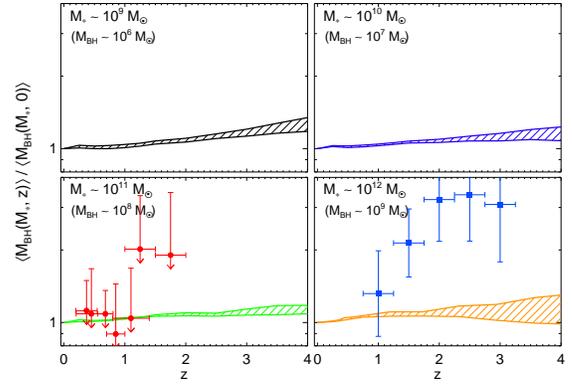}
    \caption{Predicted evolution in $\mbh$ at fixed stellar mass with redshift, 
    as Figure~\ref{fig:mbh.ms.z}, but in a model with no dissipation. 
    The predicted evolution for massive black holes is almost entirely 
    a consequence of dissipation. 
    \label{fig:mbh.ms.z.nodiss}}
\end{figure}

A similar 
(relative) effect is seen for the evolution in $\mdyn$ and the FP, as well as 
BH masses. We show the evolution in black hole mass at fixed stellar mass 
predicted without dissipation in Figure~\ref{fig:mbh.ms.z.nodiss}. As expected 
based on the evolution in $R_{e}$, the evolution here is essentially 
eliminated without dissipation.

\subsection{Effects of Dry Mergers} 
\label{sec:evol:dry}

\begin{figure}
    \centering
    %\scaleup
    %\plotone{r_z_mbins_nodry.ps}
    \plotone{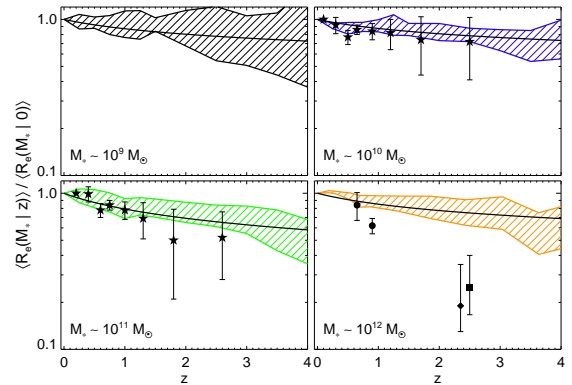}
    \caption{Predicted evolution in spheroid sizes with redshift, as 
    Figure~\ref{fig:reff.z}, but in a model with no dry mergers (structural change in 
    spheroid-spheroid mergers is suppressed). The predicted evolution is 
    essentially identical at $M_{\ast}\lesssim10^{11}\,\msun$, but weaker 
    for very massive systems, which can no longer be puffed up at 
    low redshift to match observations. 
    \label{fig:reff.z.nodry}}
\end{figure}

At the high-mass end, there is also some evolution 
because systems merging early will have dry mergers puff up their sizes at 
later times. Figure~\ref{fig:reff.z.nodry} shows the evolution in the size-mass relation, with 
dry mergers suppressed (as in \S~\ref{sec:z0:dry}). The difference here is 
real, but small. Dry mergers -- while important for the size 
evolution of individual objects -- 
do not have a dramatic effect on the overall evolution 
of the size-mass relation. If the ``pre dry-merger'' 
size-mass relation is a power-law, $R_{e}\propto\mstar^{\alpha}$, then 
by the energy conservation argument in Equation~(\ref{eqn:egy}), a merger with 
mass ratio $f$ ($f\le1$) of two systems initially on this 
correlation will have a final effective radius a factor $(1+f)^{2}/(1+f^{2-\alpha})$ 
larger than the 
primary. Relative to the size-mass relation, the system has moved above 
the relation by a factor $(1+f)^{2-\alpha}/(1+f^{2-\alpha})$. 

For observed values of $\alpha$ in disks or ellipticals ($\sim0.3$ and $\sim0.6$, respectively)
this amounts to only $\sim0.1$\,dex evolution per 
major ($f\gtrsim1/3$) merger, compared to e.g.\ $\sim0.3$ dex scatter in the correlation. 
This is why, in Figure~\ref{fig:fp.proj.nodry}, the absence 
of dry mergers makes a difference only 
at the highest end of the mass function, where the number of major dry mergers 
might be large. Most of the population did not experience substantial dry merging, and 
what did, experienced it mainly at late times. As a result, the inferred redshift evolution 
without dry mergers is largely the same, with only an offset in the rate of size evolution 
at $z\lesssim1-2$ for the most massive systems (leading to an offset in the 
relative size prediction at $z\sim3$). 

It should be emphasized that the difference seen at the highest masses and redshifts 
in Figure~\ref{fig:reff.z.nodry} pertains by definition to those ellipticals which are 
{\em already} that massive at those redshifts -- i.e.\ systems which have formed and 
assembled $\sim10^{12}\,\msun$ by $z\gtrsim2-3$. These systems obviously 
live in the most dense environments, and are very rare -- their evolution is strong 
because they are expected to undergo a large number of dry mergers, as well 
as mergers with later-forming gas-poor disks and spheroids and minor 
mergers (see \S~\ref{sec:track} 
below) that will build up an extended envelope and rapidly increase their sizes. 
The typical system with the same mass observed at lower redshifts 
was not fully assembled at these early times, but forms in less 
overdense regions and (although star formation in progenitors may have ceased 
at early times) assembled more recently, in a smaller number of dry mergers. 
So the evolution seen should not be taken to imply a large number of dry mergers 
for the average massive galaxy, but rather to imply a large number  of 
subsequent mergers (``large'' being $\sim$ a few; again see \S~\ref{sec:track} 
for discussion of how this 
can efficiently increase the spheroid size) for, in particular, galaxies 
in extreme environments that are already massive at early times.

\breaker
\section{Evolution of Individual Systems: What Happens to the Systems 
Formed at High Redshift?}
\label{sec:track}

We have predicted that high-$z$ massive ellipticals should be substantially more compact 
than low-$z$ ellipticals of the same mass, in apparent agreement with 
observations. If this is true, then what happens to these early-forming, compact ellipticals 
as they evolve to $z=0$? 

\begin{figure*}
    \centering
    %\scaleup
    %\plotone{mc_demo_hist.ps}
    \plotone{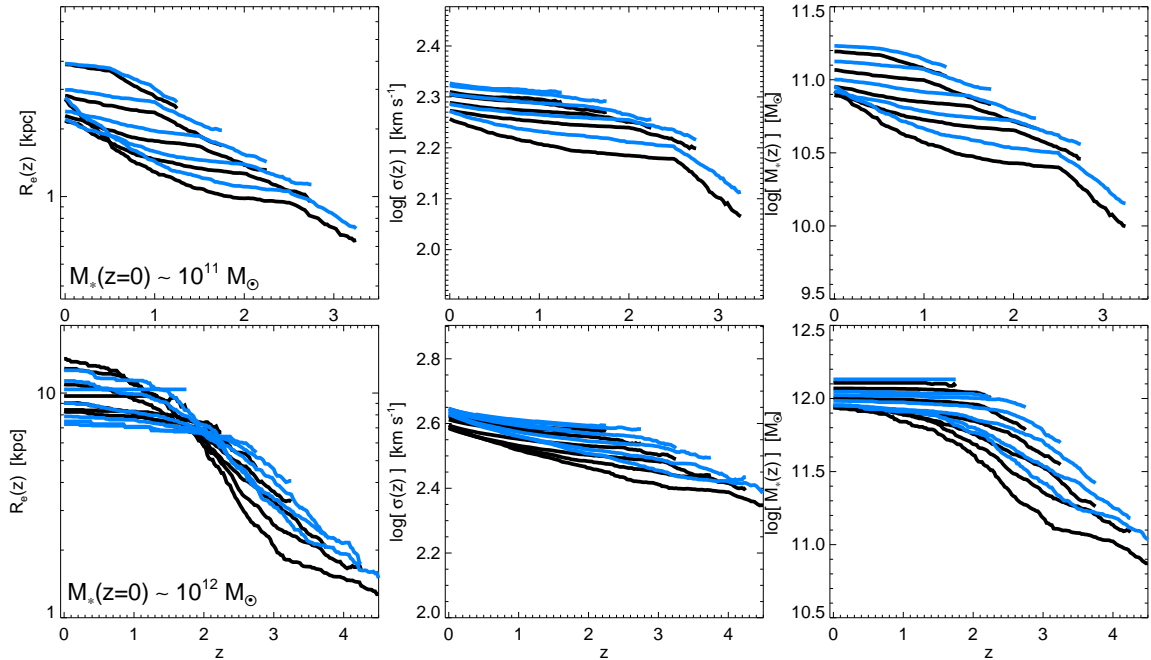}
    \caption{Median size ({\em left}), velocity dispersion ({\em middle}), and 
    instantaneous stellar mass ({\em right})
    as a function of redshift, for various galaxies 
    that have similar $z=0$ stellar mass (labeled). 
    These median tracks 
    are shown for systems that have their first major merger (that first form a 
    spheroid) within narrow intervals in $z$ (shown where each line begins: e.g.\ the line 
    beginning at $z\sim3.5$ shows the median track of all systems which have their first 
    major merger at $z=3.5\pm0.25$ and will have the same labeled stellar mass 
    at $z=0$). Different colors show different variations in our model, and reflect the 
    uncertainties in redshift evolution (as the shaded range in Figure~\ref{fig:reff.z}). 
    Systems which form early are relatively compact, but grow in mass owing to 
    subsequent mergers with gas-poor disks and other spheroids. 
    \label{fig:tracks}}
\end{figure*}

\begin{figure}
    \centering
    \scaleup
    %\plotone{mc_demo_sizemassplane.ps}
    \plotterr{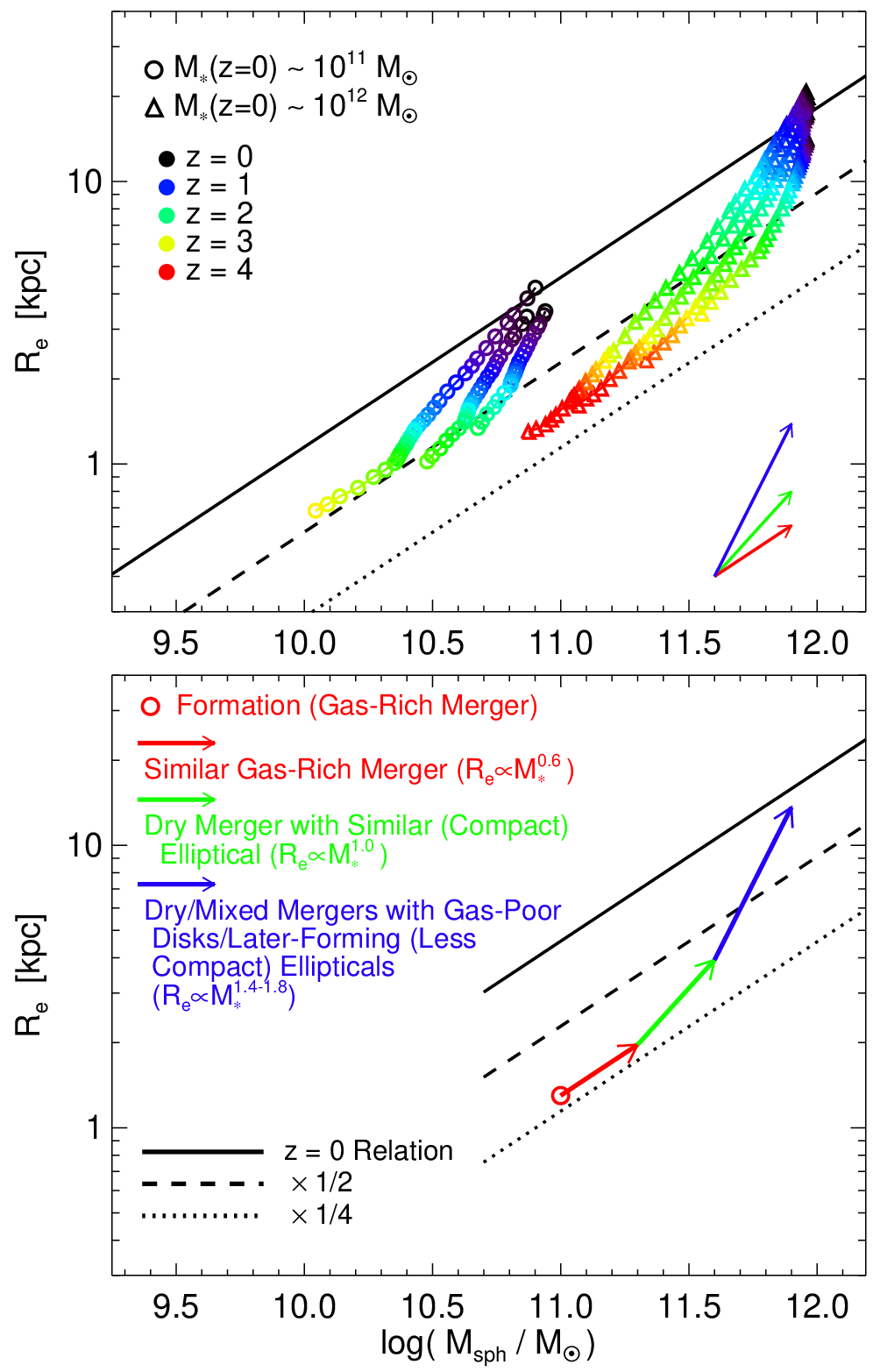}
    \caption{Tracks followed by early-forming systems from 
    Figure~\ref{fig:tracks} in the size-stellar mass plane. 
    {\em Top:}
    Symbol shape denotes the $z=0$ stellar mass (labeled), 
    and color denotes redshift (from black at $z=0$ to red at $z=4$, as labeled). 
    Points show the median tracks as in Figure~\ref{fig:tracks} for systems which 
    first formed at a given redshift (the beginning of each plotted track). 
    Lines show the $z=0$ size mass relation (solid), with slope 
    $R_{e}\propto M_{\ast}^{0.6}$, and the same divided by a factor of 2 (dashed) 
    and 4 (dotted). 
    {\em Bottom:} Vectors illustrate motion in the plane owing to different 
    types of mergers, for a system that will have a mass $M_{\ast}\sim10^{12}\,\msun$ 
    at $z=0$ but first forms as a compact $M_{\ast}\sim10^{11}\,\msun$ 
    spheroid ($\sim1/4$ the size of a typical similar-mass system at $z=0$) 
    at $z\sim3-4$ in a gas-rich merger. 
    As most systems are still gas-rich at these times, the next merger ($z\sim3-4$) 
    may be similarly gas-rich, moving the system along the 
    $R_{e}\propto M_{\ast}^{0.6}$ slope (maintaining the relative 
    compact-ness; {\em red}). In biased regions, nearby systems 
    are also rapidly transformed into ellipticals (which, having their initial mergers 
    at early times from gas-rich disks, are also compact). A 1:1 merger with 
    an identical elliptical will double $M_{\ast}$ and $R_{e}$
    ($R_{e}\propto M_{\ast}^{1.0}$; {\em green}). At later times, systems 
    merging will be newly accreted (into the halo), 
    evolved (relatively gas-poor) disks, or ellipticals made 
    at later times from such evolved disks (which, being formed in more 
    gas-poor mergers, will be less compact). Such a dry/mixed merger 
    (with a less compact system) moves the system up more rapidly than 
    dry mergers with identical systems ($R_{e}\propto M_{\ast}^{1.4-1.8}$; {\em blue}). 
    \label{fig:tracks.sizemass}}
\end{figure}

We are now interested in the evolution of individual systems, from their formation redshift 
to $z=0$, rather than the evolution of the population at the same mass (recall, 
the systems with $\mstar\sim10^{11}\,\msun$ at $z\sim2-3$ are {\em not}, generally, 
the same galaxies that have $\mstar\sim10^{11}\,\msun$ at $z=0$). 
Figure~\ref{fig:tracks} shows such a case study, in fact a set of them, tracking the 
median (averaged over many Monte Carlo realizations) evolution in sizes of 
individual systems that are formed at a given redshift and will have a 
specific stellar mass at $z=0$. 
For two $z=0$ stellar masses where evolution is significant, we 
show the evolution in effective radii, velocity dispersions, and stellar masses. 
Figure~\ref{fig:tracks.sizemass} shows the same, but projects a 
few of the representative median 
tracks (for early-forming systems) 
into the size-stellar mass plane, to show how the systems move as a 
function of time and merger history
in this space relative to e.g.\ the observed $z=0$ correlation.

The systems which have their first major merger at some early time are indeed much 
more compact than their $z=0$ descendents. Partly, this is because they begin life 
as lower-mass systems, but even for their (instantaneous) 
stellar mass they are more compact than $z=0$ objects of that mass, 
reflecting the higher degree of dissipation (the evolution in Figure~\ref{fig:reff.z}). 
However, by $z=0$, systems of the same stellar mass from different formation 
epochs all have effective radii within a narrow range (a factor of $\sim2$). 
The systems which form late ($z\lesssim1$) 
evolve weakly -- they rarely experience any dry mergers, so their $z=0$ sizes 
largely reflect their sizes at formation. However, the systems which form 
early ($z\gtrsim3$) evolve strongly -- they experience a number of 
both mergers with lower-redshift, larger and much less gas-rich disks, 
as well as other spheroids (true spheroid-spheroid dry mergers).  
Such mergers increase the sizes of the ellipticals both in absolute 
terms and in relative terms (moving them above the size-mass relation 
representative of the time when they formed). Although the 
number of major mergers for most systems is relatively small, 
it is a function of their formation time. The systems which first merge at 
earliest times do so because they live in highly biased environments -- they 
are precisely those expected to undergo the most dry merging. 
As a result, the effects of evolving dissipational content and the effects of 
dry mergers appear to conspire such that, despite a significant evolution in the 
size-mass relation with redshift, by $z=0$ ellipticals formed 
at early times will have broadly similar sizes to those forming around $z\sim0$. 

If we examine Figures~\ref{fig:tracks}-\ref{fig:tracks.sizemass}, we can plot out the likely 
course of evolution for the compact, massive passive galaxies seen at 
$z\sim3$ in \citet{kriek:drg.seds,
zirm:drg.sizes,labbe05:drgs,wuyts:irac.drg.colors,
daddi05:drgs,vandokkum:z2.sizes,franx:size.evol,damjanov:red.nuggets}. 
At the redshifts observed, these are systems which have formed 
recently (systems which form earlier will be even more rare) in a 
gas-rich merger \citep[in agreement with their young observed ages, 
relative to $z=0$ spheroids of the same mass;][]{kriek:drg.seds}, 
very high gas fractions $\gtrsim30-40\%$ characteristic of 
disks at these and higher redshifts. 
The large dissipational fraction 
yields a massive galaxy ($M_{\ast}\gtrsim10^{11}\,\msun$) 
with a very small effective radius $\sim1$\,kpc. As discussed in \S~\ref{sec:evol} 
above, however, the velocity dispersion is not 
extremely large (if we took $\sigma(z)=\sigma(\mstar,0)\times\,R_{e}(0)/R_{e}(z)$, 
we would estimate $\sigma\sim600-800\,{\rm km\,s^{-1}}$; 
however, the predicted velocity dispersion is in fact $\sim250-300\,{\rm km\,s^{-1}}$). 
Because the system is so much more compact, the dark matter fraction within 
$R_{e}$ is much smaller than in a typical $z=0$ spheroid of the same mass, 
so its contribution to the velocity dispersion is much smaller. 

The system, 
having formed in a high density environment \citep[consistent with 
their observed clustering and number densities; e.g.][]{quadri:highz.color.density,
vandokkum06:drgs}, 
will undergo a 
significant number of mergers between this time and $z\sim1-2$, where 
the merger rate begins to level off. These mergers are with a mix of 
disks and ellipticals. At times shortly after formation, successive mergers 
may be with other gas-rich disks, yielding similar amounts of 
dissipation and keeping the system relatively compact. 
This is illustrated in Figure~\ref{fig:tracks.sizemass} as a 
``similar gas-rich merger''; which will move the system parallel to 
the size-mass relation ($R_{e}\propto M_{\ast}^{0.6}$). 
Soon, nearby systems will (given the biased environment) be 
largely elliptical -- but they may be similar, compact ellipticals. 
Spheroid-spheroid dry mergers will puff up the system, 
moving it up somewhat relative to the initial size-mass relation at 
formation. This is shown as a ``dry merger with similar 
(compact) elliptical'' in Figure~\ref{fig:tracks.sizemass} -- 
here merging two identical (equally compact) ellipticals doubles 
both $R_{e}$ and $M_{\ast}$ (see \S~\ref{sec:model:sims:dry}), 
moving the system ``up'' relative to the size-mass 
relation along the steeper axis $R_{e}\propto M_{\ast}^{1.0}$. 

As time goes by (in a hierarchical scenario), 
the galaxies merging will have formed in a more extended 
physical region over longer periods of time (being recently accreted into 
the parent halo); merging disks at later times will be progressively 
more gas-poor (characteristic of low-$z$ disks, which have consumed their 
gas in star formation), and thus contribute dissipationless stars with 
very large effective radii, building up the stellar component 
(an extended, violently relaxed envelope) at large radii. Likewise, pure 
spheroid-spheroid mergers will include spheroids that formed 
later from such less gas-rich disks (and are therefore less compact), 
and these will serve to puff up the system even more efficiently 
and build up an envelope consistent with that typically observed in 
BCGs at $z=0$ (the expected location of such systems at $z=0$, 
based on their clustering properties and inferred halo masses). 
This is denoted in Figure~\ref{fig:tracks.sizemass} by the label 
``dry/mixed mergers with gas-poor disks/later-forming (less compact) 
ellipticals,'' and allows the system to move more rapidly 
``up'' in size relative to the size-mass relation (along an 
axis $R_{e}\propto M_{\ast}^{1.4-1.8}$, depending on the details of the 
type of mergers involved). This enables systems that form compact 
for their mass to migrate to the $z=0$ size-mass relation in a 
relatively small number of mergers 
(in the illustrative example in Figure~\ref{fig:tracks.sizemass}, the 
system moves from being $1/4$ the size of a $z=0$ analogue to 
lying on the local relation in just two equal-mass 
dry mergers, one with a similarly compact elliptical, one 
with a later-forming, less compact elliptical; the latter merger 
builds a more extended envelope and therefore moves the 
system more rapidly towards agreement with the $z=0$ relation). 

At sufficiently large masses (especially for galaxies at the center of 
massive clusters), minor mergers will increasingly dominate the 
growth of the galaxy \citep[see e.g.][]{maller:sph.merger.rates,zheng:hod.evolution,
masjedi:cross.correlations}; these will have similar effects to those described 
here, and may even further increase the size at $z=0$ (especially if small 
merging systems are disrupted as they merge; in this case they will 
not contribute much to the central structure of the galaxy, but will increasingly 
build up an extended envelope at large radii). More detailed modeling of the structure 
of e.g.\ BCGs and the most massive ($M_{\ast}\gtrsim10^{12}\,M_{\sun}$) 
galaxies should account for both major and minor mergers in dense environments. 

The effective radius therefore rapidly grows, and by $z=0$ the system 
lies on (or even somewhat above) the median size-mass relation for 
systems of the same $z=0$ stellar mass. The velocity dispersion grows 
slightly as mass is accumulated, but since most of this mass is contributed 
from dissipationless components to the extended envelope (recall, 
a merger of two identical spheroids will leave $\sigma$ unchanged), it 
does not contribute much, and the $z=0$ galaxy has a peak 
(central) velocity dispersion $\approx400\,{\rm km\,s^{-1}}$, large but only 
$30\%$ larger than its initial central velocity dispersion and 
completely consistent with those observed in the most massive galaxies 
today \citep{bernardi:most.massive}. 
In detail, in fact, the predicted descendants of 
early compact systems here have similar abundances, 
velocity dispersions, and remarkably similar predicted locations in e.g.\ the 
$z=0$ size-mass, fundamental plane, and Faber-Jackson relations 
to the sample in \citet{bernardi:most.massive}, suggesting that 
many of those systems may be the products of this process. 
For the most part, then, these extreme systems 
are completely consistent in all the properties we can predict here 
(effective radius, mass, velocity dispersion, black hole mass, and 
profile shape/Sersic index) with their constituting the central $\sim10-30\%$ of 
the mass in a significant fraction of the brightest group or cluster galaxies
today.

\begin{figure}
    \centering
    %\scaleup
    %\plotone{MC_z_last.ps}
    \plotone{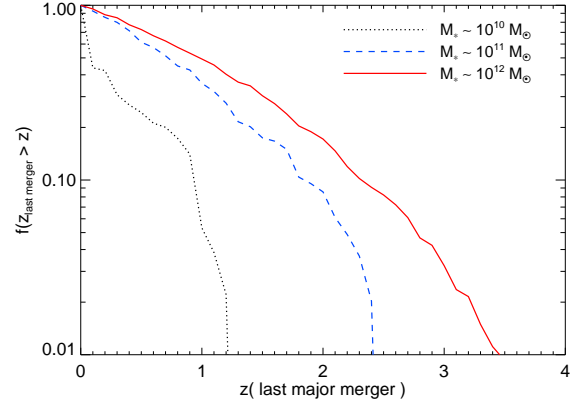}
    \caption{Fraction of $z=0$ spheroids of a given stellar mass which had their 
    last major merger (gas rich or gas poor) above a given redshift. 
    A few percent of massive systems today are expected to have survived 
    from a much earlier formation time $z\sim2-3$ without any subsequent 
    mergers modifying their structure -- i.e.\ be remnants of the era where 
    scaling relations were different. These ellipticals would generally be compact 
    for their mass, with old ages, and strongly clustered. 
    \label{fig:surviving.frac}}
\end{figure}

This can explain why such objects are not ubiquitous in the
local Universe. However, we could still 
ask what fraction of such systems might be expected to survive to $z=0$, 
without disappearing into a larger $z=0$ galaxy. In 
short, should any of this massive and compact population exist at $z=0$, which 
might represent the direct remnants of high-redshift, compact ellipticals? 

Figure~\ref{fig:surviving.frac} shows the fraction (at a given $z=0$ stellar mass) of objects 
which had their last major merger of any kind (which will largely 
set their observed size) at (or above) a given redshift. At 
masses corresponding to those observed in high-redshift compact galaxy populations 
($\sim10^{11}\,\msun$), the fraction of the population today which has survived 
relatively intact since $z\gtrsim2$ is expected to be small, $\sim1-10\%$. 
This translates to a $z=0$ space density of such systems 
$\sim10^{-4}\,{\rm Mpc^{-3}}$: not large, but not completely negligible. 
This is an upper limit to the number density of such systems: if they have experienced a 
sufficiently large number of minor mergers, they may be significantly altered 
by $z=0$ (although many of the same reasons that allow some systems to avoid 
major mergers will allow them to avoid a large number of minor mergers). 
At any redshift, however, there is a scatter in elliptical sizes owing to a scatter in 
e.g.\ the dissipational fractions at the time of their formation (reflecting scatter 
in disk gas fractions). Since disk gas fractions have significant (factor $\sim1.5-2$) 
scatter, that is actually expected to dominate the scatter in e.g.\ the size-mass relation 
at $z=0$ (rather than scatter in the formation times of ellipticals). That is 
to say, there could be a similar fraction of
ellipticals that are {\it too compact} present at $z=0$ 
which formed recently (not at $z>2$), from the most anomalously gas-rich disks 
at lower redshifts (say the $\sim2\,\sigma$ outliers 
in the disk gas fraction-mass relations). This probably will not explain systems 
as small as $\sim1\,$kpc at these stellar masses, but it illustrates that some additional 
indicator is needed to distinguish those possibilities. If stellar population ages could 
be measured, this is straightforward: the system which formed at low redshift will have 
significant recent star formation (in fact, quite a lot, since we are interested in 
compact systems with large dissipational fractions). So it should be possible, in principle, 
to identify the descendents of the early-forming ellipticals. 

One important caveat to this discussion is that we have focused here on the 
role of major mergers with mass ratios below 1:3. 
Our calculation and other halo occupation models \citep{zheng:hod.evolution}, 
clustering measurements \citep{masjedi:cross.correlations}, 
and cosmological simulations \citep{maller:sph.merger.rates} 
suggest that as galaxies approach the extreme $M_{\ast}\gg 10^{12}\,M_{\sun}$ 
BCG populations, growth by a large series of minor mergers becomes 
progressively more 
important than growth by major mergers. Our predictions in this limit should 
therefore be treated with some degree of caution, 
in contrast with the bulk of the elliptical population at masses 
$\lesssim$ a few $L_{\ast}$, the vast majority of which still live in less
extreme environments \citep[e.g.][]{blanton:env,wang:sdss.hod,
masjedi:merger.rates} and are expected to 
have experienced a small number ($\sim$ a couple) of major re-mergers 
since redshifts $z\sim2-4$. Our modeling can, in principle, be extended to 
include minor mergers, and doing so we find qualitatively similar conclusions 
(because merger histories are roughly self-similar, the relative growth 
and courses of evolution seen in Figure~\ref{fig:tracks} are conserved). 
However, as demonstrated in \citet{hopkins:groups.ell}, our predictions 
(and those of other halo occupation-based approaches) become considerably 
less robust in the regime of minor mergers (mass ratios $\sim$1:10 or so), 
owing to greater uncertainty in the satellite mass function at these 
mass ratios and weaker observational constraints on the halo occupation of 
small galaxies. 

In any case, because early-forming, compact systems were massive and experienced 
mergers at such early times, many will represent the primary galaxies 
in the highest density peaks, being BCGs today. Indeed, this is true for most of 
the massive high-$z$ spheroid population. However, in order to 
{\em survive} to $z=0$ from this population, mergers must be avoided (by our 
definition) -- something very hard for BCGs. We would still expect to find 
such objects around high-density peaks (there is no getting around 
the fact that they will only form so early and so massive in high-density peaks), 
but those that survive, in many cases, would be expected to do so 
precisely because they were {\em not} the central galaxy or BCG. (In detail, they 
probably were the central galaxy or BCG at their time of formation, but 
were displaced from this position in subsequent halo-halo mergers, 
becoming a satellite in a still larger system, suppressing their merger history). 

There may be some interesting candidates for such objects 
among the so-called ``compact elliptical'' family. For example 
(although we intend this only as an illustrative case), NGC 4486B is a 
satellite of NGC 4486 in Virgo. It has a stellar mass $\gtrsim10^{10}\,\msun$ 
(\papertwo) and 
velocity dispersion $\sigma=200\,{\rm km\,s^{-1}}$, but an effective 
radius of only $200$\,pc \citep[see e.g.][]{jk:profiles}. This is somewhat smaller than 
the $\sim10^{11}\,\msun$ which we just described, but the scenario 
can be generalized to masses about this low. 
The stellar population age is quite old, $10.5$\,Gyr 
\citep[perhaps even up to $\sim13$\,Gyr, 
depending on the indicator used;][]{caldwell:ssps}, implying a time of 
last significant star formation $z\gtrsim2$. 
This is further supported by a corresponding $\alpha$-enrichment and metallicities. 
The system has high rotation and ellipticity suggestive of a very gas-rich formation 
\citep{bender:88.shapes,bender:ell.kinematics}. 
The estimated dynamical friction time is sufficiently long that the system could have become 
a satellite in the Virgo progenitor as early as $z\sim2$ and still not merged with the 
central galaxy. All of these lines of evidence suggest it may have 
survived from an early formation 
time as a very compact system. Unfortunately, except for the direct stellar population 
age measurements, these arguments are mostly circumstantial, and those 
stellar population estimates are still quite uncertain. Better determinations of the 
star formation history in this object would be an important test of this hypothesis. 

\breaker
\section{Conclusions}
\label{sec:discuss}

We have combined the results from a large library of numerical simulations with 
simple, observationally constrained models for the halo occupation 
(and corresponding merger history) of galaxies in order to predict 
how spheroid scaling laws evolve. We demonstrate that, despite 
a variety of progenitor properties, complex merger histories, and evolution in 
e.g.\ progenitor sizes and gas content with redshift, the most important 
property driving the structural scaling relations of spheroids at each redshift 
(and, in particular, making these scaling relations {\em different} from those 
obeyed by disk galaxies) is the amount of dissipation involved in the 
formation of the spheroid. At the highest masses, dry mergers at relatively 
late times are also important, but are few, consistent with 
observational constraints. Together, these predict a significant mass-dependent 
evolution in spheroid scaling laws, which has important consequences 
for the fundamental plane, spheroid densities, and the formation histories of 
massive galaxies. 

At $z=0$, a simple model accounting for the effects of dissipation in mergers 
is able to explain the size-mass, velocity dispersion-mass, and fundamental 
plane (dynamical-stellar mass) correlations, the 
relation between galaxy profile shape (Sersic index) and mass, 
the correlation between black hole mass and various host properties 
(bulge binding energy, mass, velocity dispersion, and profile shape), 
the relative mass fractions in compact dissipational components in ellipticals, 
and the abundance of cuspy/disky/rapidly rotating versus cored/boxy/slowly 
rotating ellipticals \citep[see][]{hopkins:groups.ell}. 
Dissipation in the initial disk-disk mergers that form the spheroid 
progenitors is the most important parameter controlling these relations. 

In particular, at high masses, where disks are gas-poor, the sizes, densities, 
and dynamical mass to stellar mass ratios of ellipticals reflect those of their 
progenitor disks. At lower masses, where disks are gas-rich, mergers involve a 
significant amount of new star formation in dissipational nuclear starbursts triggered 
by loss of energy and angular momentum in the gas in the merger, building up a 
compact central stellar distribution (the ``dissipational'' component) 
that yields a smaller, more dense remnant. This mass-dependence in the 
dissipational content of progenitor disks makes the size-mass relation of ellipticals 
steeper than that of disks and gives the observed tilt in the fundamental plane 
(essentially the statement that low-mass disks, being more dominated by this 
nuclear dissipational component, have higher baryon-to-dark matter fractions within 
their {\em stellar} effective radii). Black holes are formed with a mass that, in a simple 
feedback-regulated model, corresponds to the binding energy of the 
dissipational component of bulge at the time of this dissipation, which the black hole 
must work against in order to self-regulate its growth. This matches the 
detailed observations of a black hole fundamental plane in \citet{hopkins:bhfp.obs} 
and gives rise to the secondary (indirect) correlations between black hole 
mass and spheroid velocity dispersion, mass, size, and Sersic index (all of which 
agree with those observed). 

To lowest order, the dissipational fractions that give rise to these trends at 
$z=0$ reflect the observed gas fractions of disk galaxies at 
moderate redshifts $z\sim0-2$. This is because most ellipticals are formed in a 
relatively small number of major mergers at low redshifts in this range. 
It is only when one considers the most massive populations of e.g.\ BCGs and 
massive cored, boxy, slowly rotating ellipticals that dry mergers become important. 
Their effect is to puff up these systems, producing extended envelopes and 
raising their outer Sersic indices. This is sufficient to explain the observed weak dependence 
of outer Sersic index on mass in elliptical galaxies, following the 
detailed multi-component decompositions of observed systems 
in \papertwo\ and \paperthree. When systems are fitted to a single 
Sersic index, the result is a steeper dependence that reflects a complex 
combination of e.g.\ the outer Sersic index (primarily driven by merger history), 
the mass fraction in a dissipational component (yielding rising densities 
at small radius), the shape of the two components where they are comparable, 
and the dynamic range of the fit (since real ellipticals are not perfect Sersic profiles, 
there is some dependence, and in particular when fitting the entire profile to a single 
Sersic law, the dependence can be significant). 

The detailed interplay of dissipation and dry mergers also gives rise to 
important residual trends that can be used as tests of these models. For example, 
at a given stellar mass, the oldest systems will have the largest effective 
radii and dynamical masses. This is driven by two effects: first, disks which have 
high star formation efficiencies (earlier star formation times) 
will be more gas-poor at the time of their mergers -- i.e.\ will have less mass in a 
dissipational component. Second, systems which may be gas-rich but form very 
early (i.e.\ have a high redshift of their last gas-rich merger) tend to do so because 
they are in dense environments, and will therefore experience subsequent 
mergers with both more gas-poor disks and other spheroids, increasing their 
size and dynamical-to-stellar mass ratios. 

At higher redshifts, disk gas fractions are expected and observed to be 
systematically larger. As a result, spheroids observed at these redshifts 
are expected to have formed in more gas-rich mergers and 
to have a larger mass fraction in their dissipational component, making 
them more compact. 
We predict that elliptical sizes evolve in a mass-dependent fashion: 
since low-mass disks are still quite gas-rich at $z=0$, 
there is little room for them to become more so at high redshift, so 
they cannot evolve much. High mass disks, on the other hand, could in principle 
be much more gas rich than their observed $z=0$ $\lesssim10\%$ values. 
As well, at high masses, dry mergers become important in the 
predicted size evolution. Consequently, the highest mass ellipticals 
are expected to evolve to become a factor $\sim2-3$ smaller at $z\gtrsim2$. 
In terms of the mean effective radius of spheroids of a given 
stellar mass, relative to that at $z=0$, we find evolution of the form 
$(1+z)^{-\beta}$ where $\beta\approx0.23\,\log{(M_{\ast}/10^{9}\,\msun)}$ 
(of course enforcing $\beta>0$ or whatever minimum evolution is set 
by observed evolution in progenitor disk sizes at these masses). 

This is consistent with 
observations from e.g.\ \citet{trujillo:size.evolution,
mcintosh:size.evolution,zirm:drg.sizes,vandokkum:z2.sizes}, 
although we highlight important biases arising from 
stellar population gradients which are relevant when attempting to estimate 
the sizes of the most massive galaxies at $z\gtrsim2$. 
We demonstrate that this size evolution owes almost entirely to 
the change in dissipation with redshift, although dry mergers 
are of some importance for the most massive systems. Whether 
we explicitly include it or not, the evolution cannot be explained 
by simple scaling of disk sizes with redshift \citep[which, in any case, is 
observed to be weak; see e.g.][and references therein]{somerville:disk.size.evol}. 

This evolution has important implications for the other fundamental plane correlations. 
Because the fundamental plane tilt (at $z=0$) arises owing to systematically 
higher dissipational fractions in low-mass systems (making them more 
compact and yielding lower dark matter fractions within the stellar $R_{e}$), 
at high redshift (where even high-mass systems must be highly dissipational, 
since disk gas fractions are much higher) we expect the tilt to be weaker. 
Note that we refer here to the {\em stellar mass} fundamental plane, 
i.e.\ $M_{\rm dyn}$ versus $\mstar$ -- there will of course be stellar population 
effects introducing normalization and tilt changes in various observed bands. 
The predicted change is within the $z=0$ uncertainties at $z\le1$, but by $z\gtrsim3$, 
the tilt is predicted to largely disappear. This is important both as a test of these 
models, but also as a caution to observational attempts to use the fundamental plane 
as a detailed stellar population probe. It is certainly much less significant than the 
expected evolution in mass-to-light ratios in optical bands, but could be important 
for detailed stellar population probes such as that in \citet{vandokkum:imf.evol}
which assume a fixed stellar mass fundamental plane (although selection of 
e.g.\ satellite galaxies in clusters, which are less likely to undergo significant future 
merging relative to central galaxies in the field or groups, may be sufficient to mitigate 
these evolutionary effects). 
% and compare different 
% stellar population indicators and mass-to-light ratios (basically this physical evolution could 
% be mistaken for an ``excess'' or ``deficit'' of $\sim0.1$\,dex in stellar population 
% mass-to-light ratio evolution). 

We argue that velocity dispersions will 
change weakly with redshift, despite the evolution in 
effective radii, because systems with higher dissipational fractions (smaller $R_{e}$) 
are more baryon-dominated in their stellar effective radii (i.e.\ the velocity dispersion 
includes a weaker contribution from the halo). This is important, because velocity dispersion 
is in general preserved or increases in subsequent re-mergers. If high-redshift 
massive galaxies had extreme velocity dispersions $\gtrsim600\,{\rm km\,s^{-1}}$ 
(corresponding to the simple assumption that $\sigma$ is inversely 
proportional to just the stellar effective radius), then the appropriate number of 
systems with equal or higher velocity dispersions would have to exist at $z=0$ 
\citep[and they do not, even in survey volumes of the SDSS;][]{bernardi:most.massive}. 
However, accounting for the interplay between 
dark matter and baryonic mass, our predicted evolution in velocity dispersions is 
completely consistent with observational constraints. 

Furthermore, because of the increasing dissipational components and 
deepening potential wells at high redshift, we expect that, in any feedback-regulated 
model of black hole growth, black holes must become at least somewhat more 
massive (relative to their host spheroid stellar mass) at high redshift. We predict a 
similar mass-dependent evolution in $M_{\rm BH}/M_{\ast}$, with little or 
no significant evolution at $M_{\ast}\lesssim10^{10}\,\msun$ (corresponding to 
black holes with masses $\lesssim10^{7}\,\msun$), and factor 
$\sim2$ increase to $z\sim2-3$ in the masses of the black holes in the most massive 
systems ($M_{\ast}\gg10^{11}\,\msun$, corresponding to $\mbh\gtrsim10^{9}\,\msun$). 
Likewise, we predict the evolution in black hole mass relative to host velocity dispersion, 
which exhibits a more complex behavior. 
The physical details of driving these trends are discussed in more detail in 
\citet{hopkins:bhfp.theory}, but here we embed those models in a more fully 
cosmological context. The evolution predicted is consistent with recent 
observations \citep{peng:magorrian.evolution} 
and with indirect constraints based on the relative 
evolution of black hole mass density and spheroid mass density \citep{hopkins:old.age,
merloni:magorrian.evolution}, however it is still relatively weak. 
Selection effects \citep{lauer:mbh.bias} 
would be expected to explain the difference between some observational 
estimates that infer very strong \citep[e.g.][]{walter04:z6.msigma.evolution} or rapid 
low-redshift \citep{woo06:lowz.msigma.evolution} evolution 
and our predictions. 

We outline the history of these massive systems which are different 
when first formed at high redshift (where they are very dissipational) and 
observed at low redshift. Most of these compact, early forming ellipticals will 
undergo a significant number of dry mergers, or mergers with lower-redshift, 
more gas-depleted and larger disks, which will serve to puff up the remnant, building 
up some stellar mass and an extended stellar envelope (although changing 
the central velocity dispersion by a relatively small amount). This raises 
their effective radii more rapidly than their stellar mass, in the sense of 
moving them up in the size-mass relation. Interestingly, the evolution in disk 
gas fractions and the effects of such dry mergers almost exactly offset one another 
in massive systems. In other words, a $\sim10^{12}\,\msun$ galaxy which 
first formed from a merger at $z=3$ (which had a high gas content, leading to a very 
compact $\gtrsim10^{11}\,\msun$ initial galaxy with $R_{e}\sim1$\,kpc) 
will have almost the same size at $z=0$ as a 
galaxy of the same observed stellar mass which formed at much later time, 
say $z=1$, from a less dissipational merger (because disk gas fractions have 
decreased in that time interval). The system which formed earlier was initially more 
compact, but experienced more mergers in the intervening time, so for a given 
stellar mass, the scatter in $R_{e}$ is not large. Likewise, velocities dispersions 
and other fundamental plane properties have small scatter at a given stellar mass 
at any observed redshift because of this cancellation effect. 

This is not to say that most systems will undergo many dry mergers -- the overall 
density of dry mergers predicted here is low, consistent with observational 
constraints from e.g.\ \citet{bell:dry.mergers,vandokkum:dry.mergers,lin:mergers.by.type}, 
and as we noted above most 
(especially $\lesssim L_{\ast}$) ellipticals today formed in a few 
(or just one) gas-rich mergers at relatively low redshifts. However, the systems 
which form earliest do so because they live in the most dense environments, 
which evolve the most rapidly, and are expected to undergo the most subsequent 
merging activity. We predict, as a result, 
that most of these compact systems observed at high redshifts 
will end up constituting some of the central dense stellar mass in 
observed brightest cluster or group galaxies at $z=0$; this is consistent with 
the observed sizes \citep{vonderlinden:bcg.scaling.relations}, 
surface brightness profile shapes (extended envelopes 
build up through their subsequent dry merging) \citep[see e.g.\ 
\paperthree\ and][]{gallagherostriker72,
naab:dry.mergers}, 
and masses and velocity dispersions \citep{bernardi:bcg.scalings} of these $z=0$ systems, 
and also expected given the clustering properties of high redshift massive 
systems \citep{quadri:highz.color.density}. 

We estimate what 
fraction, as a function of stellar mass, of 
the massive galaxy population today may have survived from these early times un-remerged 
and thus reflect the smaller sizes and higher dissipational content characteristic of 
systems formed at that time, finding that e.g.\ $\sim1-5\%$ of $z=0$ 
$\sim10^{11}\,\msun$ systems may have survived since $z\sim2$. This is small, 
but not negligible, and by identifying compact ellipticals with old stellar population 
ages, it should be possible to locate the surviving remnants of this population. 
We predict that these will be massive, compact (with other indicators of gas-rich 
origin and high inferred dissipational fractions from their stellar profiles), 
with old stellar population ages. Because they still generally would initially form in 
high-density environments, but have avoided merging since then, we 
would roughly expect them to be in dense environments (rich groups and 
clusters) today, but {\em not} to be the central galaxy (again, most early-forming 
systems will be the central galaxy -- but if they are all the way until $z=0$, 
it is unlikely that they will have avoided further mergers, and thus would
not be 
un-remerged). There may be some observed candidate members of this 
population in the so-called ``compact elliptical'' class, galaxies similar to 
NGC 4486B in Virgo which appears to satisfy these criteria. Further observations 
to test this hypothesis could prove extremely valuable as nearby probes of 
high-redshift galaxy evolution.

Future work should extend these models to include the (at present 
more uncertain) role of particularly minor mergers 
(where e.g.\ satellite disruption may be important to explain the extended halos in 
massive BCGs) and other sources of dissipation, such as stellar mass 
loss \citep[see e.g.][]{ciottiostriker:recycling}. However, 
we show that other effects, e.g.\ evolution in the sizes of disk galaxies, 
dry mergers alone, and evolution in galaxy host halo properties, are insufficient to  
explain effects like those predicted here, in particular to account for 
e.g.\ the tilt of the fundamental plane, the relative sizes of ellipticals (versus those of 
disks) as a function of mass, the scaling of elliptical galaxy profile shapes with mass, 
and the redshift evolution in the fundamental plane 
correlations. Therefore, these predictions are important tests of any model 
in which dissipational star formation is a significant influence on galaxy formation. 

Because dissipation is necessary in a basic sense to reconcile the 
densities of ellipticals and spirals, this represents an important test of 
the merger hypothesis itself, as well as a test of our understanding of galaxy merger 
histories and the evolution in different galaxy components with redshift. 
In addition, we provide a unified 
theoretical lens through which to interpret a number of 
observations of spheroid and black hole-host 
scalings at both $z=0$ and high redshift, in which spheroids are fundamentally 
multi-component objects (with dissipational components originally formed 
in central starbursts triggered in the initial gas-rich interactions that formed 
the system, and dissipationless components representing the scattered, 
violently relaxed stars from the pre-initial merger stellar disks). 
As multi-component objects, it is primarily the relative mass fraction in the 
dissipational component, reflecting the gas content of spheroid progenitors, 
that drives their structural properties and the evolution of those structural 
properties with redshift.

\acknowledgments We thank Tod Lauer, John Kormendy, Sadegh Khochfar, 
Marijn Franx, Ivo Labbe, Norm Murray, Chien Peng, 
and Barry Rothberg for helpful discussions and contributed data sets 
used in this paper. This work
was supported in part by NSF grants ACI 96-19019, AST 00-71019, AST
02-06299, and AST 03-07690, and NASA ATP grants NAG5-12140,
NAG5-13292, and NAG5-13381. Support for 
TJC and SW was provided by the W.~M.\ Keck 
Foundation.

\bibliography{/Users/phopkins/Documents/lars_galaxies/papers/ms}

%\clearpage

\begin{appendix}
%\appendixcolumns

\section{Implementation of the Halo Occupation Model}
\label{sec:appendix}

Here, we present a condensed outline of the simplest implementation of the model 
used to make the predictions in this paper. As described in the text 
(\S~\ref{sec:model}), we have experimented with a wide variety of 
modifications to the model elements and methodology -- but our intention here 
is to summarize
the basic framework upon which these modifications represent increasing 
layers of complexity. We therefore leave the 
description of these experiments to \S~\ref{sec:model} and 
\citet{hopkins:groups.qso}. 

As outlined in \S~\ref{sec:model:sims:summary}, our methodology consists of a 
few key steps: 

We construct a Monte Carlo sample of halos at some observed redshift 
$z_{\rm obs}$, sampling according to the halo mass function at that 
redshift \citep[constructed in standard fashion for the adopted cosmology 
following][]{shethtormen}. 

For each halo, we determine a mock growth history, i.e.\ 
$M_{\rm halo}(z)$ for all $z>z_{\rm obs}$ up to some 
initial (maximum) redshift $z_{\rm init}$. There are several 
ways to do this: using the extended Press-Schechter formalism, 
adopting fits to individual halo growth histories in simulations 
\citep[taken from e.g.][]{delucia:ell.formation,
stewart:mw.minor.accretion} or directly tracking the main-branch 
progenitor halo mass of each $z=0$ halo in the simulations, 
integrating over analytic fits to mean merger histories from 
initial seed populations \citep{fakhouri:halo.merger.rates}, or adopting the mean 
$M_{\rm halo}(z | M_{\rm halo}[z=z_{\rm obs}])$ for a given $z=z_{\rm obs}$ 
population of halos 
fitted in e.g.\ \citet{wechsler:concentration,wechsler:assembly.bias,
neistein:natural.downsizing}. 
Because we are considering the population in a 
{\em statistical} fashion, it makes no difference which of these 
approaches we adopt so long as they yield a similar 
median (and the scatter in galaxy properties in a given halo is, 
in any case, larger than the scatter at different times between 
these methodologies). The simplest approach, then, is to 
consider each halo in the Monte Carlo sample to have a mass 
at each $z>z_{\rm obs}$ given by the mean growth 
history \citep[adopting the analytic fits in][in 
their Equation~11 and Appendices]{neistein:natural.downsizing}.  

Now we have an average $M_{\rm halo}(z)$ for each halo in 
our representative Monte Carlo sample. 
We then begin at some initial redshift $z_{\rm initial}$ 
and evolve the system forward in timesteps of 
some $\Delta z$.\footnote{Here, we choose $z_{\rm initial}=6$ and 
$\Delta z=0.01$, but in general we find that 
our results converge with respect to $\Delta z$ for values 
$\Delta z \lesssim 0.1$, and the predictions at any given 
$z_{\rm obs}$ converge rapidly once $z_{\rm initial}$ 
is larger by some difference $z_{\rm initial}-z_{\rm obs}\gtrsim 1-2$. 
The reasons for this are discussed in the text (\S~\ref{sec:model:hod}) and below.}
We ignore halos until they reach a mass (our effective resolution limit, 
corresponding to resolution limits in many of the simulations
on which the HOD calculations are based) of 
$M_{\rm halo}=10^{10}\,\msun$. This corresponds, given a typical HOD, to 
an extremely small stellar mass $M_{\ast}\lesssim10^{7}\,\msun$, so 
is irrelevant for the final mass of all the galaxies considered here, and 
is a source of little uncertainty. 

At each timestep, all galaxies that have not yet experienced a major merger 
above the resolution limit are initialized according to the halo occupation 
model (specifically, we 
initialize the galaxy stellar mass $M_{\ast}(M_{\rm halo})$ and 
then determine $R_{e}(M_{\ast})$ and $f_{\rm gas}(M_{\ast})$). 
As described in the text, we have experimented with a variety of 
observational constraints regarding the implementation of the HOD 
\citep[see][for a more detailed comparison]{hopkins:groups.qso}. It is possible, 
for example, to use the quoted fits from \citet{conroy:monotonic.hod} 
at each of several redshifts, and linearly interpolate between each redshift 
where the HOD was fitted to apply it to our model. 
It is also possible to use the methodology in that 
paper and in \citet{valeostriker:monotonic.hod} -- monotonically ranking galaxies 
in stellar mass and halos in either mass or circular velocity, and assigning 
them to one another in one-to-one correlation -- together with 
an analytic redshift-dependent fit to the galaxy stellar mass function 
(extending to high redshift), such as that in \citet{fontana:highz.mfs}, to 
obtain the HOD at each redshift. In practice, all of these applications yield 
similar results -- these observational comparisons and other 
direct measurements \citep[see e.g.][]{yan:clf.evolution,cooray:highz,
conroy:monotonic.hod,heymans:mhalo-mgal.evol,
conroy:mhalo-mgal.evol,brown:hod.evol} 
imply that the evolution in $M_{\ast}(M_{\rm halo})$ with 
redshift is weak (and only significant at the highest masses, 
where number densities drop sufficiently rapidly at high redshift 
as to make them a negligible contribution to the lower-redshift population). 
For this reason, we obtain nearly identical results using the simplest possible 
prescription: assuming $M_{\ast}(M_{\rm halo})$ is redshift-independent 
and applying the observed $z=0$ relation, assigning each halo 
galaxy a stellar mass in Monte Carlo fashion
\citep[compare the redshift-dependent fits in][who reach a similar conclusion]{conroy:hod.vs.z}. Specifically, we consider the fits from \citet{wang:sdss.hod}, for central galaxies
\begin{equation}
M_{\ast} =  M_{1}\,{\Bigl [}(M_{\rm halo}/M_{0})^{-\alpha} + (M_{\rm halo}/M_{0})^{-\beta} {\Bigr ]}^{-1}
\label{eqn:hod.fit}
\end{equation}
with $(M_{0},\,M_{1},\,\alpha,\,\beta)=(3.16\times10^{11}\,h^{-1}\,\msun,\,
4.48\times10^{10}\,\msun,\,0.39,\,1.96)$ and a 
lognormal scatter with $\sigma=0.148$\,dex dispersion at each $M_{\rm halo}$
\footnote{The ``turnover'' in $M_{\ast}(M_{\rm halo})$ in Equation~(\ref{eqn:hod.fit}) mainly 
corresponds to quenched, spheroid galaxies, and so one could argue 
should not be applied to a sample of strictly un-merged galaxies. 
We have experimented with not including this turnover, i.e.\
adopting the low-mass slope of the HOD with $M_{\ast}\propto M_{\rm halo}^{1.96}$, 
with a maximum at a stellar mass equal to the halo mass times the 
universal baryon fraction. Because very few systems survive to such large 
masses without merging, however, this makes no difference to our predictions.}.

We then assign each disk an effective 
radius $R_{e}(M_{\ast}\,|\,z)$ 
according to the methodology in \S~\ref{sec:model:hod}, where 
$R_{e}(M_{\ast}\,|\,z=0)\approx3.4\,(M_{\rm disk}/10^{10}\,\msun)^{0.3}$\,kpc 
from \citet{shen:size.mass} (converted 
to our adopted cosmology and stellar IMF) 
and $R_{e}(M_{\ast}\,|\,z)=(1+z)^{-\beta_{d}}\,R_{e}(M_{\ast}\,|\,z=0)$ 
as per Equation~(\ref{eqn:rdisk.evol}) (and we consider 
both $\beta_{d}=0$ -- no evolution -- and $\beta_{d}=0.4$ -- the 
evolution suggested by observations -- showing the 
difference between the two in the range of uncertainties in 
our model). The gas fractions are assigned in similar fashion, 
with the $z=0$ observed relation in Equation~(\ref{eqn:fgas.z0}) and 
appropriate redshift evolution in Equation~(\ref{eqn:fgas.z}) (where 
we again consider a range to allow for observational uncertainties 
as described in the text, specifically $\beta=0.5-2.0$ in 
Equation~\ref{eqn:fgas.z}). 

These assignments are re-initialized at each redshift, if 
the halo remains un-merged. For this reason (and because at 
any redshift, most galaxies of a given mass have assembled 
in the last redshift interval $\Delta z\sim1-2$), the large uncertainties 
at high redshift are quickly suppressed at any lower redshift. 
For example, even though the uncertainty in $R_{e}$ of a typical 
disk at high redshift (e.g.\ $z=4$) is large (say, $0.5\,$dex), and this 
will enter as such in the uncertainty in predicted size of an 
elliptical forming at this redshift (although the uncertainty is 
somewhat suppressed since the system is likely to be gas-rich; see 
the discussion in \S~\ref{sec:model}), by $z=0$, 
the original uncertainty will be suppressed by a typical 
factor $\sim16-32$ (depending on how much the galaxy has grown 
by low-redshift mergers), introducing only $\sim 0.02-0.03$\,dex 
uncertainty in the low-redshift prediction (in other words, at lower 
redshifts, the uncertainties are dominated by the uncertainties in 
the HOD around those redshifts, rather than by the propagation of 
uncertainties from higher redshifts). 

The critical step at each redshift is then determining whether or 
not the galaxy experiences a merger. 
Because we are interested in galaxy-galaxy mergers, 
rather than e.g.\ a subhalo merging into the parent halo 
(and the two can be very different, as many subhalos, especially 
those which are small relative to the parent halo, 
may not merge for a Hubble time), we do not wish to 
adopt an extended Press-Schechter or 
simulation-based merger tree, but opt for a more sophisticated 
approach. At each timestep, we consider the halo to have a subhalo 
population according to that fitted in simulations 
\citep{kravtsov:subhalo.mfs} 
\citep[alternatively, following the fits in][who adopt a semi-analytic 
methodology but reach similar conclusions]{vandenbosch:subhalo.mf} 
-- this subhalo mass function is, in units of 
$M_{\rm subhalo}/M_{\rm halo}$, only weakly halo mass 
or redshift-dependent, and it makes no difference whether 
we adopt the fits from different authors 
\citep[we have also compared the subhalo mass functions in][and reach similar 
conclusions]{springel:cluster.subhalos,
tormen:cluster.subhalos,delucia:subhalos,gao:subhalo.mf,
zentner:substructure.sam.hod,nurmi:subhalo.mf}. 
We then calculate the time for each such subhalo and 
its contained galaxy to merge with the central galaxy, according 
to one of several methodologies: either assuming a simple 
dynamical friction timescale, employing a more sophisticated 
timescale based on the cross section for resonant galaxy-galaxy 
interaction and orbital capture, or a 
timescale calculated based on a similar calculation in 
angular momentum space; the probability of a merger with 
each subsystem is then taken as the ratio of the timestep 
to the merger timescale. We compare these methodologies 
in detail in \citet{hopkins:groups.qso} and \citet{hopkins:groups.ell} 
and discuss the details of the derivation for each. For our purposes 
here, as demonstrated in those papers, they give similar results, 
and we note that the results are also similar to what is commonly 
adopted in many semi-analytic models based on N-body simulations, 
which follow subhalos until they can no longer be resolved and then 
approximate the remaining merger timescale with a dynamical 
friction timescale estimate. 

In \citet{hopkins:groups.qso} we show the detailed PDFs for 
the merger rate as a function of halo mass, galaxy mass, 
and redshift (see specifically Figures~2-4 therein), 
that arise from various combinations of these 
assumptions. We show there that 
the different approaches agree reasonably well, and agree 
with independent estimates such as sufficiently high resolution 
N-body experiments \citep[e.g.][]{maller:sph.merger.rates} 
and semi-analytic models with similar merger timescale 
calibrations from numerical tests \citep{somerville:new.sam}. 
For our purposes here, the important thing is that
\citet{hopkins:groups.qso} show that the model predictions 
agree well with the observed mass function of 
galaxy mergers at all redshifts observed; as well as
the observed galaxy merger rate as a function of 
environmental density, halo mass (estimated from 
clustering or group dynamics), and redshift; 
and the clustering (both large-scale and small-scale) 
and therefore halo occupation of mergers and 
recent merger remnants (see their Figures~6-12). 
This gives us some confidence that the calculation yields a 
reasonable approximation to the merger rate. 
For the details and summary of the adopted calculation, 
we refer to \S~2.1 of \citet{hopkins:groups.qso}. 
For our fiducial model in this paper, we employ their ``default'' 
model: we construct the subhalo mass function for each halo 
adopting the analytic fits in \citet{kravtsov:subhalo.mfs}. 
We populate each subhalo as if it were a random member of 
our halo population (of the same mass) -- specifically, we 
draw a random halo of the same mass from our Monte Carlo 
sample and assign the subhalo the same properties\footnote{
So long as the Monte Carlo sample is sufficiently large, 
this is identical to tracing the growth and merger history of the 
subhalo itself. As discussed in the text, we have explicitly chosen 
to ignore effects unique to satellites such as ram-pressure 
stripping. Although these are important for low-mass satellites 
which are long-lived in e.g.\ massive groups and clusters, 
they do not generally apply to the cases of interest: namely major 
mergers, for which the merger times are short (too short for 
long-timescale effects such as stripping and harassment) and the halos are of comparable 
mass (making e.g.\ ram-pressure relatively unimportant). Moreover, 
for the typical $\lesssim$ a few $L_{\ast}$ galaxies of interest here, 
the ``subhalo'' phase is very short in major mergers -- it represents a brief 
intermediate merging stage between field or small/loose group galaxies 
on their way to merging. For more massive cluster systems, however, the 
caveat should apply, along with the others discussed in \S~\ref{sec:track}.}. 
We then estimate the merger timescale using the ``group capture'' 
gravitational cross-section timescales, specifically using 
the fitting formulae from \citet{krivitsky.kontorovich} calibrated to a 
large set of numerical simulations of different encounters in group environments 
\citep[see also][]{white:cross.section,makino:merger.cross.sections,mamon:groups.review}. 
This ratio of the timestep to this timescale gives the probability of each merger, 
and we determine whether each occurs in Monte Carlo fashion. 

When a merger occurs above the cutoff mass ratio threshold (a ``major'' merger), 
the properties of the remnant are determined as a function of the 
properties of both progenitors, as described in \S~\ref{sec:model:sims}. 
The final stellar mass is the sum of the two progenitor baryonic masses; 
the two constituent components, the dissipational and dissipationless 
stellar mass, of the remnant are the sum of the dissipational and 
dissipationless masses of the progenitors. For un-merged 
progenitors, the dissipational and dissipationless mass are the 
pre-merger gas and stellar mass, respectively. The effective radii of the 
remnants are calculated as described in the text, for the dissipationless 
component as a function of the mass ratio and scale radii of the progenitors 
(Equation~\ref{eqn:re.dry}),
and then with the appropriate correction applied for the dissipational 
mass fraction (Equation~\ref{eqn:re.wet}). Other properties follow 
as described in \S~\ref{sec:model:sims}. Properties such as the dark matter 
fraction within a given radius (for dynamical masses, etc) are calculated 
assuming a \citet{hernquist:profile} profile for the dark matter halo, with a 
redshift-dependent concentration from \citet{bullock:concentrations}, and 
projecting the profile along with that of the galaxy. 

The stellar mass and stellar mass profile of major merger remnants 
are taken to be those calculated at the previous merger, although the 
halo grows continuously with time. In other words, major merger 
remnants are effectively ``quenched'' in some sense (although new gas can and 
does come into the galaxy through subsequent mergers with 
gas-rich galaxies) -- we ignore details of the re-growth of disks around 
early-forming ellipticals through direct cosmological accretion of cold gas. 
Although this may be important for especially the most high-redshift 
systems, a proper treatment requires a much more complete cosmological 
model (ideally highly resolved cosmological simulations that can form 
proper low-redshift disks, still a major theoretical challenge). Moreover, 
it has been argued that feedback from quasar and starburst activity in 
major mergers may actually be an important physical agent of 
``quenching,'' \citep{scannapieco:sam,granato:sam,monaco:feedback,
hopkins:red.galaxies,hopkins:transition.mass,hopkins:clustering,naab:dry.mergers,
bundy:agn.lf.to.mf.evol}, 
in which case this is a more accurate assumption than allowing for 
new cooling. 

In any event, we show in \citet{hopkins:groups.ell} that 
such a prescription, with a major merger 
mass ratio threshold set to the value used in this paper (1:3), 
yields very good agreement with the observed mass function and mass density of 
quenched galaxies as a function of redshift, as well as 
quenched central galaxy fraction as a bivariate function of 
stellar and halo mass, the clustering of red 
galaxies as a function of mass and redshift, the distribution of stellar 
population ages (and implicitly, ``quenching times'') as a function of 
elliptical stellar mass, and the distribution of elliptical structural properties 
related to cooling. This argues that the current prescription is at least a 
good approximation to the actual cooling histories of the galaxies 
of interest here. We have also experimented with alternative 
prescriptions such as those found in semi-analytic models 
\citep[with halo cooling suppressed by low-level AGN feedback, or 
truncated above a certain halo or stellar mass threshold, 
see e.g.][]{croton:sam,bower:sam,cattaneo:sam,delucia:sam,somerville:new.sam}, and find 
similar results \citep[this is because, it turns out, the halo mass threshold 
for the transition to ``hot mode'' accretion and effective quenching in these 
models, and empirically constrained stellar/halo mass ranges for 
quenching, correspond quite closely with the regime where most galaxies 
experience their first major merger; see e.g.][]{cattaneo:sam}. 

This process is repeated until the desired $z_{\rm obs}$ is reached. 
We then discard the systems that remain un-merged (as these will not, 
predominantly, be elliptical galaxies) and construct the mock observed 
sample of interest from our Monte Carlo population.

\end{appendix}

\end{document}